\documentclass[a4paper,11pt]{article}
\makeatletter
\g@addto@macro\bfseries{\boldmath}
\makeatother
\usepackage[english]{babel}
\usepackage{jheppub}
\pdfoutput=1
\usepackage[T1]{fontenc} 
\usepackage{bbold}
\usepackage{amsmath}
\usepackage{empheq}
\usepackage{booktabs}
\usepackage{color}
\usepackage[utf8]{inputenc}
\usepackage{xspace}
\usepackage{scalerel}
\usepackage[most]{tcolorbox}
\usepackage{multirow}
\usepackage{scalefnt}
\usepackage{bold-extra}

\DeclareRobustCommand{\ensuremathrm}[1]{\ensuremath{\mathrm{#1}}\xspace}

\DeclareRobustCommand{\rd}{\ensuremathrm{d}} 

\newcommand\tS{\tilde{S}}
\newcommand\F{${\rm F}$}
\newcommand\FJ{${\rm FJ}$}
\newcommand\FJJ{${\rm FJJ}$}
\newcommand\PhiB{\Phi_{\scriptscriptstyle \rm F}}
\newcommand\PhiBres{\Phi_{\scriptscriptstyle \rm F,res}}
\newcommand\PhiBJ{\Phi_{\scriptscriptstyle \rm FJ}}
\newcommand\PhiBJbar{{\bar \Phi}'_{\scriptscriptstyle \rm FJ}}
\newcommand\PhiBJJ{\Phi_{\scriptscriptstyle \rm FJJ}}
\newcommand{\Fcorr}{F^{\tmop{corr}}_\ell}
\newcommand{\order}[1]{{\cal O}\left(#1\right)}
\newcommand{\as}{\alpha_s}

\newcommand{\kt}[1]{k_{\scaleto{\rm T}{4pt},#1}}
\newcommand{\veckt}[1]{\vec{k}_{\scaleto{\rm T}{4pt},#1}}
\newcommand{\dk}[1]{\langle \mathd k_{#1}\rangle}

\newcommand{\pt}{{p_{\text{\scalefont{0.77}T}}}}
\newcommand{\ptrad}{{p_{\text{\scalefont{0.77}T,rad}}}}
\newcommand{\pth}{{p_{\text{\scalefont{0.77}T,H}}}}
\newcommand{\ptz}{{p_{\text{\scalefont{0.77}T,Z}}}}
\newcommand{\ptj}{{p_{\text{\scalefont{0.77}T,J}}}}
\newcommand{\ptl}{{p_{\text{\scalefont{0.77}T,$\ell^-$}}}}
\newcommand{\yh}{{y_{\text{\scalefont{0.77}H}}}}
\newcommand{\yz}{{y_{\text{\scalefont{0.77}Z}}}}
\newcommand{\yl}{{y_{\text{\scalefont{0.77}$\ell^-$}}}}
\newcommand{\mh}{{m_{\text{\scalefont{0.77}H}}}}
\newcommand{\mz}{{m_{\text{\scalefont{0.77}Z}}}}
\newcommand{\mw}{{m_{\text{\scalefont{0.77}W}}}}
\newcommand{\mt}{{m_{\text{\scalefont{0.77}t}}}}
\newcommand{\muF}{{\mu_{\text{\scalefont{0.77}F}}}}
\newcommand{\muR}{{\mu_{\text{\scalefont{0.77}R}}}}
\newcommand{\KF}{K_{\text{\scalefont{0.77}F}}}
\newcommand{\KR}{K_{\text{\scalefont{0.77}R}}}

\newcommand{\noun}[1]{{\scshape #1}}

\newcommand{\POWHEG}{\noun{POWHEG}}

\newcommand{\POWHEGBOX}{\noun{POWHEG-BOX}}

\newcommand{\minlobare}{{\noun{MiNLO}}}
\newcommand{\minlo}{{\noun{MiNLO$^{\prime}$}}}
\newcommand{\minnlo}{{\noun{MiNNLO$_{\rm PS}$}}}
\newcommand{\Matrix}{{\noun{Matrix}}}
\newcommand{\PYTHIA}[1]{\noun{Pythia{#1}}}
\newcommand{\Herwigpp}{\noun{Herwig++}}

\newcommand{\nnlops}{NNLO+PS}
\newcommand{\nlops}{NLO+PS}

\newcommand{\abar}{\frac{\as}{2\pi}}
\newcommand{\abarmu}[1]{\frac{\as(#1)}{2\pi}}

\newcommand{\citere}[1]{ref.\,\cite{#1}}
\newcommand{\citeres}[1]{refs.\,\cite{#1}}
\newcommand{\eqn}[1]{eq.\,(\ref{#1})}

\newcommand{\fig}[1]{figure\,\ref{#1}}

\newcommand{\tab}[1]{table\,\ref{#1}}
\newcommand{\sct}[1]{section~\ref{#1}}

\newcommand{\app}[1]{appendix~\ref{#1}}

\newcommand{\LambdaPWG}{\Lambda_{\rm pwg}}

\usepackage{xcolor}
\newcommand{\mathd}{\mathrm{d}}
\newcommand{\tmop}[1]{\ensuremath{\operatorname{#1}}}

\newtcolorbox{empheqboxed}{colback=white!35, 
 colframe=black,
 width=\textwidth,
 sharpish corners,
 top=-2mm, 
 bottom=0pt
}

\title{{\sc MiNNLO$_{\rm PS}$}: A new method to match NNLO
  QCD to parton showers}
\title{{M{\scalefont{0.77}I}NNLO$_{\text{PS}}$}: A new method to match NNLO QCD to parton showers}

\author[a]{Pier Francesco Monni,}
\author[b]{Paolo Nason,}
\author[a,c]{Emanuele Re,}
\author[a,d]{Marius Wiesemann}
\author[d]{and Giulia Zanderighi}

\emailAdd{pier.monni@cern.ch}
\emailAdd{paolo.nason@mib.infn.it}
\emailAdd{emanuele.re@lapth.cnrs.fr}
\emailAdd{marius.wiesemann@cern.ch}
\emailAdd{zanderi@mpp.mpg.de}

\affiliation[a]{CERN, Theoretical Physics Department, CH-1211 Geneva 23, Switzerland}

\affiliation[b] {Universit\`a di Milano\,-\,Bicocca and INFN, Sezione di
  Milano\,-\,Bicocca, Piazza della Scienza 3, 20126 Milano, Italy}

\affiliation[c]{LAPTh, Universit\'e Grenoble Alpes, Universit\'e Savoie Mont Blanc, CNRS, 74940 Annecy, France}

\affiliation[d]{Max-Planck-Institut f\"ur Physik, F\"ohringer Ring 6,
  80805 M\"unchen, Germany}

\abstract{We present a novel method to combine QCD calculations at
  next-to-next-to-leading order (NNLO) with parton shower (PS)
  simulations, that can be applied to the production of heavy systems 
  in hadronic collisions, such as colour singlets or a
  $t\bar{t}$ pair. 
  The NNLO corrections are included by connecting the \minlo{} method
  with transverse-momentum resummation, and they are calculated at
  generation time without any additional reweighting, making the
  algorithm considerably efficient.
  Moreover, the combination of different jet multiplicities does not
  require any unphysical merging scale, and the matching preserves the
  structure of the leading logarithmic corrections of the Monte Carlo
  simulation for parton showers ordered in transverse momentum. We
  present proof-of-concept applications to hadronic Higgs production
  and the Drell-Yan process at the LHC.}

\keywords{Perturbative QCD, NLO computations, Resummation}

\preprint{\\CERN-TH-2019-117\\ LAPTH-042/19\\ MPP-2019-177}

\begin{document}

\maketitle

\section{Introduction}
\label{sec:intro}

Particle phenomenology at the Large Hadron Collider (LHC) has entered
the precision era.  After the landmark discovery of the Higgs boson
\cite{Aad:2012tfa,Chatrchyan:2012xdj}, which explains the electroweak
(EW) symmetry breaking and completes the particle content predicted by
the Standard Model (SM), the LHC is now focussing upon the search for
hints of new-physics phenomena. Despite several indications that there
must be physics beyond the SM (BSM) at relatively low scales, as of
now new-physics searches at the LHC have been unsuccessful.
In this scenario, precision measurements have become of foremost
importance, as they enhance the sensitivity of indirect searches
for new physics through small deviations from SM predictions.

Modifications of the electroweak sector, induced by many BSM theories,
can be unravelled through precision studies of electroweak
interactions. Both the production of a Higgs-boson and of EW vector
bosons play a crucial role in this respect.
Notably, for the Drell-Yan process the experimental uncertainties have
already surpassed the percent
level~\cite{Aaboud:2017ffb,Sirunyan:2017igm,Aaboud:2017svj}.
On the theory side, predictions have to be controlled at the same
level of precision, which demands calculations with at least
next-to-next-to-leading order (NNLO) accuracy in QCD perturbation
theory. Furthermore, the experimental measurements operate at the
level of hadronic events and require full-fledged Monte Carlo
simulations in their analyses. The inclusion of NNLO QCD corrections
in event generators is therefore mandatory to fully exploit LHC data.

In this paper, we present a novel method to perform the consistent
matching of NNLO calculations and parton showers (hereafter
\nnlops{}), based on the structure of transverse-momentum
resummation.
Our method builds upon \minlobare{}~\cite{Hamilton:2012np}, a procedure
for improving NLO multijet calculations with the appropriate choice of scales
and with the inclusion of Sudakov form factors, that is particularly suited
to be interfaced with parton-shower generators using the \POWHEG{} method.
In ref.~\cite{Hamilton:2012rf} the \minlobare{} procedure
was refined in such a way that, in processes involving the production
of a massive colour singlet system in association with one jet, the NLO
accuracy is formally retained also for observables inclusive in
the jet. Such procedure, dubbed~\minlo{}, yields an NLO multi-jet merging
method that does not use a merging scale. Notably, a numerical method
to extend the~\minlo{} procedure to more complex processes, and its application to Higgs production
in association with up to two jets, was presented
in ref.~\cite{Frederix:2015fyz}.

%
In this article we extend the \minlo{} method to achieve NNLO accuracy
at the fully differential level in the zero-jet phase space, while
retaining NLO accuracy for the one-jet configurations. Our method will
be referred to as \minnlo{} in the following, and it has the following
features:
\begin{itemize}
\item NNLO corrections are calculated directly during the generation
  of the events, with no need for further reweighting.
\item No merging scale is required to separate different
multiplicities in the generated event samples. 
\item When combined with transverse-momentum ordered parton showers,
  the matching preserves the leading logarithmic structure of the
  shower simulation.
\end{itemize}
Maintaining the logarithmic accuracy of the shower is a crucial
requirement of all NLO+PS, and {\rm a fortiori} NNLO+PS approaches.
We stress that this requirement is immediately met by the \minnlo{}
approach, that works by generating the first two hardest emissions and
letting the shower generate all the remaining ones.  We also recall
that, if the shower ordering variable differs from the \nlops{} one,
maintaining the Leading Logarithmic accuracy of the shower becomes a
delicate issue.  An example is given by a POWHEG based generator
interfaced to an angular-ordered shower. To preserve the accuracy of
the shower, not only one needs to veto shower radiation that has
relative transverse momentum greater than the one generated by POWHEG,
but also one has to resort to truncated showers to compensate for
missing collinear-soft radiation. Failing to do so spoils the shower
accuracy at leading-logarithmic level (in fact, at the
double-logarithmic level).\footnote{ Truncated shower were first
  discussed in ref.~\cite{Nason:2004rx}, and first implemented in the
  \Herwigpp{} context in ref.~\cite{Hamilton:2008pd}.  Currently
  \Herwigpp{} implements them in its internal \nlops{}, POWHEG-scheme
  processes (see ref.~\cite{Bahr:2008pv}). They have also been used in
  a slightly different context in
  refs.~\cite{Hoeche:2009rj,Hoche:2010kg}.}

Three different \nnlops{} approaches have been previously formulated
in the literature~\cite{Hamilton:2012rf,Alioli:2013hqa,Hoeche:2014aia}
and applied to the simplest LHC processes, namely
Higgs-boson production~\cite{Hamilton:2013fea,Hoche:2014dla} and the
Drell-Yan
process~\cite{Hoeche:2014aia,Karlberg:2014qua,Alioli:2015toa}. The approach
of ref.~\cite{Hamilton:2012rf} shares all
features listed above except the first one, i.e. it requires a
multi-dimensional reweighting of the MiNLO' samples in the Born phase
space to achieve NNLO accuracy. It has been recently applied to more
complicated LHC processes, such as the two Higgs-strahlung
reactions~\cite{Astill:2016hpa,Astill:2018ivh}, and the production of
two opposite-charge leptons and two neutrinos
($W^+W^-$)~\cite{Re:2018vac}.\footnote{The $W^+W^-$ simulation is
  based on the \minlo{} calculation of ref.~\cite{Hamilton:2016bfu},
  and the NNLO calculation of ref.~\cite{Grazzini:2016ctr} performed
  within the \Matrix{} framework~\cite{Grazzini:2017mhc}.}  These
computations have employed the reweighting procedure to its extreme.
Despite yielding physically sound results, the reweighting in the
high-dimensional Born phase space of these processes poses substantial
technical limitations.  Apart from the numerical demand of the
reweighting itself and certain approximations that had to be made in
these calculations, the discretisation of the Born phase space through
finite bin sizes of the reweighted observables reduces the
applicability of the results in phase-space regions with coarse
binning, usually located in the least populated regions of phase space
(e.g. in the tails of the kinematic distributions).
In fact, the numerical limitation of the reweighting constitutes a
problem already for the simpler Drell-Yan process, since the
experiments require a considerably large number of generated events
for the current and future LHC analyses.

The \minnlo{} method presented in this paper lifts these
shortcomings, while retaining the same advantages of a \minlo{}
computation. The terms relevant to achieve NNLO accuracy are obtained
by connecting the \minlo{} formula with the momentum-space resummation
of the transverse-momentum spectrum formulated
in~\citeres{Monni:2016ktx,Bizon:2017rah}. This allows us to make a
direct link between the resummation and the \POWHEG{}
procedure~\cite{Nason:2004rx}, resulting in a consistent
\nnlops{} formulation.

Due to the substantially improved numerical efficiency compared to the
reweighting approach, the \minnlo{} method allows us to tackle without
any approximations the full class of complex color-singlet final
states, such as the highly relevant four-lepton (vector-boson pair
production) processes. As proof-of-concept applications, we consider
hadronic Higgs production and the Drell-Yan process at the LHC, and
compare our results against previous predictions. These computations
are implemented and will be made publicly available within the
\POWHEGBOX{}
framework~\cite{Nason:2004rx,Frixione:2007vw,Alioli:2010xd}.\footnote{Instructions
  to download the code will be soon made available at
  \url{http://powhegbox.mib.infn.it}.}  All-order, higher-twist, and
non-perturbative QCD effects are modelled through the interface to a
parton shower generator which provides a realistic simulation of
hadronic events.

Despite the fact that the formulae presented here are limited to the
hadro-production of heavy colour-singlet systems, our formalism is quite general,
and can be applied to other processes, such as the production of heavy
quarks.

The manuscript is organized as follows: In \sct{sec:description} we
describe in general terms the main idea behind the \minnlo{} approach,
and determine the relevant corrections to the \minlo{} formulation
necessary to reach NNLO accuracy. Practical aspects of the
implementation of these new terms within the \minlo{} framework are
discussed in \sct{sec:implementation}. In \sct{sec:NNLOPS_formal} we
provide a more rigorous derivation of the \minnlo{} method by starting
from the momentum-space resummation formula for the
transverse-momentum spectrum. Our proof-of-concept computations for
Higgs and Z-boson production are presented in \sct{sec:results}, where
we provide a full validation against existing results. We summarize
our findings in \sct{sec:summary}. A number of technical details and
explicit formulae are summarized in appendices~\ref{app:spreading} to
\ref{app:bspace}.

\section{Description of the procedure}
\label{sec:description}
In this section we describe the procedure to perform a consistent
matching of a NNLO QCD calculation for the production of a heavy
colour-singlet system to a fully exclusive parton-shower
simulation.
We start by recalling the necessary elements of the
\minlo{} method in section~\ref{sec:accuracy} and~\ref{sec:minloacc0}, while in
section~\ref{sec:procedure} we derive the additional terms necessary
to achieve NNLO accuracy.

\subsection{The \minlo{} method}
\label{sec:accuracy}
We review now the basic elements of the \minlo{} method, and how it achieves NLO
accuracy. We formulate it in a way that is as independent as possible
from the details of the implementation.

We consider the production of a generic colour-singlet system \F{} of
invariant mass $Q$ and transverse momentum $\pt$ in hadronic
collisions.
We start with the \minlo{} formula~\cite{Nason:2004rx,Hamilton:2012rf} for an 
arbitrary infrared-safe observable $O$, embedded in 
the \POWHEG{} method~\cite{Nason:2004rx,Frixione:2007vw} as follows
\begin{align}
  \langle O \rangle = \int \mathd \PhiBJ{} \mathd \Phi_{\tmop{rad}} 
  \bar{B} (\PhiBJ{}) \left[\Delta_{\rm pwg} (\LambdaPWG) O (\PhiBJ)+  \Delta_{\rm pwg} (\ptrad)  \frac{R (\PhiBJ{}, \Phi_{\tmop{rad}})}{B
  (\PhiBJ{})} O (\PhiBJJ) \right], 
\label{eq:Ominlo}
\end{align}
where
\begin{align}
  \bar{B} (\PhiBJ{}) & = e^{- \tilde{S}(\pt)} \left[B (\PhiBJ{}) \left(1 + \frac{\as(\pt)}{2\pi}[\tilde{S}(\pt)]^{(1)}\right) + V
    (\PhiBJ{})\right] \nonumber \\
  & + \int \mathd \Phi_{\tmop{rad}} R (\PhiBJ{}, \Phi_{\tmop{rad}}) e^{-
  \tilde{S} (\pt)}\,.
\label{eq:bbarminnlo}
\end{align}
Equation~\eqref{eq:Ominlo} is accurate up to NLO both in the zero and
one jet configurations. $B$ denotes the differential cross section for
the production of \F{} plus one light parton (\FJ{}), and $V$ and $R$
are the UV-renormalised virtual and real corrections to this process,
respectively.  The $V(\PhiBJ{})$ term in eq.~(\ref{eq:bbarminnlo}) is
infrared divergent, and so is the integral of $R$. These divergences
cancel in their sum, so that $\bar{B}$ is infrared finite.
$\Delta_{\rm pwg}$ denotes the usual \POWHEG{} Sudakov form factor,
$\LambdaPWG$ is an infrared cutoff of the order of a typical hadronic
scale, and $\ptrad$ corresponds to the transverse momentum of the
secondary emission associated with the radiation variables
$\Phi_{\tmop{rad}}$.  $\tilde{S}(\pt)$ stands for the \minlo{} Sudakov
form factor~\cite{Hamilton:2012rf}, that is evaluated using the
kinematics $\PhiBJ$ in the Born and virtual terms, and with the full
real kinematics $\PhiBJJ$ in the real term.  The factor
$(1 + \frac{\as(\pt)}{2\pi}[\tilde{S}(\pt)]^{(1)})$ is the first order expansion of the
inverse of the \minlo{} Sudakov form factor, necessary to avoid any
source of double counting in eq.~\eqref{eq:Ominlo}.  The \minlo{}
procedure specifies that the scale at which the strong coupling
constant and the parton densities are evaluated should be equal to
that contained in the Sudakov form factor, that we take to be
the transverse
momentum $\pt$ of the colour-singlet system \F{}.\footnote{If the colourless system is produced
  via strong interactions, as it is the case for Higgs-boson
  production, the extra powers of $\as$ are evaluated at a scale
  related to the mass of the heavy colourless system. For the sake of
  simplicity, and to avoid confusion, for the time being we will focus
  upon cases in which the production is of electroweak origin.}  It
also specifies that the transverse momentum appearing in the Sudakov
form factor that multiplies $R$ in eq.~(\ref{eq:bbarminnlo}) should be
the one of the real kinematics configuration $\PhiBJJ$, which differs
from the one appearing elsewhere in the formula, that is relative to
the underlying Born kinematics $\PhiBJ$. It turns out that in the
singular regions of the secondary emission the transverse momentum of
\F{} in the real and underlying Born kinematics become identical, so
that the cancellation between the collinear and soft singularities can
occur.

For simplicity of notation, eqs.~\eqref{eq:Ominlo}
and~\eqref{eq:bbarminnlo} refer to the case in which there is only one
singular region for the secondary emission.  In general, there are
singularities both in the initial-state (\POWHEG{} handles the two
initial-state regions together), and in the final-state.  \POWHEG{}
deals with the multiple singular regions by partitioning the real
matrix elements, as discussed in detail in \citere{Frixione:2007vw}.

To simplify the discussion that follows, without loss of generality,
we ignore the first (Sudakov suppressed) term on the right-hand side
of \eqn{eq:Ominlo} (this is simply done for the sake of clarity; this
term is always included in the \POWHEG{} implementations), and rewrite
the equation as
\begin{eqnarray}
  \langle O \rangle & = & \int \mathd \PhiBJ{} \mathd \Phi_{\tmop{rad}} 
  \bar{B} (\PhiBJ{}) \Delta_{\rm pwg} (\ptrad)  \frac{R (\PhiBJ{}, \Phi_{\tmop{rad}})}{B
  (\PhiBJ{})} O (\PhiBJ{}) \nonumber\\
  & + & \int \mathd \PhiBJ{} \mathd \Phi_{\tmop{rad}}  \bar{B} (\PhiBJ{}) \Delta_{\rm pwg}
  (\ptrad)  \frac{R (\PhiBJ{}, \Phi_{\tmop{rad}})}{B (\PhiBJ{})}  \{ O (\PhiBJJ) - O
  (\PhiBJ{}) \}, 
\end{eqnarray}
or equivalently
\begin{eqnarray}
  \langle O \rangle & = & \int \mathd \PhiBJ{}  \bigg\{ e^{- \tilde{S}(\pt)}
          \left [B (\PhiBJ{}) \left (1 + \frac{\as(\pt)}{2\pi}[\tilde{S}(\pt)]^{(1)}\right) + V (\PhiBJ{})\right] \nonumber \\
          &+& \int \mathd \Phi_{\tmop{rad}} R
  (\PhiBJ{}, \Phi_{\tmop{rad}}) e^{- \tilde{S} (\pt)} \bigg\} O (\PhiBJ{})
  \nonumber\\
  & + & \int \mathd \PhiBJ{} \mathd \Phi_{\tmop{rad}} e^{- \tilde{S} (\pt)} R
  (\PhiBJ{}, \Phi_{\tmop{rad}})  \{ O (\PhiBJJ) - O (\PhiBJ{}) \} +\mathcal{O}(\as^3).
 \label{eq:rsplit}
\end{eqnarray}
In the first two lines of eq.~\eqref{eq:rsplit}, the radiation
integral has been evaluated according to the usual unitarity
condition
\begin{equation}
\int \mathd \Phi_{\tmop{rad}} \Delta_{\rm pwg} (\ptrad)  \frac{R (\PhiBJ{},
   \Phi_{\tmop{rad}})}{B (\PhiBJ{})} = 1, 
\end{equation}
that is possible because the observable $O(\PhiBJ)$ does not depend
upon the radiation phase space $\Phi_{\tmop{rad}}$.
Owing to the infrared safety of the observable,
the last line of \eqn{eq:rsplit} has no singularities. This is obvious as far as
the secondary emission is concerned, since the difference between the
observables vanishes when it becomes unresolved. 
Singularities associated with the first emission, on the other hand, are suppressed
by the fact that the separation of regions in
\POWHEG{} \cite{Frixione:2007vw} 
ensures that the secondary emission is always more singular
than the first one. Therefore, the contribution of the last line of
eq.~\eqref{eq:rsplit} is of pure order $\as^2$, and it is
dominated by large scales.

We conclude that in order to achieve NNLO accuracy it is sufficient
to correct \eqn{eq:Ominlo} in such a way that it remains
unaltered at large $\pt$ (where it already has ${\cal O}(\as^2)$
accuracy), and is NNLO accurate for observables of the form $O(\Phi)=g(\PhiB(\Phi))$,
where $g$ is an arbitrary function, and
$\PhiB(\Phi)$ represents an infrared-safe projection of the 
kinematic configuration corresponding to a generic phase space $\Phi$
to the one where the transverse momentum of the colour singlet
vanishes ($\PhiB$). For instance, $\PhiB$ involves the rapidity of the 
colour-singlet system and its internal variables.

According to the \minlo{}
method, NLO accuracy is guaranteed if the $\bar{B}(\PhiBJ)$ function in \eqn{eq:bbarminnlo} 
is defined as a total derivative up to the relevant perturbative
order~\cite{Hamilton:2012rf}. As we will show in the next
section, achieving NNLO accuracy will require the inclusion of
additional terms in the $\bar{B}(\PhiBJ)$ function.

\subsection{\minlo{} accuracy}\label{sec:minloacc0}

In the \nnlops{} approach of ref.~\cite{Hamilton:2013fea}, NNLO accuracy is achieved by a reweighting procedure.
This is carried out by first computing the inclusive cross section at fixed kinematics of the colourless system
in the \minlo{} approach and at NNLO, and then by reweighting the events by the ratio
of the latter result to the former. This procedure works
regardless of the corrections that the \minlo{} approach already provides at the NNLO
level, since this is eventually divided out and replaced by the correct one.

In the present work we are not relying upon a reweighting procedure, and thus we need
to develop an analytic understanding of what \minlo{} provides at the NNLO order.
We do this by noticing that eq.~(\ref{eq:rsplit}) is equivalent to the following
equation:
\begin{eqnarray}
  \langle O (\Phi) \rangle &=& \int \mathd \PhiBJ{} e^{- \tilde{S}(\pt)} \left[B (\PhiBJ)
                               \left(1 + \frac{\as(\pt)}{2\pi}[\tilde{S}(\pt)]^{(1)}\right) + V (\PhiBJ)\right] O (\PhiBJ)
\nonumber \\
  &+& \int \mathd \PhiBJJ R (\PhiBJJ)
  e^{- \tilde{S}(\pt)} O (\PhiBJJ) . \label{eq:minloacc1}
\end{eqnarray}
up to terms of N$^3$LO order. This result follows
from the fact that the term involving $R$ in the second line of~(\ref{eq:rsplit}) cancels
exactly the last term in the curly bracket in the last line of~(\ref{eq:rsplit}).
On the other hand, eq.~(\ref{eq:rsplit}) is derived from the full \minlo{} result using
only the exact unitarity of the shower and of the POWHEG radiation implementation, and thus does not
introduce any fixed order approximation.

Therefore eq.~(\ref{eq:minloacc1})
represents analytically what \minlo{} provides at the $\as^2$ level.
It has unavoidably a formal character, with the virtual and real contributions that are separately
infrared divergent. As such, it is independent of the specific method used to
cancel infrared divergences. In particular, it does not depend upon the details of the
POWHEG implementation, such as the mapping between the real cross section and the underlying Born,
and it can thus be used to make direct contact with analytic resummation formulae.

\subsection{Reaching NNLO accuracy: the \minnlo{} method}
\label{sec:procedure}
In this section we present a simple derivation of the missing terms
needed to reach NNLO accuracy in the \minlo{} formula. For the
interested reader, we report a detailed and more rigorous derivation
in section~\ref{sec:NNLOPS_formal}.

As it will be shown in section~\ref{sec:NNLOPS_formal} (and also
appendix~\ref{app:bspace}), up to the second perturbative order, the
differential cross section in $\pt$ and in the Born phase space $\PhiB$
is described by the following formula
\begin{align}
\label{eq:start}
  \frac{\mathd\sigma}{\mathd\PhiB\mathd \pt} &= \frac{\mathd}{\mathd \pt}
     \bigg\{ \exp[-\tilde{S}(\pt)] {\cal L}(\pt)\Bigg\} + R_f(\pt)\,,
\end{align}
where $R_f$ contains terms that are non-singular in the small $\pt$
limit. We notice that $\PhiB$ on the left-hand side of
eq.~\eqref{eq:start} is defined through a projection of the full phase
space with multiple emissions, in particular $\PhiBJ$ and $\PhiBJJ$,
onto the $\PhiB$ phase space. We denote this projection by
\begin{equation}\label{eq:projres}
\PhiBres(\Phi)\,,
\end{equation}
and $\Phi$ stands for $\PhiBJ$, $\PhiBJJ$, and so on.  The suffix
``res'' in $\PhiBres$ stands for ``resummation'', to make clear that
the projection is relative to how the recoil of the colour-singlet
system is treated in the resummation approach.
The Sudakov form factor $\tilde{S}$ reads
\begin{equation}\label{eq:Rdef}
\tilde{S}(\pt) = 2\int_{\pt}^{Q}\frac{\mathd q}{q}
                    \left(A(\as(q))\ln\frac{Q^2}{q^2} +
                    \tilde{B}(\as(q))\right),
\end{equation}
with
\begin{align}
A(\as)=& \left(\abar\right) A^{(1)} + \left(\abar\right)^2 A^{(2)}+ \left(\abar\right)^3 A^{(3)}\,,\notag\\
\tilde{B}(\as)=& \left(\abar\right) B^{(1)} + \left(\abar\right)^2 \tilde{B}^{(2)}\,.
\end{align}
where all coefficients are defined in
section~\ref{sec:NNLOPS_formal} and appendix~\ref{app:formulae}.
The factor ${\cal L}$, defined in eq.~\eqref{eq:luminosity} of
section~\ref{sec:NNLOPS_formal}, involves the parton luminosities, the
Born squared amplitude for the production of the colour-singlet system
\F{}, the hard-virtual corrections up to two loops and the collinear
coefficient functions up to second order. These constitute some of the
ingredients necessary for the next-to-next-to-next-to-leading
logarithm (N$^3$LL) resummation.
Here, for ease of notation, we do not indicate explicitly the $\PhiB$
dependence of ${\cal L}$ and $R_f$.

As it stands, eq.~\eqref{eq:start} is such that its integral over
$\pt$ between an infrared cutoff $\Lambda$ (more precisely, the scale
value when the Sudakov form factor $\tilde{S}(\pt)$ vanishes) and $Q$
reproduces the NNLO total cross section for the production of the
colour-singlet system. We can recast eq.~\eqref{eq:start} as
\begin{equation}
\label{eq:startDiff}
  \frac{\mathd\sigma}{\mathd\PhiB\mathd \pt}  =
  \frac{\mathd\sigma^{\rm sing}}{\mathd\PhiB\mathd
    \pt}+R_f(\pt),\qquad \frac{\mathd\sigma^{\rm
      sing}}{\mathd\PhiB\mathd \pt} = \exp[-\tilde{S}(\pt)] D(\pt)\,,
\end{equation}
with
\begin{equation}
\label{eq:Dterms}
  D(\pt)  \equiv -\frac{\mathd \tilde{S}(\pt)}{\mathd \pt} {\cal L}(\pt)+\frac{\mathd {\cal L}(\pt)}{\mathd \pt}\,,
\end{equation}
and
\begin{equation}
\frac{\mathd \tilde{S}(\pt)}{\mathd \pt}  = -\frac2{\pt} \left(A(\as(\pt))\ln\frac{Q^2}{\pt^2} +
                    \tilde{B}(\as(\pt))\right)\,.
\end{equation}

We now make contact with the \minlo{} procedure. We start by writing
the regular terms $R_f$ to second order as
\begin{equation}
\label{eq:Rf}
R_f(\pt) = \frac{\mathd\sigma^{\rm (NLO)}_{\scriptscriptstyle\rm FJ}}{\mathd\PhiB\mathd
      \pt}-
  \abarmu{\pt}\left[\frac{\mathd\sigma^{\rm sing}}{\mathd\PhiB\mathd
      \pt}\right]^{(1)} -
  \left(\abarmu{\pt}\right)^2\left[\frac{\mathd\sigma^{\rm sing}}{\mathd\PhiB\mathd
      \pt}\right]^{(2)}\,,
\end{equation}
where the notation $[X]^{(i)}$ stands for the coefficient of the
$i$-th term in the perturbative expansion of the quantity $X$. The
first term on the right-hand side of the above equation is the NLO
differential cross section for the production of the singlet \F{} in
association with one jet $J$, namely
\begin{equation}
\frac{\mathd\sigma^{\rm (NLO)}_{\scriptscriptstyle\rm FJ}}{\mathd\PhiB\mathd
      \pt} = \abarmu{\pt}\left[\frac{\mathd\sigma_{\scriptscriptstyle\rm FJ}}{\mathd\PhiB\mathd
      \pt}\right]^{(1)} + \left(\abarmu{\pt}\right)^2\left[\frac{\mathd\sigma_{\scriptscriptstyle\rm FJ}}{\mathd\PhiB\mathd
      \pt}\right]^{(2)}\,.
\end{equation}
As a second step, we factor out the Sudakov exponential in
eq.~\eqref{eq:startDiff} and obtain
\begin{equation}
\label{eq:startDifffact}
  \frac{\mathd\sigma}{\mathd\PhiB\mathd \pt}  =
  \exp[-\tilde{S}(\pt)]\left\{
                                  D(\pt)+\frac{R_f(\pt)}{\exp[-\tilde{S}(\pt)]}\right\}\,.
\end{equation}
We notice that in order to preserve the
perturbative accuracy of the integral of eq.~\eqref{eq:startDifffact},
it is sufficient to expand the curly bracket in powers of $\as(\pt)$
up to a certain order.  In fact, when expanded in powers of
$\as(\pt)$, all terms in the curly brackets of
eq.~\eqref{eq:startDifffact} contain at most a $1/\pt$ singularity and 
(for the terms arising from the derivative of $\tilde{S}$) a single logarithm
of $\pt$. The contribution of the terms of order
$\as^m(\pt)\ln^n\frac{Q}{\pt}$ to the total integral of
eq.~\eqref{eq:startDifffact} between the infrared scale $\Lambda$ and
$Q$ is of order~\cite{Hamilton:2012rf}
\begin{equation}
    \int_{\Lambda}^{Q} \mathd \pt \frac{1}{\pt} \as^m(\pt) \ln^n\frac{Q}{\pt}
\exp(-\tilde{S}(\pt))    \approx {\cal O}\left(\as^{m-\frac{n+1}{2}}(Q)\right)\,.
\end{equation}

This crucially implies that, for the integral to be NLO accurate, i.e.
${\cal O}(\as(Q))$, one has to include all terms up to order
$\as^2(\pt)$ in the curly brackets of
eq.~\eqref{eq:startDifffact}. This guarantees that the perturbative
left-over is of formal order ${\cal O}(\as^2(Q))$ in the total cross
section. After performing this expansion in
eq.~\eqref{eq:startDifffact} we obtain
\begin{align}
\label{eq:minlo}
  \frac{\mathd\sigma}{\mathd\PhiB\mathd \pt}  =
  \exp[-\tilde{S}(\pt)]&\bigg\{ \abarmu{\pt}\left[\frac{\mathd\sigma_{\scriptscriptstyle\rm FJ}}{\mathd\PhiB\mathd
      \pt}\right]^{(1)} \left(1+\abarmu{\pt} [\tilde{S}(\pt)]^{(1)}\right)
  \notag\\
&+ \left(\abarmu{\pt}\right)^2\left[\frac{\mathd\sigma_{\scriptscriptstyle\rm FJ}}{\mathd\PhiB\mathd
      \pt}\right]^{(2)}\bigg\}\,.
\end{align}
We notice that this formula is what we would obtain by integrating formula~(\ref{eq:minloacc1})
for an observable of the type
\begin{equation}
  O(\Phi)=\delta(\PhiBres(\Phi)-\PhiB)\delta(\pt(\Phi)-\pt),
\end{equation}
where $\PhiBres(\Phi)$ was introduced in eq.~(\ref{eq:projres}).
It is thus equivalent to the \minlo{} formula. This equivalence
relies upon the fact that the \minlo{} result can
be cast in the form of eq.~(\ref{eq:minloacc1}), that
is independent of any particular phase space projection.

We stress again that, in order for eq.~\eqref{eq:minlo} to have NLO
accuracy, $S$ must include correctly terms of order up to $\as^2$
which exactly reproduce the singular part of the cross section and
hence ensure that eq.~\eqref{eq:minlo} can be reassembled back as a
total derivative to the desired perturbative order.

In order to achieve NNLO accuracy, it is now sufficient to guarantee that 
eq.~\eqref{eq:minlo} has ${\cal O}(\as^2(Q))$ accuracy at fixed $\PhiB$ after integration
over $\pt$.  This requires the inclusion of all
terms up to ${\cal O}(\as^3(\pt))$ in the curly brackets of
eq.~\eqref{eq:startDifffact}, and we obtain
\begin{align}
\label{eq:minnlo}
  \frac{\mathd\sigma}{\mathd\PhiB\mathd \pt}  &=
  \exp[-\tilde{S}(\pt)]\bigg\{ \abarmu{\pt}\left[\frac{\mathd\sigma_{\scriptscriptstyle\rm FJ}}{\mathd\PhiB\mathd
      \pt}\right]^{(1)} \left(1+\abarmu{\pt} [\tilde{S}(\pt)]^{(1)}\right)
  \notag\\
&+ \left(\abarmu{\pt}\right)^2\left[\frac{\mathd\sigma_{\scriptscriptstyle\rm FJ}}{\mathd\PhiB\mathd
      \pt}\right]^{(2)} + \left(\abarmu{\pt}\right)^3 [D(\pt)]^{(3)} +{\rm regular~terms}\bigg\}\,,
\end{align}
where $[D(\pt)]^{(3)}$ is the third-order term in the expansion of the
$D(\pt)$ function~\eqref{eq:Dterms}. The regular terms that we omitted
in eq.~\eqref{eq:minnlo} arise from the ${\cal O}(\as^3(\pt))$
expansion of the term $R_f(\pt)/\exp[-\tilde{S}(\pt)]$ in
eq.~\eqref{eq:startDifffact}, which vanish in the limit $\pt\to 0$.
The absence of a $1/\pt$ singularity ensures that such terms give a
N$^3$LO contribution to the total cross section, and therefore can be
ignored. We explicitly verified that their inclusion yields a
subleading numerical effect.
Equation~\eqref{eq:minnlo} constitutes the reference formula to build the
\minnlo{} generator. This simply amounts to adding to the \minlo{}
formula the new term
\begin{align}
\label{eq:D3}
  [D(\pt)]^{(3)} &= 
  -\left[\frac{\mathd \tilde{S}(\pt)}{\mathd \pt}\right]^{(1)}[{\cal L}(\pt)]^{(2)}
  -\left[\frac{\mathd \tilde{S}(\pt)}{\mathd \pt}\right]^{(2)}[{\cal
    L}(\pt)]^{(1)}\notag\\
& -\left[\frac{\mathd \tilde{S}(\pt)}{\mathd \pt}\right]^{(3)}[{\cal L}(\pt)]^{(0)}
                   + \left[\frac{\mathd {\cal L}(\pt)}{\mathd \pt}\right]^{(3)} \\
      & = 
   \frac{2}{\pt} \left( A^{(1)} \ln
  \frac{Q^2}{\pt^2} + B^{(1)} \right) [{\cal L}(\pt)]^{(2)}     + \frac{2}{\pt} \left( A^{(2)} \ln \frac{Q^2}{\pt^2} + \tilde{B}^{(2)} \right)
  [{\cal L}(\pt)]^{(1)} \notag\\
& + \frac{2}{\pt}  A^{(3)} \ln \frac{Q^2}{\pt^2} \,[{\cal L}(\pt)]^{(0)}+ \left[  \frac{\mathd {\cal L}(\pt)}{\mathd \pt}
  \right]^{(3)},\nonumber
\end{align}
where all coefficients are defined in appendices~\ref{app:formulae}
and~\ref{app:D3formulae}.

\section{Implementation of the $[D(\pt)]^{(3)}$ term in the \minlo{}
  framework}
\label{sec:implementation}
The \minlo{} method based on eq.~\eqref{eq:minlo} has been implemented
within the \POWHEGBOX{} framework~\cite{Alioli:2010xd} and it has been
thoroughly tested. In order to achieve NNLO accuracy, we therefore
include the new terms discussed in the previous section as a
correction to the existing implementation. 

We recall that all terms in the \minlo{} formula~\eqref{eq:minlo} are
directly related to the phase space of the production of the colour
singlet \F{} together with either one ($\PhiBJ$) or two jets
($\PhiBJJ$). Conversely, in the \minnlo{} master
formula~\eqref{eq:minnlo}, the new term $[D(\pt)]^{(3)}$ arises from a
resummed calculation in the $\pt \to 0$ limit where the information
about the rapidity of the radiation has been integrated out
inclusively. As such it depends on the phase space $\PhiB$ of the
colour singlet with no additional radiation, and carries an explicit
dependence on the $\pt$ of the system.
This dependence, however, does not correspond to a well-defined
phase-space point for the full event kinematics (neither $\PhiBJ$ nor
$\PhiBJJ$), since the presence of a $\pt$ requires at least one parton
recoiling against \F{}, but we have no information on the kinematics
of such a parton. This has no consequence on the accuracy of the
\minnlo{} formula, since at finite transverse momentum
$[D(\pt)]^{(3)}$ contributes with a ${\cal O}(\as^3(Q))$ correction to
the integrated cross section. It follows that at large values of
$\pt$ the
kinematics associated with the $[D(\pt)]^{(3)}$ terms can be completed
in an arbitrary way, implying variations beyond NNLO accuracy. 
In particular, we observe that a $\PhiB$ phase-space point can be
obtained from a $\PhiBJ$ phase-space point through a suitable
mapping, while the $\pt$ corresponds to that of the $\PhiBJ$
kinematics. The mapping should project $\PhiBJ$ to $\PhiB$ smoothly
when $\pt \to 0$.

In order to embed the new \minnlo{} formulation of
eq.~\eqref{eq:minnlo} into the \minlo{} framework, one must therefore
associate each value of $[D(\pt)]^{(3)}$ to a specific point in the
$\PhiBJ$ phase space. This requires supplementing the $\PhiB$ and
$\pt$ information of the $[D(\pt)]^{(3)}$ term with the remaining
kinematics of the radiation that has been previously lost. In other
words, $[D(\pt)]^{(3)}$ should be spread over the $\PhiBJ$ phase space
in such a way that, upon integration, eq.~\eqref{eq:minnlo} is eventually
reproduced.

The most obvious way to spread the $[D(\pt)]^{(3)}$ term in the
$\PhiBJ$ phase space is either uniformly, or according to some
distribution of choice.
To this end, we multiply $[D(\pt)]^{(3)}$ by the following factor
\begin{equation}
  \Fcorr(\PhiBJ)= \frac{ J_\ell (\PhiBJ)}{  \sum_{l'} \int
  \mathd \PhiBJ' J_{l'} (\PhiBJ') \delta (\pt - \pt')
  \delta (\PhiB - \PhiB')}, \label{eq:fcorr}
\end{equation}
where $\PhiB$ ($\PhiB'$) is a projection of the $\PhiBJ$ ($\PhiBJ'$)
phase space into the phase space for the production of the colour singlet
\F{} alone (for instance performed according to the FKS mapping for 
initial-state radiation (ISR) discussed in section 5.5.1 of
ref.~\cite{Frixione:2007vw}, that preserves the rapidity of the colour-singlet 
system). $J_\ell$ is an arbitrary function of $\PhiBJ$, and
$\ell$ labels the flavour structure of the \FJ{} production
process. Finally, $\pt$ is the transverse momentum of the radiation
(hence that of the colour singlet) in the $\PhiBJ$ phase space.
The factor $\Fcorr$ is such that upon integration over the $\PhiBJ$
phase space together with a function that depends only on $\pt$ and $\PhiB$,
the result reduces to the integral of that function over $\PhiB$ and $\pt$. In
formulae, for an arbitrary function $G(\PhiB,\pt)$, we have
\begin{align}
 \sum_\ell   \int  \mathd &\PhiBJ' G(\PhiB',\pt') F_\ell^{\tmop{corr}}(\PhiBJ')
  =\int \mathd \PhiB\, \mathd \pt G(\PhiB,\pt) \notag\\
&\times
     \sum_\ell \int  \mathd \PhiBJ' \delta(\PhiB-\PhiB') \delta(\pt-\pt') \Fcorr(\PhiBJ')=\int \mathd \PhiB\, \mathd \pt G(\PhiB,\pt)\,.
\end{align}
The full phase-space parametrisation and the $\PhiBJ\to\PhiB$ mapping
are given in appendix~\ref{app:spreading}.

The function $ J_\ell (\PhiBJ)$ in eq.~\eqref{eq:fcorr} gives us some
freedom in choosing how to spread $[D(\pt)]^{(3)}$ in the radiation
phase space. Among the sensible choices, one could simply use a
uniform distribution by setting
\begin{equation}
 J_\ell (\PhiBJ) = 1\,.
\end{equation}
For this trivial choice an analytic solution for the integral in the denominator of
eq.~\eqref{eq:fcorr} is given for illustration in
appendix~\ref{app:spreading}. However, we found that this choice
generates a spurious behaviour when the jet is produced at very large
rapidities.
A more natural choice is to spread $[D(\pt)]^{(3)}$ according to the
actual rapidity distribution of the radiation, by setting
\begin{equation}
\label{eq:full_matrix}
 J_\ell (\PhiBJ) = |M^{\scriptscriptstyle\rm FJ}_\ell(\PhiBJ)|^2 (f^{[a]} f^{[b]})_\ell\,,
\end{equation}
where $|M^{\scriptscriptstyle\rm FJ}_\ell(\PhiBJ)|^2$ is the tree-level
matrix element squared for the \FJ{} process, and the quantity
$(f^{[a]} f^{[b]})_\ell$ represents the product of the parton
densities in the initial-state flavour configuration given by the
index $\ell$.  This choice provides a more physical distribution of
$[D(\pt)]^{(3)}$, but it can become computationally expensive for
complex processes with several degrees of freedom as the integral in
the denominator of \eqn{eq:fcorr} has to be evaluated for every
phase-space point numerically.
A convenient compromise is to take the collinear limit of the squared
amplitude of eq.~\eqref{eq:full_matrix}, namely
\begin{equation}
|M^{\scriptscriptstyle\rm FJ}_\ell(\PhiBJ)|^2 \simeq |M^{\scriptscriptstyle\rm F}(\PhiB)|^2 P_\ell(\Phi_{\rm rad})\,,
\end{equation}
where $|M^{\scriptscriptstyle\rm F}(\PhiB)|^2$ is the Born matrix element squared for the
production of the colour singlet \F{}, and $P_\ell(\Phi_{\rm rad})$ is the
collinear splitting function.
After noticing that the Born squared amplitude $ |M^{\scriptscriptstyle\rm F}(\PhiB)|^2$
cancels in the ratio of eq.~\eqref{eq:fcorr}, we can simply set
\begin{equation}
 J_\ell (\PhiBJ) = P_\ell(\Phi_{\rm rad}) (f^{[a]} f^{[b]})_\ell\,,
\end{equation}
where the full expression is reported in
eqs.~\eqref{eq:APsplitting-quark},~\eqref{eq:APsplitting-gluon}.  This
prescription is computationally faster, since the integral in the
denominator of \eqn{eq:fcorr} has a better convergence, and it does
not change for more involved processes.

With these considerations,
eq.~\eqref{eq:minnlo} can be recast in a way that is differential in
the entire $\PhiBJ$ phase space as
\begin{align}
\label{eq:minnlo-PhiBJ}
  \frac{\mathd\sigma}{\mathd\PhiBJ}  &=
  \exp[-\tilde{S}(\pt)]\bigg\{ \abarmu{\pt}\left[\frac{\mathd\sigma_{\scriptscriptstyle\rm FJ}}{\mathd\PhiBJ}\right]^{(1)} \left(1+\abarmu{\pt} [\tilde{S}(\pt)]^{(1)}\right)
  \\
&+ \left(\abarmu{\pt}\right)^2\left[\frac{\mathd\sigma_{\scriptscriptstyle\rm FJ}}{\mathd\PhiBJ}\right]^{(2)} + \left(\abarmu{\pt}\right)^3 [D(\pt)]^{(3)}  F^{\tmop{corr}}(\PhiBJ)\bigg\}\,,\notag
\end{align}
where the sum over flavour configurations is understood, and $\pt$ is
meant to be defined in the $\PhiBJ$ phase space.

A second aspect relevant to the implementation of the \minnlo{}
procedure is related to how one switches off the Sudakov form factor,
as well as the terms $[\tilde{S}(\pt)]^{(1)}$ and $[D(\pt)]^{(3)}$ in
eq.~\eqref{eq:minnlo-PhiBJ}, in the large $\pt$ region of the
spectrum. We stress that the details of this operation do not modify
the accuracy of the result. This is because in the large $\pt$ region
eq.~\eqref{eq:minnlo-PhiBJ} differs from the NLO \FJ{} distribution
only by ${\cal O}(\as^3)$ corrections relative to the Born.
This implies that one has some freedom in choosing how to turn off
the logarithmic terms at scales $\pt\gtrsim Q$.
One important constraint to keep in mind is that in the regime
$\pt \ll Q$ the logarithmic structure has to be preserved in order to
retain the NNLO accuracy in the total (inclusive) cross section.

There are of course different sensible ways to switch off the
logarithmic terms at large $\pt$. One possibility is to set the
quantities $\tilde{S}(\pt)$, $[\tilde{S}(\pt)]^{(1)}$, and $[D(\pt)]^{(3)}$ to zero at
$\pt\geq Q$. This prescription is adopted in the original \minlo{}
implementation of ref.~\cite{Hamilton:2012rf}.
A second possibility, closer in spirit to what is done in resummed
calculations, is to modify the logarithms contained in $\tilde{S}(\pt)$,
$[\tilde{S}(\pt)]^{(1)}$, and $[D(\pt)]^{(3)}$,  
so
that they vanish in the large $\pt$ limit. This is done by means of
the following replacement
\begin{equation}
\label{eq:modlog}
\ln\frac{Q}{\pt} \to \frac{1}{p}\ln\left(1+\left(\frac{Q}{\pt}\right)^p\right)\,,
\end{equation}
where $p$ is a free positive parameter. Larger values of $p$
correspond to logarithms that tend to zero at a faster rate at large $\pt$.  
%
With this modification we also include
the Jacobian factor
\begin{equation}
\label{eq:jacob}
\frac{\left(Q/\pt\right)^p}{1+\left(Q/\pt\right)^p}\,.
\end{equation}
in front of the $[D(\pt)]^{(3)}$ term, which has the effect of switching it off its $1/\pt$
terms when $\pt$ goes above~$Q$.

It is easy to convince ourselves that this modification does not alter the \minnlo{} accuracy.
In fact, the modified logarithms only play a role when $\pt \gtrsim Q$. In this regime, the
counting of the orders in the \minnlo{} formula simplifies, since there are no enhancements
due to large logarithms. Under these circumstances, the $[D(\pt)]^{(3)}$ term is subleading, and the
Sudakov form factor also leads to subleading effects once combined with its first order expansion
in formula~(\ref{eq:minnlo-PhiBJ}). Thus, as long as the modified logarithms are used
consistently in both $\tilde{S}$
and $[\tilde{S}(\pt)]^{(1)}$, only terms beyond the relevant accuracy are generated.

\section{Derivation of the \minnlo{} master formula}
\label{sec:NNLOPS_formal}
In this section we present a more rigorous derivation of the
\minnlo{} formalism that has been outlined in section~\ref{sec:procedure}.
Our starting point is the calculation of the cumulative transverse-momentum spectrum 
\begin{equation}
\frac{\mathd\sigma(\pt)}{\mathd \PhiB} = \int_0^{\pt} \mathd p_\perp
\frac{\mathd\sigma}{\mathd \PhiB\mathd p_\perp} 
\,, 
\end{equation}
for a colour singlet produced in the collision of two hadrons.
More precisely, we consider the second-order perturbative expansion of
the above cumulative cross section in the limit $\pt \to 0$ (i.e.\ the singular part), with up to
two emissions. This information can be directly accessed in the momentum
space formulation of transverse-momentum resummation presented in
\citeres{Monni:2016ktx,Bizon:2017rah}.\footnote{Our starting point
  follows from eqs.~(2.58) and (2.59) of
  \citere{Bizon:2017rah}, by considering only the first two emissions.}
The singular part expressed in this way reads\footnote{The convolution between two
  functions $f(z)$ and $g(z)$ is defined as
  $(f\otimes g)(z) \equiv \int_z^1 \frac{\mathd x}{x} f(x)
  g\left(\frac{z}{x}\right)$.}
\begin{align}
\frac{\mathd\sigma^{\rm sing}(\pt)}{\mathd\PhiB} &= \Bigg \{\int \dk{1} e^{-S(\kt{1})} \left[
  S'(\kt{1}) \, \mathbb{1} + \frac{\as(\kt{1})}{\pi} \hat{P}  +
  2\beta(\as(\kt{1}))\frac{\mathd\ln C}{\mathd \as
  }  \right] \notag\\
&\otimes \bigg[\Delta^{(C)}(\kt{1},\Lambda) {\cal L}^{(C)}(\kt{1})  \,\Theta(\pt - \kt{1}) + \int \dk{2} \left[
  S'(\kt{2}) \,\mathbb{1} + \frac{\as(\kt{2})}{\pi} \hat{P} \right.\notag\\
&\left.+
  2\beta(\as(\kt{2}))\frac{\mathd\ln C}{\mathd \as
 } \right] \otimes \left(\Delta^{(C)}(\kt{1},\kt{2}) {\cal L}^{(C)}(\kt{1})\right) \notag\\
&\times\Theta(\kt{1}-\kt{2})\Theta(\pt - |\veckt{1}
  + \veckt{2}|) \bigg]\Bigg \}+ \left\{C\rightarrow G;  {\cal L}^{(C)}\rightarrow  {\cal L}^{(G)} \right\}+O(\as^3)\,,
\label{eq:starting}
\end{align}
where we have defined
\begin{align}
\kt{i}  &=|\veckt{i}|\,,\\
\dk{i} &= \frac{\mathd\kt{i}}{\kt{i}}\frac{\mathd\phi_i}{2\pi}\,,\\
S'(\kt{i}) &= \frac{\mathd S (\kt{i})}{\mathd L},\qquad L = \ln\frac{Q}{\kt{i}}\,.
\end{align}
The various terms of eq.~\eqref{eq:starting} are explained in the
following.
We defined the following notation in terms of the initial-state legs $a$ and $b$:
\begin{align}
\mathbb{1} &= \mathbb{1}^{[a]} \, \mathbb{1}^{[b]}\notag\,,\\
\hat{P} &= \hat{P}^{[a]}\mathbb{1}^{[b]} +
          {P}^{[b]}\mathbb{1}^{[a]} \notag\,,\\
{C} &= {C}^{[a]}\mathbb{1}^{[b]} +
  {C}^{[b]}\mathbb{1}^{[a]} \notag\,,\\
{G} &= {G}^{[a]}\mathbb{1}^{[b]} +
 {G}^{[b]}\mathbb{1}^{[a]} \,,
\end{align}
where the identity matrix indicates a trivial dependence on the momentum fraction $z$, i.e.
\begin{align}
\mathbb{1}^{[a/b]} \equiv \delta(1-z^{[a/b]})\,.
\end{align}
The regularised splitting function $\hat{P}^{[a/b]}$ and the
coefficient functions $C^{[a/b]}$ and $G^{[a/b]}$ are defined as
\begin{align}
  \hat{P}^{[a/b]}(z) &= \hat{P}^{(0)}(z) + \frac{\as(\kt{i})}{2\pi}
            \hat{P}^{(1)}(z) + \left(\frac{\as(\kt{i})}{2\pi}\right)^2
            \hat{P}^{(2)}(z) + \dots\\
  C^{[a/b]}(z) &= \delta(1-z) + \frac{\as(\kt{i})}{2\pi}
            C^{(1)}(z) + \left(\frac{\as(\kt{i})}{2\pi}\right)^2
            C^{(2)}(z) + \dots\\
 G^{[a/b]}(z) &= \frac{\as(\kt{i})}{2\pi}
            G^{(1)}(z) + \dots
\label{eq:unity_decomposition}
\end{align}

The ${\cal L}$ factors contain the parton luminosities convoluted with
the coefficient functions and multiplied by the virtual
corrections. They read
\begin{align}
\label{eq:luminosity-C}
&{\cal L}^{(C)}(\kt{i})=\sum_{c,
  c'}\frac{\mathd|M^{\scriptscriptstyle\rm F}|_{cc'}^2}{\mathd\PhiB} \sum_{i, j} \left(C^{[a]}_{c
  i}\otimes f_i^{[a]}\right) H(Q) \left(C^{[b]}_{c'
  j}\otimes f_j^{[b]}\right)
\,,\\
\label{eq:luminosity-G}
&{\cal L}^{(G)}(\kt{i})=\sum_{c,
  c'}\frac{\mathd|M^{\scriptscriptstyle\rm F}|_{cc'}^2}{\mathd\PhiB} \sum_{i, j} \left(G^{[a]}_{c
  i}\otimes f_i^{[a]}\right) H(Q) \left(G^{[b]}_{c'
  j}\otimes f_j^{[b]}\right)
\,.
\end{align}
The last term in eq.~\eqref{eq:starting} accounts for azimuthal
correlations and it is non-zero only for processes that are
$gg$-initiated at the Born level. Its structure is identical to the
one of the first term, provided one replaces the coefficient functions
$C$ with the corresponding $G$ functions~\cite{Catani:2010pd}.
The quantity $\Delta^{(C)}(Q_1,Q_2)$ represents the no emission
probability
between the scales $Q_1$ and $Q_2 < Q_1$, and it is given by
\begin{equation}
\Delta^{(C)}(Q_1,Q_2) = \frac{ e^{-S(Q_2)}{\cal L}^{(C)}(Q_2) }{e^{-S(Q_1)}  {\cal L}^{(C)}(Q_1)},
\end{equation}
and an analogous definition holds for $\Delta^{(G)}$.

All considerations beyond this point hold identically for both the $C$
and $G$ terms, and therefore we omit the latter in the following
equations for the sake of simplicity (but its contribution is
understood). The function $H$ contains the contribution of the virtual
corrections to the colour-singlet process under consideration
\begin{align}
\label{eq:hard_function}
H(Q) = 1 + \frac{\as(Q)}{2\pi}H^{(1)} +
  \left(\frac{\alpha_s(Q)}{2\pi}\right)^2 H^{(2)} + \dots
\end{align}
Note that, when working directly in momentum space, the term $H^{(2)}$
differs from the second-order coefficient of the form factor in the
$\overline{\rm MS}$ scheme~\cite{Bizon:2017rah}
(cf. eq.~\eqref{eq:delta_H2} below).
The various coefficients used in the above equations are reported in
appendix~\ref{app:formulae}.

Finally, the Sudakov radiator $S$ is defined as in
eq.~\eqref{eq:Rdef}, but with the anomalous dimension $\tilde{B}(\as)$
replaced by
\begin{equation}
B(\as) = \left(\abar\right) B^{(1)} + \left(\abar\right)^2 B^{(2)}\,.
\end{equation}
The first order coefficient $B^{(1)}$ is identical to the one of
eq.~\eqref{eq:Rdef}, while the difference between the second order
coefficients $B^{(2)}$ and $\tilde{B}^{(2)}$ will be explained shortly
in this section. All explicit formulae are reported in
appendix~\ref{app:formulae}.
Equation~\eqref{eq:starting} reproduces the correct logarithmic
structure at ${\cal O}(\as^2)$, including the NNLO constant terms. The
evolution, in principle, continues with extra emissions down to the
infrared cutoff of the theory $\Lambda\ll Q$, that is the scale
at which the Sudakov form factor vanishes. However, for the time
being we are only considering the first two of such emissions, as the
additional radiation will be included by the parton shower via a
consistent matching at a later stage.

\begin{figure}[tp!]
  \centering
  \includegraphics[trim={0 -0.2cm 0 0},width=0.55\linewidth]{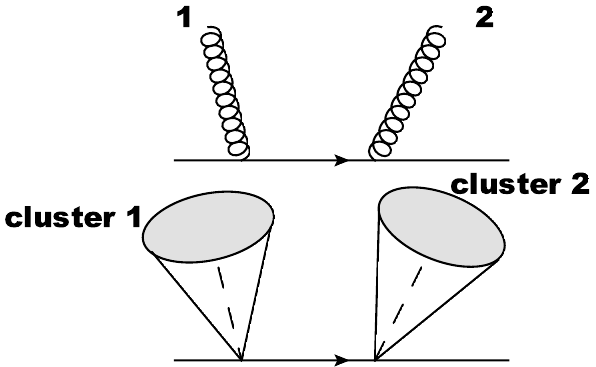}\hspace{1cm} 
  \includegraphics[trim={0 -0.2cm 0 0},width=0.34\linewidth]{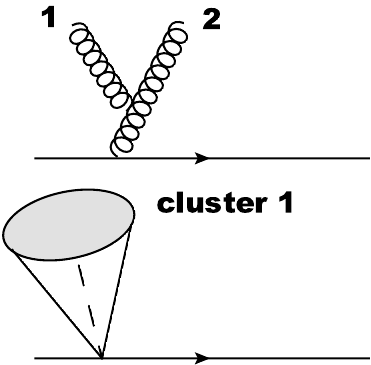} 
  \caption{Example of decomposition into correlated clusters for
    emissions off a quark line. The left configuration corresponds to
    the emission of two independent gluons, and it is mapped onto two
    separate clusters. The right configuration corresponds to the
    emission of the second gluon off the first one, and it is mapped
    onto a single cluster. The correlated clusters are not to be
    confused with partonic jets.}
  \label{fig:clusters}
\end{figure}

The formulation that led to eq.~\eqref{eq:starting} is based on an
organisation of the perturbative series that aims at the resummation
of the logarithmic terms to all orders.
In particular, each evolution step in eq.~\eqref{eq:starting}
describes the emission of an {\it inclusive correlated cluster} of
emissions~\cite{Bizon:2017rah}. In an illustrative picture, it is
convenient to think of each inclusive cluster as describing the
emission of an initial-state radiation and its subsequent
branchings. The cluster is defined by integrating inclusively over the
branching variables and retaining only the information about the total
transverse momentum of the radiation emitted within. An illustrative
example of the separation into correlated clusters for two typical
configurations is reported in fig.~\ref{fig:clusters}.

In a slightly more technical picture, limiting ourselves to the
${\cal O}(\as^2)$ case we are interested in, the correlated clusters
originate from the fact that, in the soft limit, the squared amplitude
for the emission of up to $2$ partons can be decomposed as
\begin{align}
\label{eq:clusters}
\left|{\cal M}(k_1)\right|^2 &\equiv \tilde{\cal M}^2(k_1)\,, \nonumber  \\ 
\left|{\cal M}(k_1, k_2)\right|^2 &= \tilde{\cal M}^2(k_1) \tilde{\cal M}^2(k_2) + \tilde{\cal M}^2(k_1,k_2)\,. 
\end{align}
The term $\tilde{\cal M}^2(k_1,k_2)$ is defined as the correlated part
of the squared amplitude that cannot be decomposed as the product of
two squared amplitudes for the emission of a single
parton.\footnote{These coincide with the webs in the soft
  limit~\cite{Gatheral:1983cz,Frenkel:1984pz}.} We point out that, in
the collinear limit, a decomposition of the type~\eqref{eq:clusters}
requires keeping track of all the possible flavour structures of a
$1\to 3$ collinear branching.
Each of the correlated terms $\tilde{\cal M}^2$ admits a perturbative
expansion in powers of $\as$ defined by including the virtual
corrections while keeping the number of emissions fixed. The inclusive
correlated cluster is defined as
\begin{align} 
\label{eq:inclusive_cluster}
{\cal M}_{\rm incl}^2(k) = & \,{\cal M}^2(k)\\
&+\,\int [d
  k_a][d k_b]\tilde{\cal
    M}^2(k_a,k_b)\delta^{(2)}(\vec{k}_{ta}+\vec{k}_{tb}-\vec{k}_{t})\delta(Y_{ab}-Y)
                             + {\cal O}(\as^3)\,,\notag
\end{align}
$Y_{ab}$ denotes the rapidity of the $k_a+k_b$ system in the
centre-of-mass frame of the collision, and $[d k]$ denotes the
phase-space measure for the emission $k$. Moreover, the strong
coupling constant in eq.~\eqref{eq:inclusive_cluster} is evaluated at
the transverse momentum of the inclusive cluster.
For an inclusive observable such as $\pt$, the only quantity that
matters is the total transverse momentum of each inclusive
cluster. Therefore, one can integrate over the rapidity of each
cluster, and analytically cancel the infrared and collinear
singularities.
This results in a simplified picture, in which each inclusive cluster
of transverse momentum $\kt{i}$ is emitted according to the kernel
\begin{align}
\label{eq:prob-emsn}
S'(\kt{i}) \, \mathbb{1} &+ \frac{\as(\kt{i})}{\pi} \hat{P}  +
  2\beta(\as(\kt{i}))\frac{\mathd\ln C}{\mathd \as
  }\,,
\end{align}
and analogously for the term involving the $G$ coefficient
function. In eq.~\eqref{eq:prob-emsn} we implicitly sum over the two
initial-state legs $a$ and $b$, and
we observe that the first two terms
start at order ${\cal O}(\as)$, while the third term proportional
to $\beta(\as)$, defined in \eqn{eq:beta}, starts at ${\cal O}(\as^2)$.

Equation~\eqref{eq:starting} describes the emission of two subsequent
clusters with transverse momenta $\kt{1}$ and $\kt{2} \leq \kt{1}$.
The first line of eq.~\eqref{eq:starting} encodes the emission of the
first cluster, while the remaining lines describe the emission of the
second one. The latter is split into a no-emission probability term
(that excludes configurations with more than one cluster), and a term
proportional to the real kernel~\eqref{eq:prob-emsn} that actually
describes the second emission.
We stress that, in the present treatment, we are not interested in
retaining a given logarithmic accuracy in the $\pt$ spectrum, but
rather in describing the correct singular structure at
${\cal O}(\as^2)$. However, one should bear in mind that the evolution
beyond the second emission will be subsequently generated by a parton
shower generator, which will guarantee a fully exclusive treatment of
the radiation.
We finally observe that the term 
\begin{equation}
2\beta(\as(\kt{2}))\frac{\mathd\ln C}{\mathd \as
 },
\end{equation}
in the second emission's probability contributes at most to
${\cal O}(\as^3)$ and therefore can be ignored.

We now proceed by adding and subtracting $\Theta(\pt - \kt{1})$ to the
second $\Theta$ function in eq.~\eqref{eq:starting} as
\begin{align}
\Theta(\pt - |\veckt{1}
  + \veckt{2}|) &\to \Theta(\pt - \kt{1}) \notag\\
& + \left(\Theta(\pt - |\veckt{1}
  + \veckt{2}|) - \Theta(\pt - \kt{1})\right)\,.
\end{align}
We can then recast eq.~\eqref{eq:starting}, with ${\cal O}(\as^2)$
accuracy, as
\begin{align}
\frac{\mathd\sigma^{\rm sing}(\pt)}{\mathd\PhiB} &= \int \dk{1} e^{-S(\kt{1})} \left[
  S'(\kt{1}) \, \mathbb{1} + \frac{\as(\kt{1})}{\pi} \hat{P}  +
  2\beta(\as(\kt{1}))\frac{\mathd\ln C}{\mathd \as
  }  \right] \notag\\
&\otimes \bigg\{\Delta^{(C)}(\kt{1},\Lambda) {\cal L}^{(C)}(\kt{1})  + \int \dk{2} \left[
  S'(\kt{2}) \,\mathbb{1} + \frac{\as(\kt{2})}{\pi} \hat{P} \right]\Theta(\kt{1}-\kt{2})
  \notag\\
&\otimes \left(\Delta^{(C)}(\kt{1},\kt{2}) {\cal L}^{(C)}(\kt{1})\right)  \bigg\}\Theta(\pt - \kt{1})\notag\\
& + \int \dk{1} \int \dk{2} \Theta(\kt{1}-\kt{2})\notag\\
&\times\bigg\{ S'(\kt{1}) S'(\kt{2})  {\cal L}^{(C)}(\kt{2}) +
  S'(\kt{1}) \frac{\as(\kt{2})}{\pi} \hat{P} \otimes  {\cal
  L}^{(C)}(\kt{2}) \notag\\
& + S'(\kt{2}) \frac{\as(\kt{1})}{\pi} \hat{P} \otimes  {\cal
  L}^{(C)}(\kt{2}) + \frac{\as(\kt{1})}{\pi}
  \frac{\as(\kt{2})}{\pi} \hat{P} \otimes \hat{P} \otimes {\cal
  L}^{(C)}(\kt{2})\bigg\}\notag\\
&\times
 \left(\Theta(\pt - |\veckt{1}
  + \veckt{2}|) - \Theta(\pt - \kt{1})\right) + {\cal O}(\alpha^3_s)\,,
\label{eq:starting-1}
\end{align}
where we neglected the Sudakov form factors in the integral containing
the difference between the two $\Theta$ functions, as the first
non-trivial contribution generates ${\cal O}(\as^3)$ corrections. The
resulting integral is finite at fixed $\pt$. In order
to evaluate such an integral, and since we are only interested in the
${\cal O}(\as^2)$ result, we can expand the content of the curly
brackets about $\kt{1}=\pt$ and $\kt{2}=\pt$ as follows
\begin{align}
&\bigg\{S'(\kt{1}) S'(\kt{2})  {\cal L}^{(C)}(\kt{2}) +
  S'(\kt{1}) \frac{\as(\kt{2})}{\pi} \hat{P} \otimes  {\cal
  L}^{(C)}(\kt{2}) \notag\\
& + S'(\kt{2}) \frac{\as(\kt{1})}{\pi} \hat{P} \otimes  {\cal
  L}^{(C)}(\kt{2}) + \frac{\as(\kt{1})}{\pi}
  \frac{\as(\kt{2})}{\pi} \hat{P} \otimes \hat{P} \otimes {\cal
  L}^{(C)}(\kt{2})\bigg\}\notag\\
& = \bigg\{S'(\pt) S'(\pt)  {\cal L}^{(C)}(\pt) +
  S'(\pt) \frac{\as(\pt)}{\pi} \hat{P} \otimes  {\cal
  L}^{(C)}(\pt) \notag\\
& + S'(\pt) \frac{\as(\pt)}{\pi} \hat{P} \otimes  {\cal
  L}^{(C)}(\pt) + \frac{\as(\pt)}{\pi}
  \frac{\as(\pt)}{\pi} \hat{P} \otimes \hat{P} \otimes {\cal
  L}^{(C)}(\pt)\bigg\}\notag\\
& + \bigg\{S'(\pt) S''(\pt) {\cal L}^{(C)}(\pt)
  + \frac{\as(\pt)}{\pi} S''(\pt) \hat{P} \otimes  {\cal
  L}^{(C)}(\pt) \bigg\}\left(\ln\frac{\pt}{\kt{1}} +
  \ln\frac{\pt}{\kt{2}}\right)  \notag\\
&+ {\cal O}(\as^3)\,.
\end{align}
When performing the integration, we notice that the first term on the
right-hand side of the above equation vanishes upon azimuthal
integration, and we are left with the integral of the second term,
obtaining
\begin{align}
\frac{\mathd\sigma^{\rm sing}(\pt)}{\mathd\PhiB} &= \int \dk{1} e^{-S(\kt{1})} \left[
  S'(\kt{1}) \, \mathbb{1} + \frac{\as(\kt{1})}{\pi} \hat{P}  +
  2\beta(\as(\kt{1}))\frac{\mathd\ln C}{\mathd \as
  }  \right] \notag\\
&\otimes \bigg\{\Delta^{(C)}(\kt{1},\Lambda) {\cal L}^{(C)}(\kt{1})  + \int \dk{2} \left[
  S'(\kt{2}) \,\mathbb{1} + \frac{\as(\kt{2})}{\pi} \hat{P}
  \right]\Theta(\kt{1}-\kt{2})\notag\\
& \otimes \left(\Delta^{(C)}(\kt{1},\kt{2}) {\cal L}^{(C)}(\kt{1})\right) \bigg\}\Theta(\pt - \kt{1})\notag\\
& - \frac{\zeta_3}{4}\left( S'(\pt) S''(\pt) {\cal L}^{(C)}(\pt) +
  \frac{\as(\pt)}{\pi} S''(\pt) \hat{P} \otimes {\cal L}^{(C)}(\pt)  \right) + {\cal O}(\alpha^3_s)\,,
\label{eq:starting-2}
\end{align}
where $S''(\pt) = \mathd S'(\pt)/\mathd L$.

We now incorporate the terms generated by the latter integration into
the first term of eq.~\eqref{eq:starting-2}. Retaining
${\cal O}(\as^2)$ accuracy, this can be done by redefining some of the
resummation coefficients as follows (an alternative derivation of such
redefinitions is performed using an impact-parameter space formulation
in appendix~\ref{app:bspace})
\begin{align}
B^{(2)} &\to B^{(2)} + 2\zeta_3 (A^{(1)})^2,\notag\\
H^{(2)} &\to H^{(2)} - 2\zeta_3 A^{(1)} B^{(1)},\notag\\
C^{(2)}(z) &\to C^{(2)}(z) - 2 \zeta_3 A^{(1)} \hat{P}^{(0)}(z)\,.
\label{eq:new_coeffs}
\end{align}

In order to use eq.~\eqref{eq:starting-2} in the context of the
\minlo{} algorithm, we perform a resummation-scheme transformation to
evaluate the virtual corrections~\eqref{eq:hard_function} that appear
in $ {\cal L}^{(C)}(\kt{1})$~\eqref{eq:luminosity-C} at a scale
$\kt{1}$. This implies the further replacements~\cite{Hamilton:2012rf}
\begin{align} 
H(Q) &\rightarrow H(\kt{1})\,,\notag\\ B^{(2)} &\rightarrow
B^{(2)} + 2 \pi \beta_0 H^{(1)}\,.
\label{eq:new_coeffs_2}
\end{align} 
We thus define
\begin{align}
\label{eq:redefinition_final}
B^{(2)} \to& \,\tilde{B}^{(2)} = B^{(2)} + 2\zeta_3 (A^{(1)})^2 + 2 \pi \beta_0H^{(1)},\notag\\
H^{(2)} \to& \,\tilde{H}^{(2)} = H^{(2)} - 2\zeta_3 A^{(1)} B^{(1)},\notag\\
 C^{(2)}(z) \to& \,\tilde{C}^{(2)}(z) = C^{(2)}(z) - 2 \zeta_3 A^{(1)} \hat{P}^{(0)}(z)\,,
\end{align}
and we redefine all ingredients in our calculation as
$S\to \tilde{S}$, $\Delta^{(C)}\to \tilde{\Delta}^{(C)}$,
$C\to \tilde{C}$, and ${\cal L}^{(C)}\to {\cal \tilde{L}}^{(C)}$ to
take the above replacements into account. We therefore recast
eq.~\eqref{eq:starting-2} as
\begin{align}
\frac{\mathd\sigma^{\rm sing}(\pt)}{\mathd\PhiB} &= \int \dk{1} e^{-\tilde{S}(\kt{1})} \left[
  \tilde{S}'(\kt{1}) \, \mathbb{1} + \frac{\as(\kt{1})}{\pi} \hat{P}  +
  2\beta(\as(\kt{1}))\frac{\mathd\ln \tilde{C}}{\mathd \as
  }  \right] \notag\\
&\otimes \bigg\{\tilde{\Delta}^{(C)}(\kt{1},\Lambda) {\cal \tilde{L}}^{(C)}(\kt{1}) + \int \dk{2} \left[
  \tilde{S}'(\kt{2}) \,\mathbb{1} + \frac{\as(\kt{2})}{\pi} \hat{P}
  \right]\Theta(\kt{1}-\kt{2})\notag\\
& \otimes \left(\tilde{\Delta}^{(C)}(\kt{1},\kt{2}) {\cal \tilde{L}}^{(C)}(\kt{1})\right)  \bigg\}\Theta(\pt - \kt{1})+ {\cal O}(\alpha^3_s)\,.
\end{align}
Finally, we take the derivative in $\pt$ in
order to obtain the singular structure of the differential $\pt$
distribution, that reads
\begin{align}
\frac{\mathd\sigma^{\rm sing}}{\mathd\PhiB\mathd \pt} &= \int \dk{1} e^{-\tilde{S}(\kt{1})} \left[
  \tilde{S}'(\kt{1}) \, \mathbb{1} + \frac{\as(\kt{1})}{\pi} \hat{P}  +
  2\beta(\as(\kt{1}))\frac{\mathd\ln \tilde{C}}{\mathd \as
  }  \right] \notag\\
&\otimes \bigg\{\tilde{\Delta}^{(C)}(\kt{1},\Lambda) {\cal \tilde{L}}^{(C)}(\kt{1}) + \int \dk{2} \left[
  \tilde{S}'(\kt{2}) \,\mathbb{1} + \frac{\as(\kt{2})}{\pi} \hat{P}
  \right]\Theta(\kt{1}-\kt{2})\notag\\
& \otimes \left(\tilde{\Delta}^{(C)}(\kt{1},\kt{2}) {\cal \tilde{L}}^{(C)}(\kt{1})\right)  \bigg\}\delta(\pt - \kt{1})+ {\cal O}(\alpha^3_s)\,.
\label{eq:final}
\end{align}

In order to be accurate across the whole $\pt$ spectrum, we need to
match eq.~\eqref{eq:final} to the NLO differential cross section for
the production of the colour-singlet system in association with one
jet. This can be performed in two steps.

The first step is to observe that the second emission is distributed in a
way that closely mimics the treatment of the radiation in the
\POWHEG{} method~\cite{Nason:2004rx} discussed in
section~\ref{sec:accuracy}, that is generated according to the
probability
\begin{align}
\Delta_{\rm pwg} (\LambdaPWG) + \int\mathd \Phi_{\tmop{rad}} 
  \Delta_{\rm pwg} (\ptrad)  \frac{R (\PhiBJ{}, \Phi_{\tmop{rad}})}{B
  (\PhiBJ{})}
\,,
\label{eq:pwg-replacement}
\end{align}
where the factor $\Phi_{\tmop{rad}}$ represents the full
FKS~\cite{Frixione:1995ms} radiation phase space for the second
emission $k_2$.\footnote{We point out that the parton densities are
  included in the \POWHEG{} Sudakov $\Delta_{\rm pwg}$, yielding a
  contribution analogous to the luminosity factor ${\cal L}$.} The
quantities $R$ and $B$ represent the tree-level squared amplitudes for
\FJJ{} (double emission) and \FJ{} (single emission),
respectively. Therefore, the second emission can be directly generated
according to the \POWHEG{} method, which guarantees an accurate
description at tree level for $k_2$ over the whole radiation phase
space $\Phi_{\tmop{rad}}$.

We can then focus on the first cluster contribution. For simplicity we
can integrate eq.~\eqref{eq:final} explicitly over the second emission
$k_2$, stressing that the latter can be restored fully differentially
by closely following the \POWHEG{} procedure as previously
discussed. We obtain
\begin{align}
\frac{\mathd\sigma^{\rm sing}}{\mathd\PhiB\mathd \pt} &= e^{-\tilde{S}(\pt)} \left[
  \tilde{S}'(\pt) \, \mathbb{1} + \frac{\as(\pt)}{\pi} \hat{P}  +
  2\beta(\as(\pt))\frac{\mathd\ln \tilde{C}}{\mathd \as
  }  \right] \otimes {\cal \tilde{L}}^{(C)}(\pt)+ {\cal O}(\alpha^3_s(Q))\notag\\
& = \frac{ \mathd \left[e^{-\tilde{S}(\pt)}
                               {\cal \tilde{L}}^{(C)} (\pt)\right]}{\mathd \pt}+ {\cal O}(\alpha^3_s)\,,
\label{eq:final-first-emsn}
\end{align}
where in the second line we recast the result in a more compact form.
We can at last restore the contribution of the $G$ coefficient
functions by replacing ${\cal \tilde{L}}^{(C)}$ with the full
luminosity factor as
\begin{align} 
{\cal \tilde{L}}^{(C)}(\kt{1})\to {\cal L}(\kt{1})&=\sum_{c,
c'}\frac{\mathd|M^{\scriptscriptstyle\rm F}|_{cc'}^2}{\mathd\PhiB} \sum_{i, j}
\bigg\{\left(\tilde{C}^{[a]}_{c i}\otimes f_i^{[a]}\right) \tilde{H}(\kt{1})
\left(\tilde{C}^{[b]}_{c' j}\otimes f_j^{[b]}\right) \notag\\ & +
\left(G^{[a]}_{c i}\otimes f_i^{[a]}\right) \tilde{H}(\kt{1}) \left(G^{[b]}_{c'
j}\otimes f_j^{[b]}\right)\bigg\}\,.
\label{eq:luminosity}
\end{align}
Equation~\eqref{eq:final-first-emsn}, when expanded, correctly reproduces
up to ${\cal O}(\as^2)$ the divergent (logarithmic) structure of the
differential spectrum in the small $\pt$ limit.
However, eq.~\eqref{eq:final-first-emsn} does not yet include the regular terms
 in the $\pt$ distribution (i.e. those which vanish in the $\pt \to 0$ limit).

The second step to include the regular terms (i.e. that vanish in the
$\pt \to 0$ limit) in the above formula is to add the full NLO result
for the production of the colour singlet \F{} and one additional jet,
and subtract the NLO expansion of the total derivative in
eq.~\eqref{eq:final-first-emsn}, which leads to
\begin{align}
\frac{\mathd\sigma}{\mathd\PhiB\mathd \pt}=\frac{ \mathd \left[e^{-\tilde{S}(\pt)}
                               {\cal L}(\pt)\right]}{\mathd\pt} +
                                             R_f(\pt) + {\cal O}(\alpha^3_s)\,,
\label{eq:quasimaster}
\end{align}
where we used eq.~\eqref{eq:Rf}, namely
\begin{equation}
R_f(\pt) = \frac{\mathd\sigma^{\rm (NLO)}_{\scriptscriptstyle\rm FJ}}{\mathd\PhiB\mathd\pt} - \abarmu{\pt} \left[\frac{\mathd\sigma^{\rm
  sing}}{ \mathd\PhiB\mathd \pt}\right]^{(1)} -
\left(\abarmu{\pt}\right)^2
  \left[\frac{\mathd\sigma^{\rm sing}}{\mathd\PhiB\mathd \pt}\right]^{(2)}\,.
\end{equation}

The first term on the right-hand side of eq.~\eqref{eq:quasimaster}
constitutes the starting point~\eqref{eq:start} for the discussion in
section~\ref{sec:procedure}. In particular, following the
considerations that led from eq.~\eqref{eq:start} to
eq.~\eqref{eq:minnlo-PhiBJ}, and restoring the generation of the
second radiation via the \POWHEG{} mechanism we obtain 
\begin{align}
\frac{\mathd\sigma}{\mathd\PhiBJ}=& \exp[-\tilde{S}(\pt)]\bigg\{ \abarmu{\pt}\left[\frac{\mathd\sigma_{\scriptscriptstyle\rm FJ}}{\mathd\PhiBJ}\right]^{(1)} \left(1+\abarmu{\pt} [\tilde{S}(\pt)]^{(1)}\right)\notag
  \\
&+ \left(\abarmu{\pt}\right)^2\left[\frac{\mathd\sigma_{\scriptscriptstyle\rm FJ}}{\mathd\PhiBJ}\right]^{(2)} + \left(\abarmu{\pt}\right)^3 [D(\pt)]^{(3)}  F^{\tmop{corr}}(\PhiBJ)\bigg\}\notag\\&\times\bigg\{\Delta_{\rm pwg} (\LambdaPWG) + \int\mathd \Phi_{\tmop{rad}} 
  \Delta_{\rm pwg} (\ptrad)  \frac{R (\PhiBJ{}, \Phi_{\tmop{rad}})}{B
  (\PhiBJ{})}\bigg\}+ {\cal O}(\alpha^3_s)\,,
\label{eq:master}
\end{align}
\noindent where $\pt$ is defined in the $\PhiBJ$ phase space.
Equation~\eqref{eq:master} constitutes the master formula for the \minnlo{}
method, to match a fully differential NNLO calculation to a parton
shower.

The NNLO subtraction in eq.~\eqref{eq:master} is accomplished thanks
to the Sudakov form factor that exponentially suppresses the
$\pt\to 0$ limit. We stress that this fact does not imply that the
transverse-momentum spectrum of the colour singlet will be
exponentially suppressed at small $\pt$. The extra emissions beyond
the second one, generated by the parton shower, will eventually modify
the scaling of the transverse-momentum distribution and restore the
correct ${\cal O}(\pt)$ scaling in this
regime~\cite{Parisi:1979se,Bizon:2017rah}. This, for sufficiently
accurate parton showers ordered in transverse momentum, effectively
corresponds to leading logarithmic accuracy in the $\pt$
spectrum.\footnote{We point out that transverse-momentum-ordered
  dipole showers of the type considered in this article are
  leading-logarithmic accurate for the $\pt{}$ distribution~\cite{Dasgupta:2018nvj}.}

\section{Application to Higgs-boson and Drell-Yan production at the LHC}
\label{sec:results}

In this section we apply the \minnlo{} method to hadronic Higgs-boson
production through gluon fusion in the approximation of an infinitely
heavy top quark ($pp\rightarrow H$), and to the Drell-Yan (DY) process
($pp\rightarrow Z \rightarrow \ell^+\ell^-$) for an on-shell $Z$
boson.  Rather than presenting an extensive phenomenological study for
these two processes at the LHC, our goal is to perform a thorough
validation and numerically demonstrate the \nnlops{} accuracy of the
\minnlo{} formula. This requires us to verify two aspects of the
results: firstly, that NNLO accuracy is reached for Born-level
($\PhiB{}$) observables, in particular for the total inclusive cross
section and for distributions in the Born phase space, such as the
rapidity distribution of the colour-neutral boson or, in case of DY,
the leptonic variables. Secondly, it has to be shown that the NLO
accuracy of one-jet ($\PhiBJ$) observables is preserved, in particular
for distributions related to the leading jet.  To this end, we compare
our \minnlo{} results to \minlo{} and to NNLO predictions. We stress
that the results produced by the \minnlo{} method differ from the
previous {\noun NNLOPS}~\cite{Hamilton:2013fea,Karlberg:2014qua} by
higher-order terms. In particular, the latter agree by construction
with the full NNLO for Born variables, therefore we validate our
results by directly comparing to the nominal NNLO. After defining the
general setup, we discuss the validation in the following.

\subsection{Setup}
\label{sec:setup}

We consider 13\,TeV LHC collisions. For the EW parameters we employ
the $G_\mu$ scheme with real $Z$ and $W$ masses, since we consider
on-shell $Z$ bosons.  Thus, the EW mixing angles are given by
$\cos^2\theta_{\tiny{\mbox{W}}}=\mw{}^2/\mz{}^2$ and
$\alpha=\sqrt{2}\,G_\mu \mw{}^2\sin^2\theta_{\tiny{\mbox{W}}}/\pi$.
The following values are used as input parameters:
$G_{\tiny{\mbox{F}}} =1.16639\times 10^{-5}$\,GeV$^{-2}$,
$\mw{}=80.385$\,GeV, $\mz{} = 91.1876$\,GeV, and $\mh{} = 125$\,GeV.
We obtain a branching fraction of
$\textrm{BR}(Z \rightarrow \ell^+\ell^-) = 0.0336310$ from these
inputs for the $Z$-boson decay into massless leptons.  With an
on-shell top-quark mass of $\mt= 172.5$\,GeV and $n_f=5$ massless
quark flavours, we use the corresponding NNLO PDF set with
$\as(\mz{}) = 0.118$ of PDF4LHC15~\cite{Butterworth:2015oua} for
Higgs-boson production and of NNPDF3.0~\cite{Ball:2014uwa} for the DY
results.

For \minlo{} and \minnlo{}, the factorisation scale ($\muF{}$) and
renormalisation scale ($\muR{}$) are determined by the underlying
formalism to be proportional to the transverse momentum of the colour
singlet, as discussed in \sct{sec:description}. Upon integration over
radiation this corresponds to effective scales ($\muR, \,\muF$) of
the order of $\mh{}$ and $\mz{}$ for Higgs-boson and DY production,
respectively.  The latter scales are used to obtain the fixed-order
results. We stress that the \minlo{} and \minnlo{} methods adopt a
dynamical scale during the phase-space integration. As a consequence,
the correspondence between such scales and those used in the
fixed-order predictions presented below is only approximate, and for
this reason one does not expect a perfect agreement between the two
calculations.

Uncertainties from missing higher-order contributions are estimated
from customary $7$-point variations, i.e. through changing the scales
by a factor of two around their central values $\muF{} = \KF{}\,\pt$,
$\muR{} = \KR{}\,\pt$ ($\muF{} = \KF{}\,M$, $\muR{} = \KR{}\,M$ with
$M=\mh$ or $M=\mz$ for the fixed-order results) while requiring
$0.5 \le \KF{}/ \KR{}\le 2$. This implies taking the minimum and
maximum values of the cross section for variations
$(\KF{},\KR{})=(2,2)$, $(2,1)$, $(1,2)$, $(1,1)$, $(1,\tfrac{1}{2})$,
$(\tfrac{1}{2},1)$, $(\tfrac{1}{2},\tfrac{1}{2})$. The formulae for
the scale variation are reported in appendix~\ref{app:scaledep}.

When the scales $\muR{}$ ($\muF{}$) are too small, we freeze them
consistently everywhere, both as arguments of the strong coupling
constant and partons densities, and in the terms of the cross section
that depend explicitly on them. Our choice for the freezing scale is
$1.8$~GeV for Drell-Yan, and $2.5$~GeV for Higgs-boson production.
The choice of these freezing scales is determined from two opposite
requirements: one is to stay above the lower bound of the PDF
parametrization (which is a scale of the order of $1$ GeV), and the
other is to remain close to the region where the Sudakov form factor
is negligible, in order to make sure that the contribution at the
lower integration bound of the quantity defined in~\eqref{eq:start}
stays indeed negligible. This allows for a larger value in the Higgs
case, where the Sudakov suppression is stronger.
We note that our implementation preserves scale compensation and that
the transverse momentum is not affected by this procedure. We stress
again that the aforementioned scales are only used to freeze the
values of $\mu_R$ and $\mu_F$, and they don't act as a cutoff on the
phase space $\PhiBJ$.


We notice that, at the freezing scale, the Sudakov form factor is
already quite small: in the Higgs case, already at $\pt=3$ GeV its
value is $0.01$. In the Drell-Yan case, when $\pt=1.8$~GeV the value
of the Sudakov form factor is $0.1$, when $\pt=1$~GeV it is $0.03$,
and it reaches the value $0.01$ for $\pt=500$~MeV. We remark that, if
we were to exclude from our calculation scale variation points
with $\KF=1/2$, the freezing scale could be taken as low as $1$~GeV,
which is the PDF cutoff, without a relevant change in the result, as
shown in~\tab{freezing}.

\renewcommand\arraystretch{1.3}
\begin{table}[ht]
\begin{center}
\begin{tabular}{c l c | c c c c c c c}
\toprule
 & \minnlo{} &&& \multicolumn{2}{c}{$pp\to H$ (on-shell)}
&& \multicolumn{2}{c}{$pp\to Z \to \ell^+\ell^-$ (on-shell)} &\\
&&&& freezing scale
& $\sigma_{\rm tot}$ [pb]
&& freezing scale
& $\sigma_{\rm tot}$ [fb] &\\
\midrule
& default &&& $2.5$ GeV  & $36.52(2)$ & & $1.8$ GeV & $1849(1)$ &\\
& lower freezing scale    &&& $1.25$ GeV & $36.46(4)$ & & $1$ GeV   & $1845(1)$ &\\
\bottomrule
\end{tabular}
\end{center}
\caption{\label{freezing} Total inclusive cross sections for Higgs-boson 
  production and DY production using the \minnlo{} calculations, for different values
  of the freezing scale.}
\end{table}
Finally, we stress that the region affected by the choice of
the freezing scale is at transverse momenta where non-perturbative
effects start to play a role, and in a realistic simulation these
effects are taken into account by the parton shower Monte Carlo and
its hadronization model.

It is important to bear in mind that we implement the scale variation
in all terms of eq.~\eqref{eq:master}, including the Sudakov
$\tilde{S}$. The latter variation is not present in a standard NNLO
calculation, and therefore probes additional sources of higher-order
corrections. Hence, we expect the resulting scale dependence to be
moderately larger than the one of the NNLO fixed-order predictions,
reflecting the additional sources of perturbative uncertainties in the
\minnlo{} matching procedure. Although such extra sources could be
avoided, we reckon that their inclusion is more appropriate to reflect
the actual uncertainties of the \minnlo{} method.

We have employed the \POWHEGBOX{}
framework~\cite{Nason:2004rx,Frixione:2007vw,Alioli:2010xd} to
implement the \minnlo{} formalism for the {\tt
  HJ}~\cite{Alioli:2010qp} and {\tt ZJ}~\cite{Campbell:2012am}
processes. We compare against the original \minlo{} calculations
of~\citeres{Hamilton:2012np,Hamilton:2012rf}. Fixed-order results (LO,
NLO, NNLO) for on-shell $pp\rightarrow H$ and $pp\rightarrow Z$
production are obtained with the \Matrix{}
framework
~\cite{Grazzini:2017mhc}.
The PDFs are evaluated with the {\tt LHAPDF}~\cite{Buckley:2014ana}
package and all convolutions are handled with {\tt
  HOPPET}~\cite{Salam:2008qg}. Moreover, the evaluation of
polylogarithms is carried out with the {\tt hplog}
package~\cite{Gehrmann:2001pz}.

As a cross-check we have produced NNLO results for
$pp\rightarrow
H$~\cite{Harlander:2002wh,Anastasiou:2002yz,Ravindran:2003um,Ravindran:2002dc}
and
$pp\rightarrow Z\rightarrow
\ell^+\ell^-$~\cite{Hamberg:1990np,vanNeerven:1991gh,Anastasiou:2003yy,Melnikov:2006kv,Anastasiou:2003ds}
also with the {\sc HNNLO}~\cite{Grazzini:2008tf} and {\sc
  DYNNLO}~\cite{Catani:2009sm} codes, which we found to be fully
compatible with the \Matrix{} predictions within their respective
systematic uncertainties. For lepton-related observables we use the
results from {\sc DYNNLO} in our comparison. Unless otherwise stated,
all results presented throughout this section are subject to no cuts
in the phase space of the final-state particles. All showered results
are obtained through matching to the \PYTHIA{8} parton
shower~\cite{Sjostrand:2014zea}, and they are shown at parton level,
without hadronization or underlying-event effects.
Finally, the value of the parameter $p$ of the modified logarithms
of~\eqn{eq:modlog} has to be chosen such that the logarithmic terms
are switched off sufficiently quickly at large transverse momentum, in
order to avoid spurious effects in the region dominated by hard radiation. 
We adopt $p=6$ in the following, that is slightly larger than the
values used in standard
resummations~\cite{Banfi:2012jm,Bizon:2017rah}, and verified that
variations of $p$ lead to very moderate effects that are well within
the quoted uncertainties.  \sloppy{}

\subsection{Inclusive cross section}
\label{sec:inclusive}

\renewcommand\arraystretch{1.3}
\begin{table}[ht]
\begin{center}
\begin{tabular}{c l c | c c c c c c c}
\toprule
 &&&& \multicolumn{2}{c}{$pp\to H$ (on-shell)}
&& \multicolumn{2}{c}{$pp\to Z \to \ell^+\ell^-$ (on-shell)} &\\
 &&&& $\sigma_{\rm inclusive}$ [pb]
& $\sigma/\sigma_{\rm NNLO}$
&& $\sigma_{\rm inclusive}$ [fb]
& $\sigma/\sigma_{\rm NNLO}$ &\\
\midrule
&LO   &&& $12.89(0)_{-17.3\%}^{+23.5\%}$ & 0.325 && $1658(0)_{-12.3\%}^{+11.3\%}$  & 0.881 &\\
&NLO  &&& $29.55(0)_{-15.3\%}^{+19.8\%}$ & 0.745 && $1897(0)_{-4.7\%\phantom{1}}^{+3.0\%\phantom{1}}$ & 1.008 &\\
&NNLO  &&& $39.63(3)_{-10.4\%}^{+10.7\%}$ & 1.000 && $1882(1)_{-0.9\%\phantom{1}}^{+1.1\%\phantom{1}}$ & 1.000&\\
&\minlo{}  &&& $30.40(3)_{-15.0\%}^{+33.3\%}$ & 0.767 && $1774(1)_{-14.8\%}^{+14.2\%}$ & 0.943 &\\
&\minnlo{}  &&& $36.52(2)_{-13.4\%}^{+13.9\%}$ & 0.921 && $1849(1)_{-2.3\%\phantom{1}}^{+1.8\%\phantom{1}}$ & 0.983 &\\
\bottomrule
\end{tabular}
\end{center}
\renewcommand{\baselinestretch}{1.0}
\caption{ \label{rates} Predictions of the total inclusive cross section for Higgs-boson 
production and the DY process 
at the LO, NLO, and NNLO, as well as using the \minlo{} and \minnlo{} calculations. 
For comparison also a column with the ratio to the NNLO cross section is shown.
}
\end{table}
\renewcommand\arraystretch{1}

We report \minnlo{} results for the total inclusive cross section in
\tab{rates}, together with the LO, NLO, NNLO, and \minlo{}
predictions.  We start by discussing the Higgs cross sections:
Compared to \minlo{} we find a $+19\%$ effect by including NNLO
corrections through the \minnlo{} procedure. Furthermore, as expected
from including an additional order in the perturbative series, the
size of the uncertainties due to scale variations are reduced. In
particular the upper variation bound is almost a factor of three smaller
for \minnlo{} in comparison to \minlo{}. The \minnlo{} and NNLO
results agree well within their respective scale-uncertainty bands,
which largely overlap. 
The central values of the two calculations differ by $7.9\%$ as can
be seen from the ratio, and they are included in the uncertainty bands
of each of the two calculations.
The size of the difference is justified by the large perturbative
corrections that characterise Higgs-boson production, which implies
that subleading terms can be sizable.
For processes with smaller corrections these differences will reduce,
as we will see in the context of the DY process below.  As already
mentioned above, it is not expected that \minnlo{} reproduces the NNLO
result exactly, as subleading corrections are treated differently in
the two calculations and the the renormalisation and factorisation
scales are set differently. Consistency between the two NNLO-accurate
predictions within perturbative uncertainties is therefore sufficient,
and shows that the \minnlo{} procedure induces the expected
corrections. The scale uncertainties of \minnlo{} for the total cross
section ($\sim 13\%$) are slightly larger than the NNLO ones
($\sim 10\%$).  There are two main reasons for this behaviour: On the
one hand, \minnlo{} probes scales in both the PDFs and in $\as{}$ that
in the bulk-region of the cross section are much lower ($\sim \pt$)
than in the fixed-order computation, which naturally induces a larger
scale dependence.  On the other hand, we include additional
scale-dependent terms (as pointed out before) that originate from the
analytic Sudakov form factor in the \minnlo{} procedure, which are
absent in a fixed-order calculation, see \app{app:scaledep}. This
induces a more conservative estimate for the theory uncertainties of
the \minnlo{} predictions.

In the case of the DY results in \tab{rates}, we observe that conclusions
similar to the case of Higgs production can be drawn, albeit with
significantly smaller corrections: The effect of the \minnlo{}
procedure is to increase the \minlo{} cross section by about
$5\%$. Again the scale uncertainties are vastly reduced, in the case
of DY by almost a factor of 10.  The \minnlo{} result is only $1.7\%$
below the NNLO prediction and they are in good agreement within their
respective scale uncertainties, which are extremely small.  Roughly
speaking, scale uncertainties are $2\%$ for \minnlo{}, which is a bit
larger than the $1\%$ uncertainties at NNLO. Given the above
discussion about the formal differences between \minnlo{} and NNLO
fixed-order computations, these results are very compelling and
provide a numerical proof of the accuracy of the total inclusive cross
section of the \minnlo{} procedure.
We will now turn to validating the \minnlo{} results also for
differential observables.

\subsection{Distributions for Higgs-boson production}
\begin{figure}
  \centering
  \includegraphics[width=0.48\columnwidth]{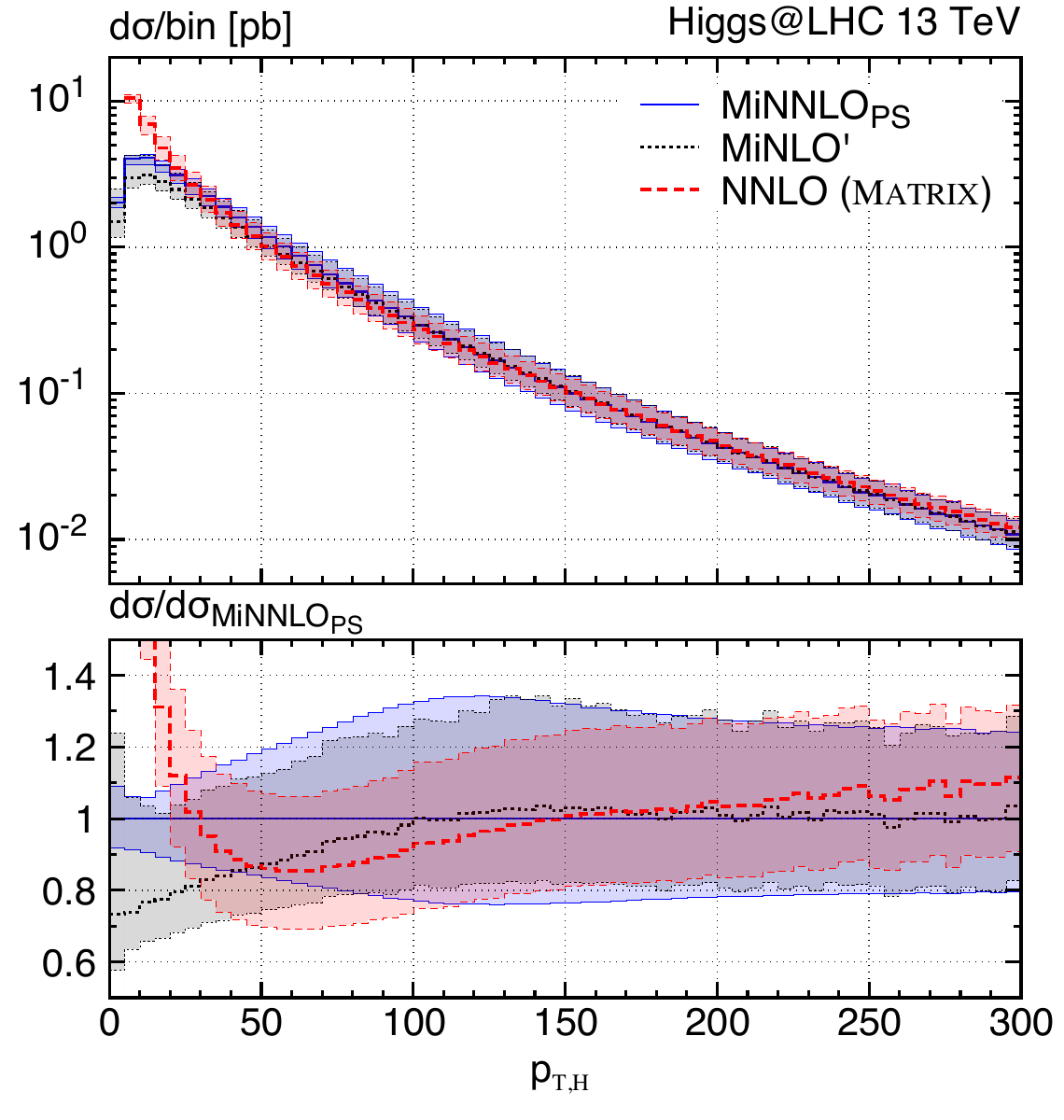}
  \includegraphics[width=0.48\columnwidth]{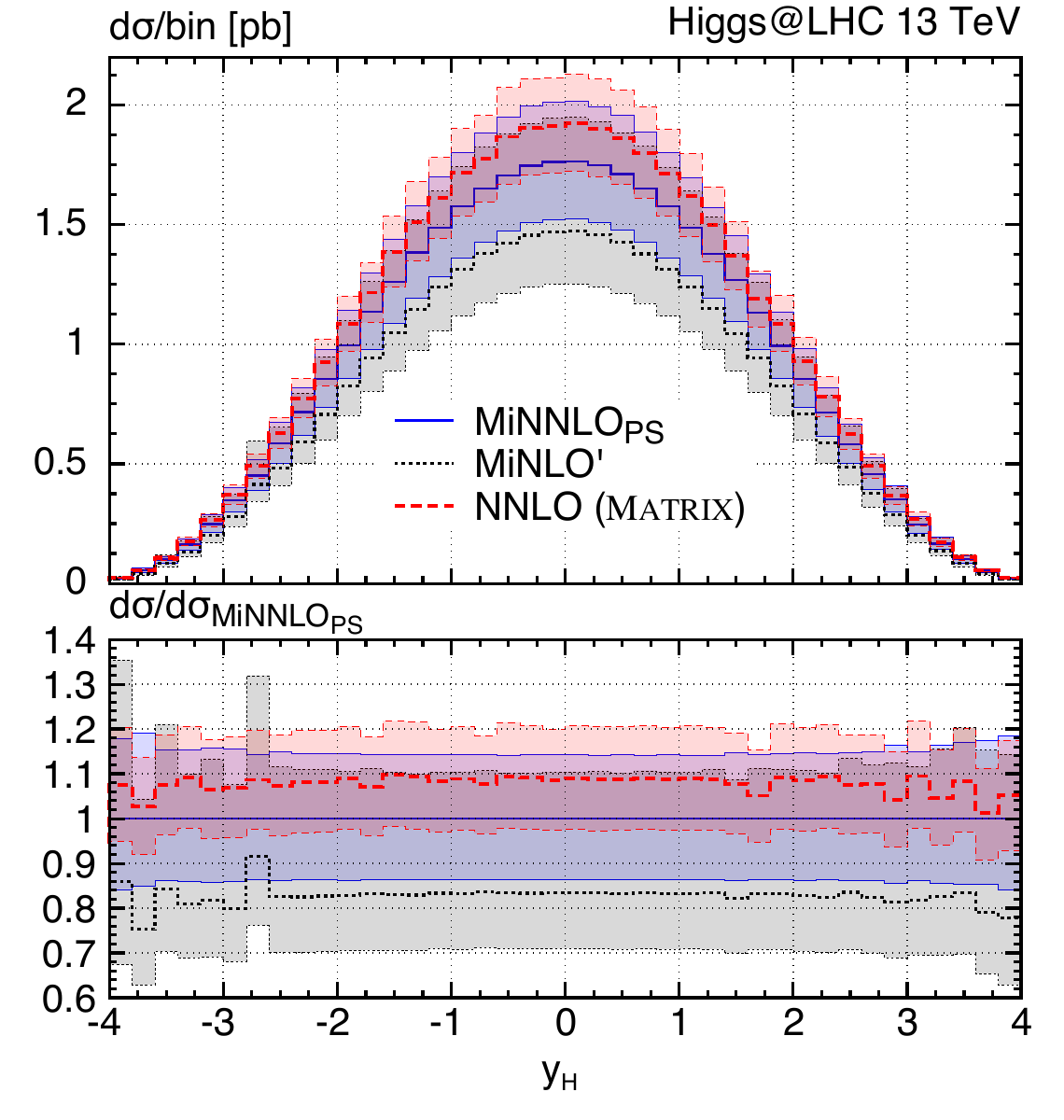}
  \caption{Distribution in the transverse momentum (left) and rapidity (right) of the Higgs boson for \minnlo{} (blue, solid), \minlo{} (black, dotted), and
    NNLO (red, dashed).}
  \label{fig:rap_distribution_Higgs}
\end{figure}

We first consider the case of Higgs-boson production. The figures of this
section are organized as follows: the main frame shows the results
from \minnlo{} (blue,solid) and \minlo{} (black, dotted) after parton showering,
as well as NNLO predictions (red, dashed), and all results are
reported in units of cross section per bin (namely, the sum of the
values of the bins is equal to the total cross section, possibly
within cuts). In an inset we display the bin-by-bin ratio of all the
histograms which appear in the main frame to the \minnlo{} curve.
The bands correspond to the residual uncertainties that are computed
from scale variations as indicated in \sct{sec:setup}.

The transverse-momentum distribution of the Higgs boson ($\pth$) is
shown in the left panel of \fig{fig:rap_distribution_Higgs}.  At fixed
order this distribution diverges in the $\pth\rightarrow 0$ limit, and
the accuracy is effectively reduced to NLO across the spectrum. By
comparing \minnlo{} and \minlo{} curves, we observe that the NNLO
corrections are included consistently in the low-$\pth$ region through
the \minnlo{} procedure.
The additional NNLO (two-loop) contributions in the \minnlo{} matching
are spread in a way that is similar in spirit to how analytic
resummations are combined with fixed order. This is enforced through
the use of the modified logarithms in \eqn{eq:modlog}. At large
$\pth$, where the \minnlo{} and \minlo{} predictions have both NLO
accuracy, we expect the \minnlo{} procedure not to alter the \minlo{}
distribution, as can be seen from the figure.
The harder tail of the NNLO curve is due to the different (less
appropriate) scale choice in the fixed-order calculation, set to the
Higgs-boson mass rather than to $\pth$.

The rapidity distribution of the Higgs boson ($\yh$) in the right
panel of \fig{fig:rap_distribution_Higgs} is the most relevant
observable for which \minnlo{} needs to be validated against the NNLO
result. Indeed, we find that up to statistical fluctuations the
NNLO/\minnlo{} ratio of the distribution is completely flat, which
shows their equivalence. Henceforth, the difference of the two results
is purely due to the normalisation, i.e. the total inclusive cross
section, which has been discussed in detail in \sct{sec:inclusive} and
requires no further comments. In particular, the conclusions about the
uncertainty bands and the size of the corrections drawn from
\tab{rates} hold also for the rapidity distribution shown in
\fig{fig:rap_distribution_Higgs}.

\begin{figure}
  \centering
  \includegraphics[width=0.48\columnwidth]{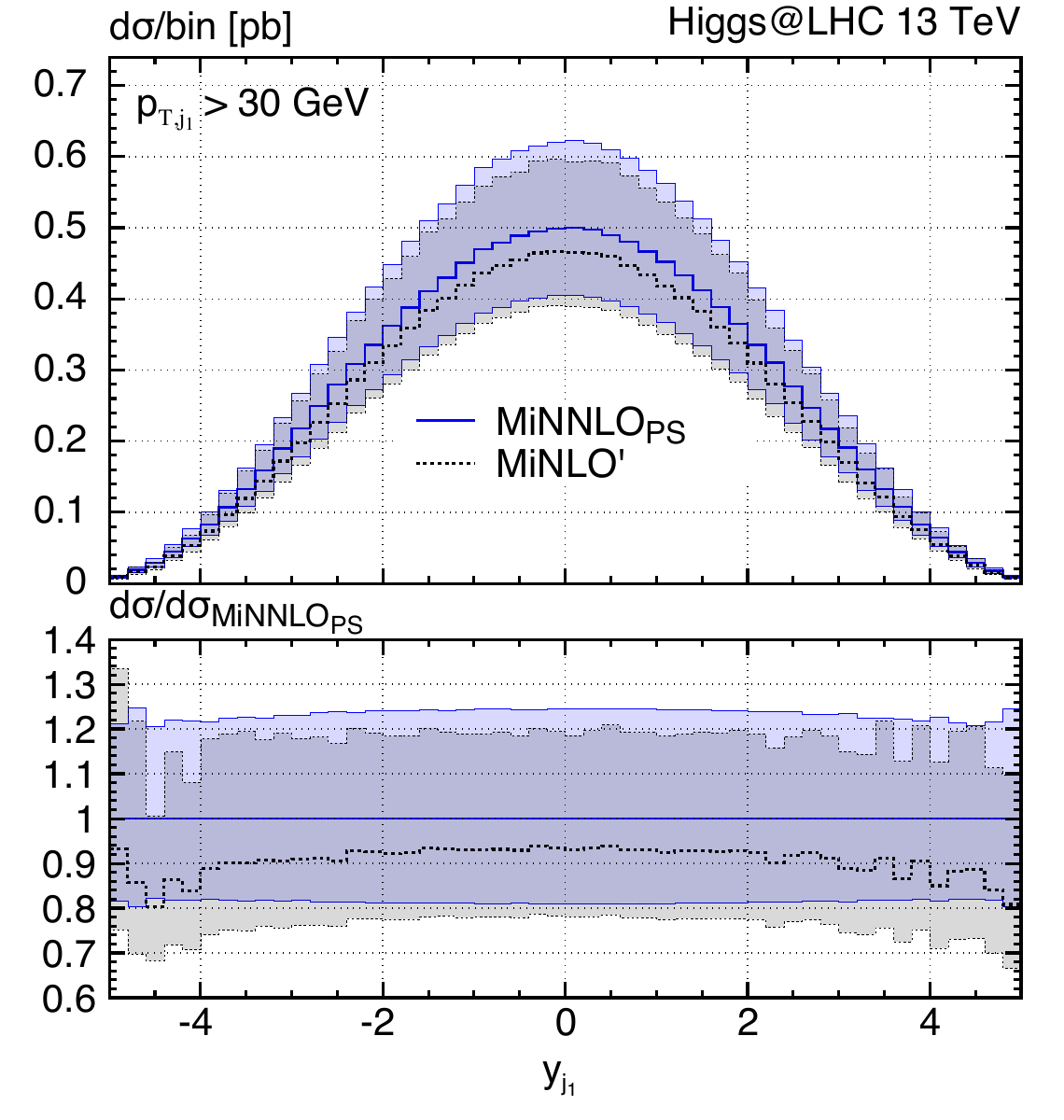}
  \includegraphics[width=0.48\columnwidth]{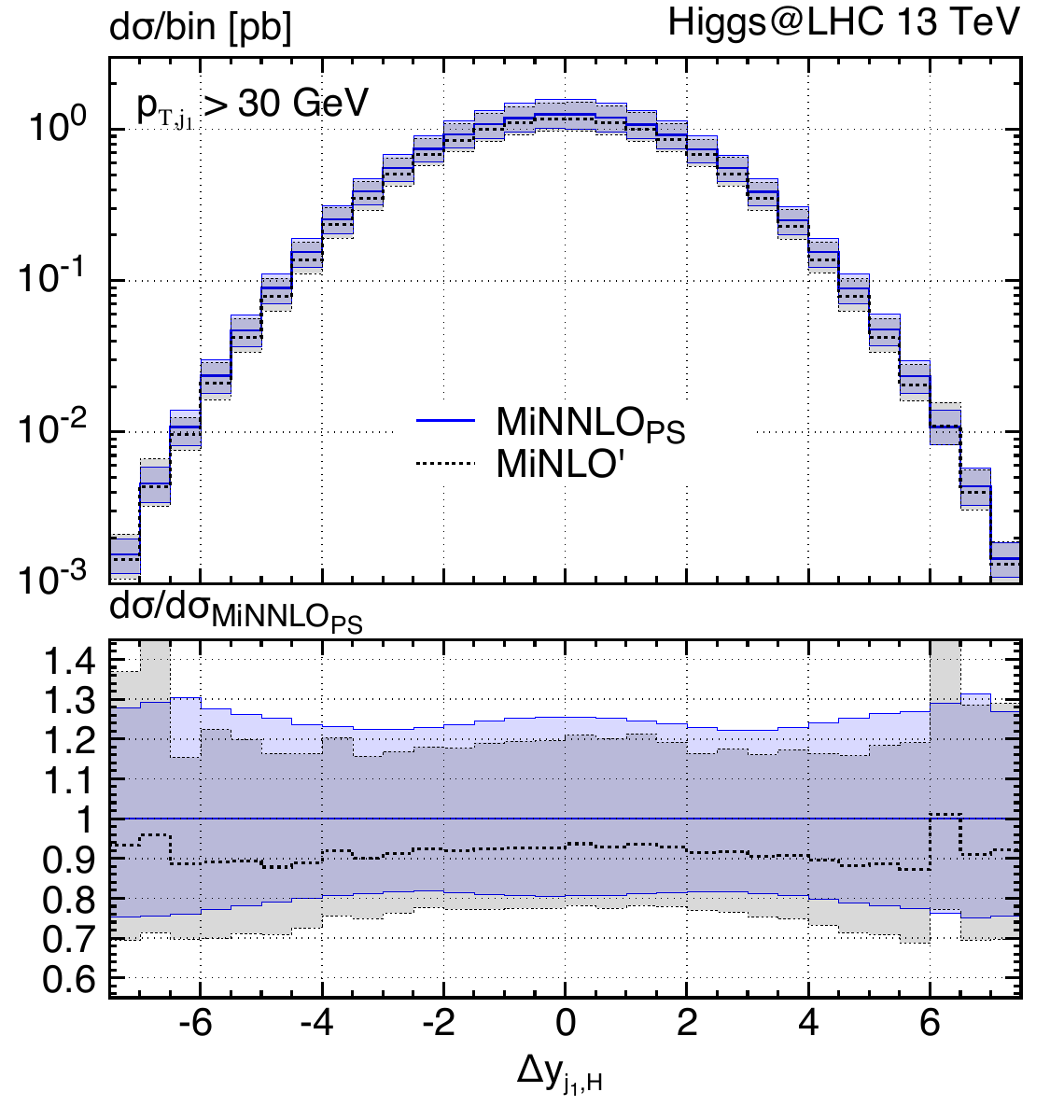}
  \caption{Distribution in the rapidity of the leading-jet (left) and its rapidity difference with the Higgs boson (right) for \minnlo{} (blue, solid) and \minlo{} (black, dotted).}
  \label{fig:rap_distribution_Higgs_jets_30}
\end{figure}

We conclude our discussion of the results for Higgs-boson production
by looking at jet-related distributions.  We note that the
transverse-momentum distribution of the leading jet is very similar to
the one of the Higgs boson which is why we refrain from showing it
here and refer to the discussion for $\pth$.
Figure~\ref{fig:rap_distribution_Higgs_jets_30} shows the rapidity
distribution of the leading jet in the left panel, and the rapidity
difference between the leading jet and the Higgs boson in the right
panel. Jets in this case are defined using the anti-$k_T$ clustering
\cite{Cacciari:2008gp} with a radius $R=0.4$, and a minimum transverse
momentum of $\ptj=30$\,GeV.  For such observables both \minnlo{}
and \minlo{} are NLO accurate, and one expects that the \minnlo{}
result does not differ from the \minlo{} prediction significantly,
i.e. beyond perturbative uncertainties.  
In particular, this numerical check is important to ensure that the
implementation and spreading of the $[D(\pt)]^{(3)}$ terms in the
$\PhiBJ$ phase space, as described in \sct{sec:implementation}, is
appropriate. We refrain from showing the NNLO curve for these
distributions, as it does not add any relevant information to these
tests. Indeed, we observe that, by and large, the \minlo{}/\minnlo{}
ratio is flat for both distributions in
\fig{fig:rap_distribution_Higgs_jets_30}, and that the two results
agree very well within perturbative uncertainties. We have repeated
these checks for various $\ptj$ thresholds in the jet definition, with
the same conclusions. In particular, we found that for hard
configurations ($\ptj\gtrsim 60$\,GeV) the \minnlo{} and \minlo{}
results become essentially identical, as expected from the fact that
\minnlo{} induces no additional corrections in phase-space regions
where the radiation is hard.  Furthermore, a similar level of
agreement is found also for the azimuthal angle between the leading
jet and the Higgs boson.

\subsection{Distributions for Drell-Yan production}

\begin{figure}
  \centering
  \includegraphics[width=0.48\columnwidth]{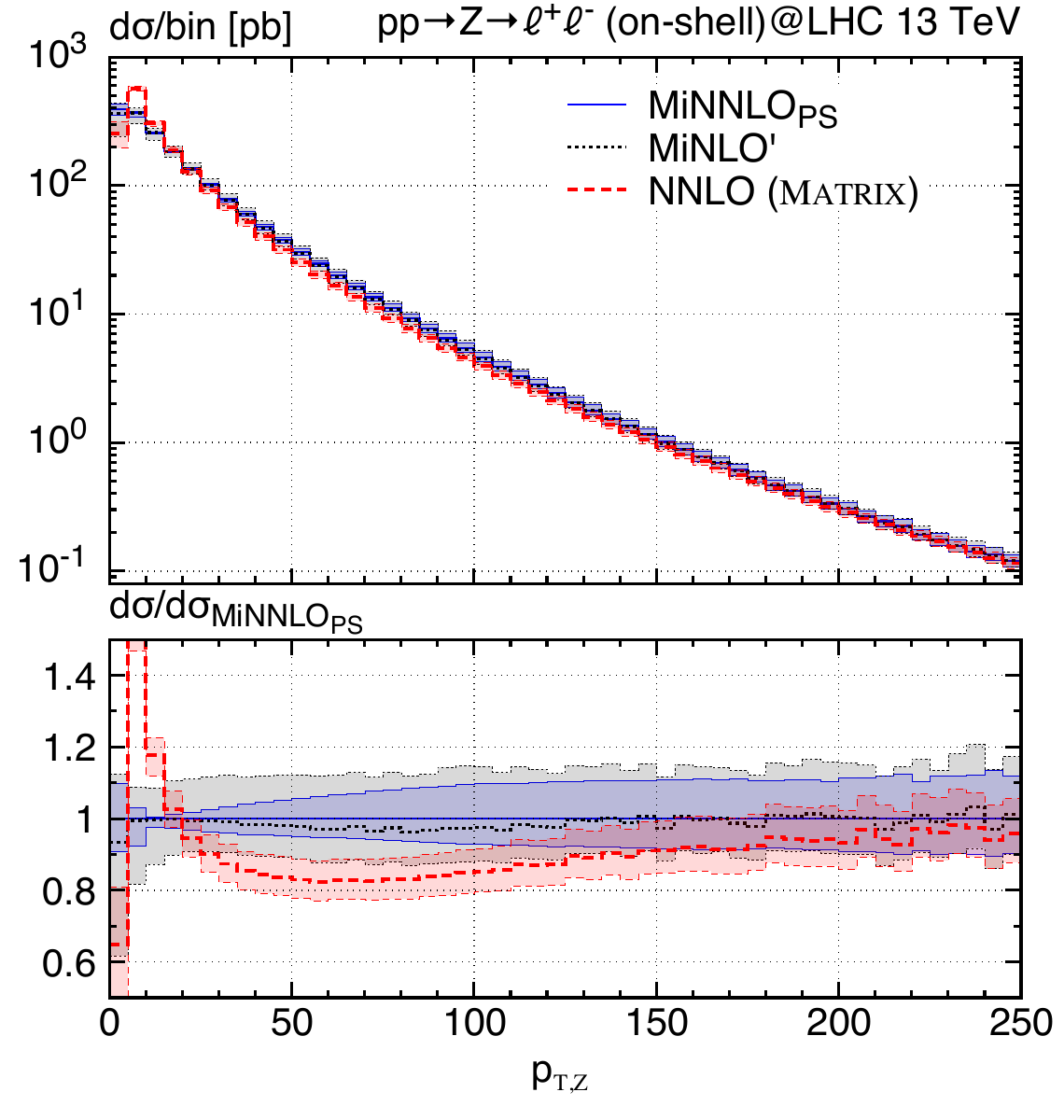}
  \includegraphics[width=0.48\columnwidth]{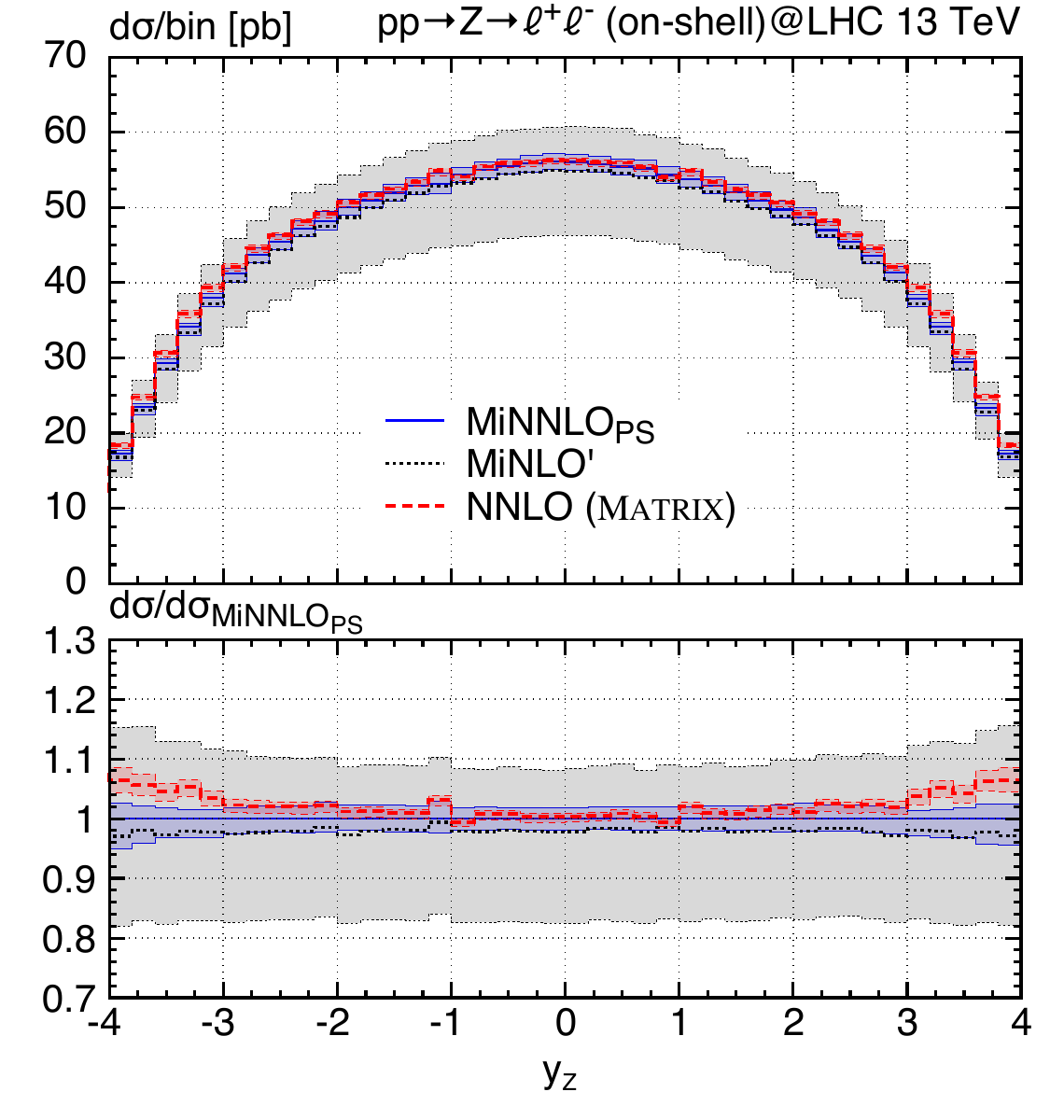}
  \caption{Distribution in the transverse momentum (left) and rapidity (right) of the $Z$ boson for \minnlo{} (blue, solid), \minlo{} (black, dotted), and
    NNLO (red, dashed).}
  \label{fig:rap_distributions_DY}
\end{figure}

We now move on to discuss distributions for the DY process. Since
most of the conclusions are similar to the ones for Higgs-boson
production, we keep the discussion rather brief. In addition to the
results discussed for the Higgs, we also study the kinematics
of the leptons arising from the decay of the $Z$ boson.

Figure~\ref{fig:rap_distributions_DY} shows the transverse-momentum
distribution of the $Z$ boson ($\ptz$) in the left panel, and its
rapidity distribution ($\yz$) in the right panel. As seen before, the
corrections are smaller in the case of the DY process, but the general
behaviour is the same as for Higgs-boson production: At large $\ptz$
the \minnlo{} result is essentially identical to the \minlo{} one,
while the additional NNLO terms enter at smaller values of $\ptz$. The
NNLO spectrum diverges at small $\ptz$, and is harder in the tail due
to the different scale setting.  For the $\yz$ distribution, the
\minnlo{} uncertainties are significantly reduced with respect to the
\minlo{} ones.  In the central region ($|\yz|<3$) the NNLO/\minnlo{}
ratio is nicely flat up to statistical fluctuations, and the two
results agree within their respective uncertainties. For very forward
$Z$ bosons ($|\yz|>3$), on the other hand, we observe a slight
increase of the NNLO/\minnlo{} ratio. We have checked explicitly that
without the \PYTHIA{8} parton shower, i.e.\ at the level of Les
Houches events, this effect is more moderate and the NNLO and
\minnlo{} uncertainty bands overlap in the forward region.  In fact,
we noticed that already for the \minlo{} prediction, \PYTHIA{8} has
the same effect, making the $Z$-boson rapidity distribution slightly
more central.\footnote{We observed that part of this effect can be
  attributed to the global recoil adopted by \PYTHIA{8} for ISR. The
  difference from the NNLO prediction is reduced if one uses a more
  local scheme for the parton-shower recoil, e.g. via the flag
  \tt{SpaceShower:dipoleRecoil=1}.}

\begin{figure}
  \centering
  \includegraphics[width=0.48\columnwidth]{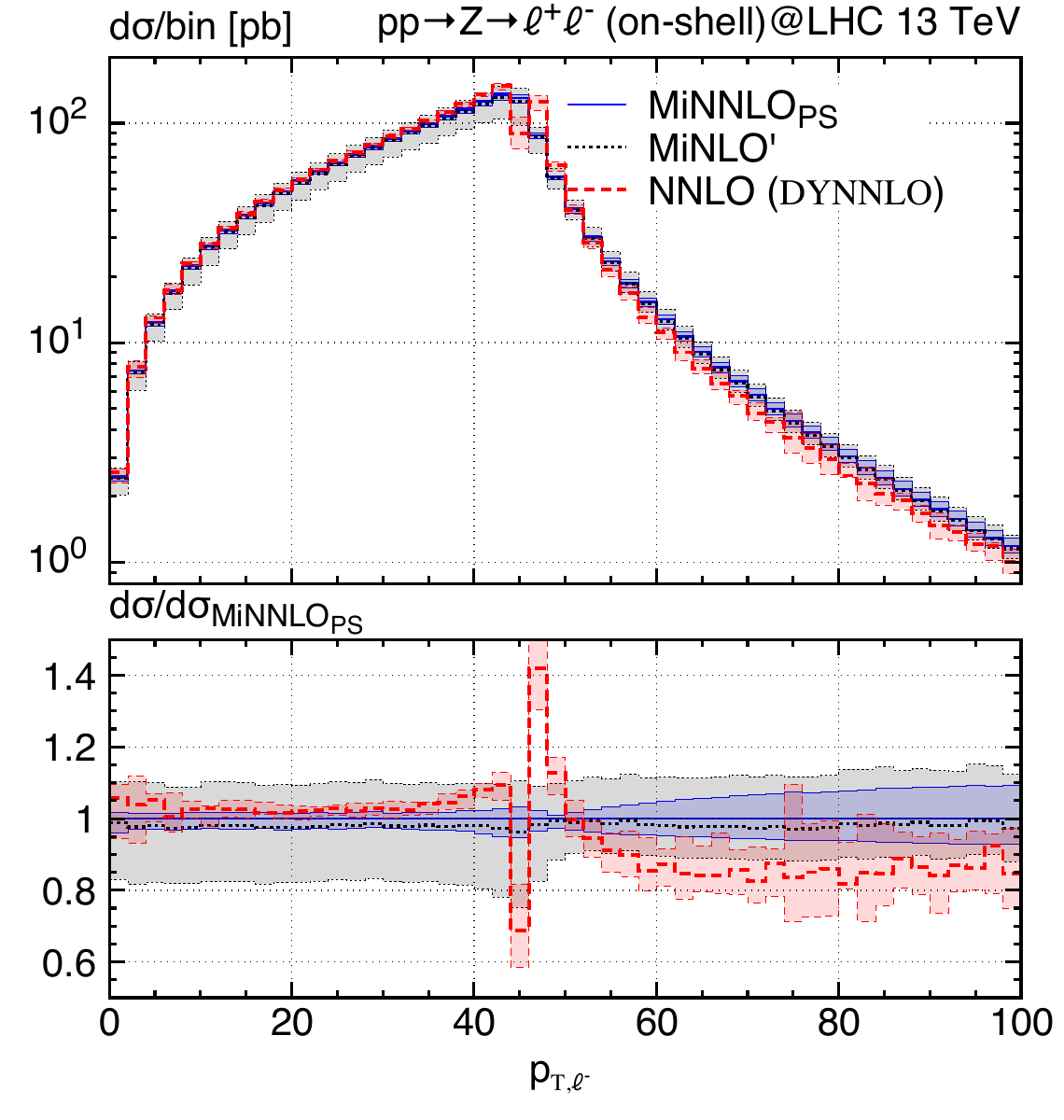}
  \includegraphics[width=0.48\columnwidth]{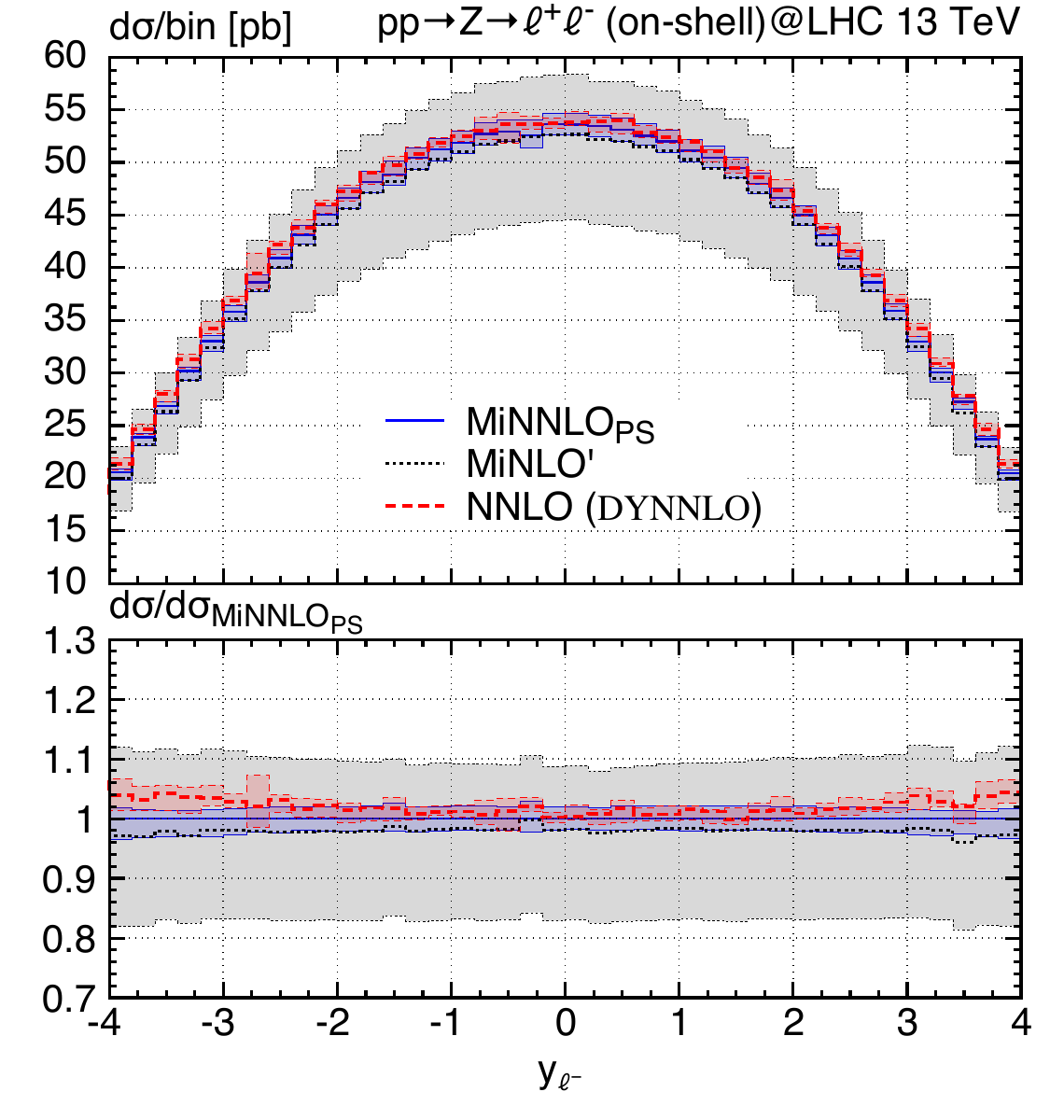}
  \caption{Distribution in the transverse momentum (left) and rapidity (right) of the negatively charged lepton for \minnlo{} (blue, solid), \minlo{} (black, dotted), and
    NNLO (red, dashed).}
  \label{fig:lept_distributions_DY}
\end{figure}

Next, we consider the transverse-momentum distribution of the
negatively charged lepton ($\ptl$) and its rapidity distribution
($\yl$) in the left and right panels of
\fig{fig:lept_distributions_DY}, respectively. For the rapidity
distribution the relative behaviour between \minnlo{}, \minlo{}, and
NNLO is essentially identical to the one of the $Z$-boson rapidity and
does not require any further discussion. As far as the
transverse-momentum spectrum is concerned, the NNLO result shows a
very peculiar behaviour for $\ptl=\mz{}/2$, which reflects the
perturbative instability associated with the fact that the leptons at
LO are back-to-back and can share only the available partonic
centre-of-mass energy $\sqrt{\hat s} = \mz{}$, so that their transverse
momenta can be at most $\ptl\leq \mz{}/2$. Beyond this value the NNLO
result is therefore effectively only NLO accurate, which can be also seen
from the increased uncertainty band. Since such an instability is
related to soft-gluon effects, this feature is cured in both the
\minnlo{} and \minlo{} results, which are in good agreement with each
other in terms of shape. Again the \minnlo{} uncertainty band is
significantly smaller than the \minlo{} one, and we observe a rather
constant correction, of the order of $\sim 5-10\%$, due to the
additional NNLO terms.

\begin{figure}
  \centering
  \includegraphics[width=0.48\columnwidth]{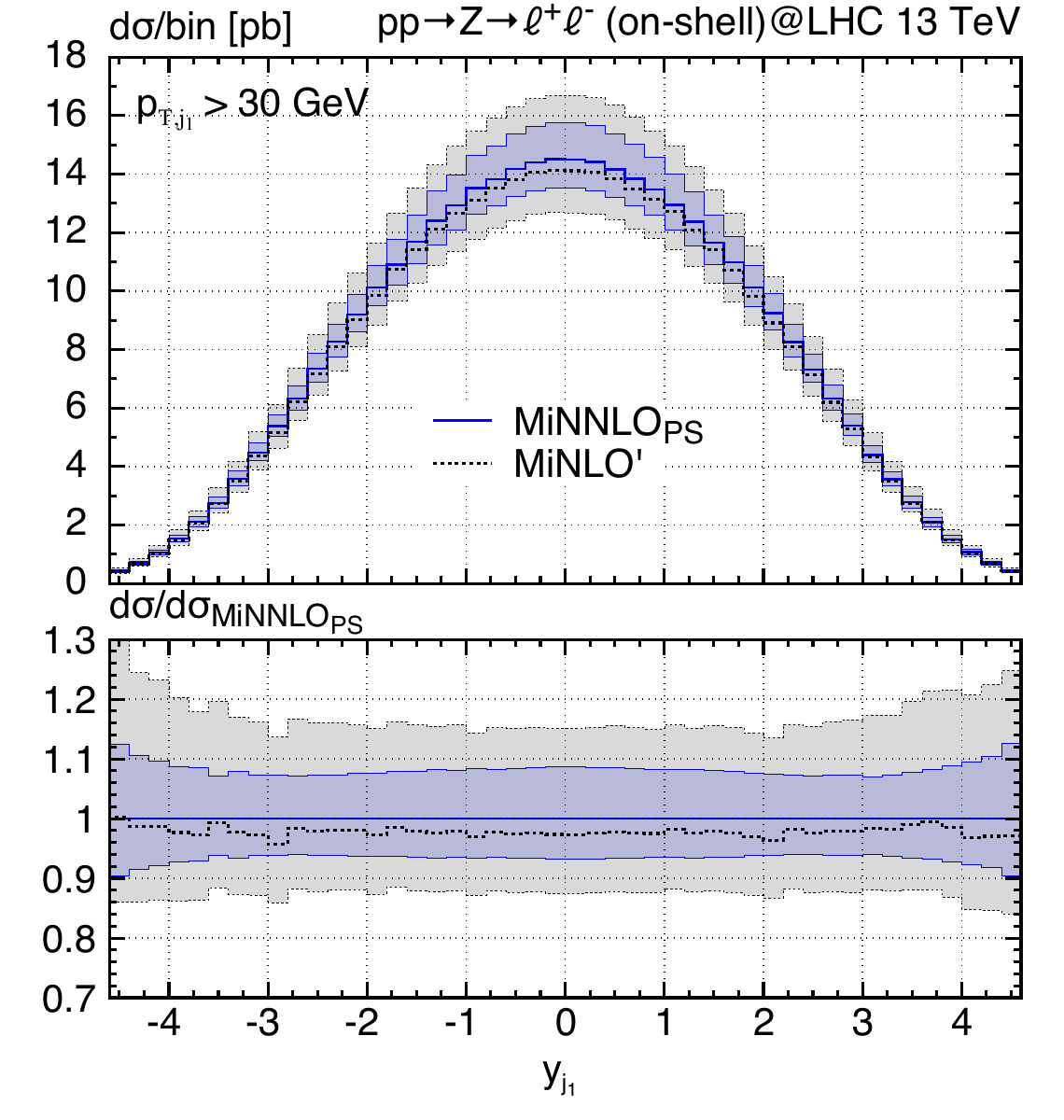}  \includegraphics[width=0.48\columnwidth]{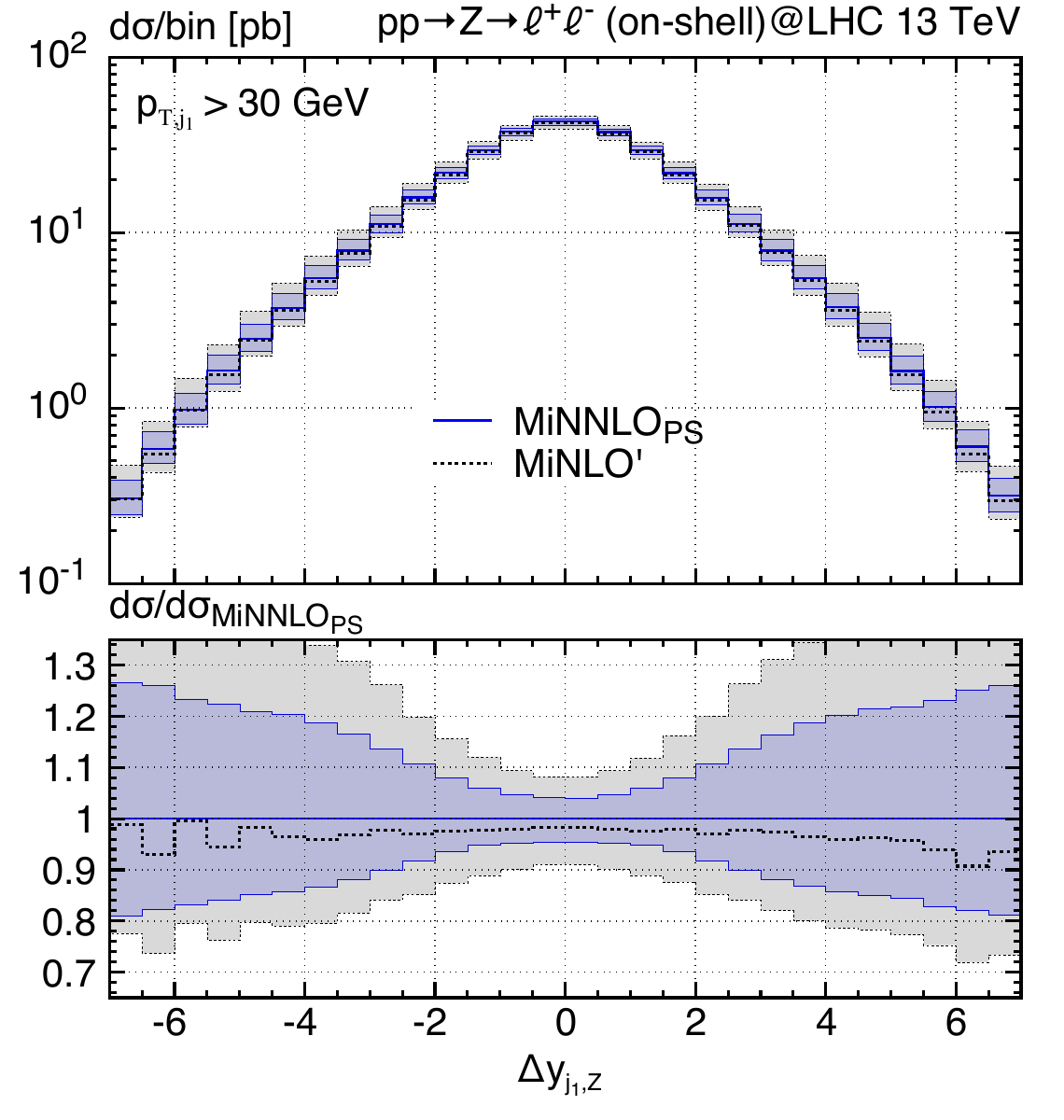}
  \caption{Distribution in the rapidity of the leading-jet (left) and its rapidity difference with the $Z$ boson (right) for \minnlo{} (blue, solid) and \minlo{} (black, dotted).}
  \label{fig:rap_distribution_DY_jets_30}
\end{figure}

Finally, also for the DY process the jet-related observables are fully
consistent within uncertainties when comparing \minnlo{} and \minlo{}
predictions, as can be seen in \fig{fig:rap_distribution_DY_jets_30}.
However, the size of their uncertainty bands is very different. This is
due to the fact that in the original \minlo{} prediction a different
prescription for the scale variation was adopted, that also involved the
integration boundaries of the Sudakov form factor.
We have checked that by using our prescription in \minlo{}
the uncertainty band becomes comparable to the \minnlo{} one.
We stress again that we have tested a variety of $\ptj$ thresholds in
the jet definition, and also looked at the azimuthal angle between the
leading jet and the $Z$ boson, and found consistent results
throughout.

\section{Summary}
\label{sec:summary}

In this article we have presented a novel approach, dubbed \minnlo{},
to combine NNLO QCD calculations with parton showers for colour-singlet 
production at the LHC.  The method is based on the \minlo{}
procedure, which achieves NLO accurate predictions simultaneously in
the zero-jet phase space $\PhiB$ and in the one-jet phase space
$\PhiBJ$. 
The necessary terms to achieve NNLO accuracy are derived by
establishing a connection of the \minlo{} and \POWHEG{} methods
with the structure of transverse-momentum resummation in direct space.
The consistent inclusion of these terms on top of a \minlo{} computation 
allows us to achieve \nnlops{}
accuracy for a variety of collider reactions.

We have discussed in detail a suitable implementation of the NNLO
corrections within the \minlo{} formalism, and their spreading in the
$\PhiBJ$ phase space.  The resulting matching preserves the leading
logarithmic structure of the shower Monte Carlo for showers ordered in
the transverse momentum, 
and the final result is NNLO accurate in the
zero-jet phase space while being NLO accurate in the one-jet phase
space. The combination of the two multiplicities does not require any
unphysical merging scale.

As a proof of concept, we have applied the approach to hadronic Higgs
production in the heavy-top limit and to the DY process, where a pair
of leptons is produced via the decay of an on-shell $Z$ boson. Our
results show that NNLO accuracy is reached both for the total
inclusive cross section and for Born-level distributions. Differences
with NNLO fixed-order results arise only from terms beyond the nominal
accuracy, and the two calculations agree well for such observables
within the respective perturbative uncertainties estimated from scale
variations.  As expected, we observe a significant reduction of the
scale dependence with respect to the \minlo{} results, in line with
the inclusion of the NNLO corrections. It was further verified that
for jet-related observables in the $\PhiBJ$ phase space, where the
accuracy of \minnlo{} and \minlo{} is formally identical, no
significant effects are induced by the \minnlo{} corrections.

The algorithm is very efficient, and NNLO accuracy is achieved
directly at generation time without any additional reweighting. The
total \minnlo{} simulation requires just $50\%$ more CPU time than the
usual \minlo{} computation. This makes it suitable for the application
to more involved colour-singlet processes, such as vector-boson pair
production, which is of significant phenomenological interest.  A
potential limitation of the algorithm concerns systems with very low
invariant mass, such as low-mass diphoton production, whose $\pt{}$
distribution is peaked towards non-perturbative scales. In this
situation, the Sudakov form factor, which is responsible for the NNLO
subtraction of the infrared singularities, can become intrinsically
non-perturbative.  Nevertheless, such scenarios are commonly not of
experimental interest.  Finally, the \minnlo{} approach could be
generalized to the production of massive coloured final states, such
as top-quark pair production. Detailed studies of further applications
of the \minnlo{} method are left for future work.

\section*{Acknowledgements}
We wish to thank Alexander Huss and Alexander Karlberg for useful
correspondence in the course of the project, and Keith Hamilton for
helpful discussions about the details of the MiNLO method.
We also would like to thank Markus Ebert for pointing out a 
sign mistake in a term entering the NNLO hard-coefficient function in \eqn{eq:new_coeffs}, which has a moderate numerical impact.
We are grateful to CERN, Max Planck Institute f\"ur Physik, and
Universit\`a Milano Bicocca for hospitality while part of this project
was carried out.
This work was supported in part by ERC Consolidator Grant HICCUP
(No. 614577).  The work of PM is supported by the Marie Sk\l{}odowska
Curie Individual Fellowship contract number 702610 Resummation4PS.
The work of ER was supported in part by a Marie Skłodowska-Curie Individual
Fellowship of the European Commission's Horizon 2020 Programme under
contract number 659147 PrecisionTools4LHC. P.N. has performed
part of this work while visiting CERN as Scientific Associate, and
also acknowledges support from Fondazione Cariplo and Regione Lombardia,
grant 2017-2070, and from INFN.

\clearpage
\appendix

\section{Phase-space parametrisation for the $[D(\pt)]^{(3)}$ term}
\label{app:spreading}

In this appendix we define the phase-space mapping from $\PhiBJ$ to
$\PhiB$ adopted for initial-state radiation in \POWHEG{}, and
discussed in section 5.5.1 of ref.~\cite{Frixione:2007vw}. The
projection is defined by performing a longitudinal boost of the \FJ{}
system to a frame where \F{} has zero rapidity, followed by a
perpendicular boost that modifies the transverse momentum of \F{} so
that it is equal to zero, followed by a longitudinal boost, exactly
opposite to the first one, that restores the original rapidity of
\F{}.  After this sequence of boosts, the rapidity of \F{} remains
unchanged, but its transverse momentum has become zero, thus yielding
a kinematic configuration in the Born phase space $\PhiB$. The
$\PhiBJ$ phase space can then be expressed in a factorised form:
\begin{equation}
  \mathd \PhiBJ = \mathd \PhiB \mathd \Phi_{\tmop{rad}},\quad
  \mathd \Phi_{\tmop{rad}} = \frac{s}{(4 \pi)^3} \frac{\xi}{1 - \xi} \mathd
  \xi \mathd \phi \mathd y,
\end{equation}
where $s$ is the square of the total incoming energy, and
\begin{align}
  \xi & = \frac{2 k^0}{\sqrt{s}},& \phantom{,}k^0&=\mbox{energy of the radiated parton,} \\
  y & = \cos \theta,& \phantom{,}\theta&=\mbox{scattering angle of the radiated parton,} \\
  \phi & , &\phantom{,}\phi&=\mbox{azimuth of the radiated parton,} 
\end{align}
which are all defined in the centre-of-mass frame of the \FJ{} system. The transverse momentum is
given by
\[ \pt^2 = \frac{s}{4} \xi^2 (1 - y^2) . \]

Denoting $\PhiBJbar \equiv \left.\PhiBJ' \right|_{\PhiB' = \PhiB} $,
the $\Fcorr{}$ factor~\eqref{eq:fcorr} becomes
\begin{equation}
  \Fcorr (\PhiBJ) = \frac{ J_\ell (\PhiBJ)}{  \sum_{l'} \int
  \mathd \Phi'_{\tmop{rad} }J_{l'} (\PhiBJbar )\delta (\pt -
  \pt')}.
\end{equation}
So, we get
\begin{equation}
  (\Fcorr (\PhiBJ))^{-1}  =  J_\ell^{-1} (\PhiBJ) \int \mathd \xi\,
  \mathd \phi \,
  \mathd y \frac{s}{(4 \pi)^3} \frac{\xi}{1 - \xi}\, J_\ell (\PhiBJbar )\, \delta \left( \sqrt{\frac{s}{4} \xi^2 (1
  - y^2)} - \pt \right).
\end{equation}
We now replace $\bar{s} = s (1 - \xi)$, where $\bar{s}$ is the
virtuality of the \FJ{} system, and multiply and divide by
$2\pt$. After rearranging the delta function we get
\begin{equation}
  (\Fcorr (\PhiBJ))^{-1}  =  J_\ell^{-1} (\PhiBJ) \frac{\bar s}{(4 \pi)^3}\int \mathd \xi
\,  \mathd \phi\, \mathd y\frac{\xi }{(1 - \xi)^2}\, J_\ell (\PhiBJbar )\, 2 \pt \delta \left( \frac{\bar{s}}{4} \frac{\xi^2 (1 - y^2)}{1 - \xi} - \pt^2\right).
\end{equation}
We introduce a variable
\begin{equation}
t = \frac{\xi^2}{1 - \xi}\,,
\end{equation}
which is a monotonically increasing function of $\xi$ in the range $[0,\infty]$ for $\xi\in[0,1]$. 
We have
\[ \mathd t = \frac{2 \xi - \xi^2}{(1 - \xi)^2} \mathd \xi, \]
and obtain
\begin{eqnarray}
\label{eq:reweight}
  (\Fcorr (\PhiBJ))^{-1} & = &J_\ell^{-1} (\PhiBJ)
                               \frac{\bar{s}}{(4 \pi)^3} \int \mathd t
                               \mathd \phi \mathd y \frac{(1 - \xi)^2}{\xi (2 -
  \xi)} \frac{\xi}{(1 - \xi)^2} J_\ell
  (\PhiBJbar ) 2 \pt \delta \left( \frac{\bar{s}}{4} t (1 - y^2) - \pt^2
  \right) \nonumber \\
& = & J^{-1}_\ell (\PhiBJ)  \frac{4 \pt}{(4 \pi)^2} \int \,
  \mathd y\, \frac{1}{1 - y^2}  \frac{1}{2 - \xi} J_\ell(\PhiBJbar )\, \Theta \left( t_{\max} (1 - y^2) -
  \frac{4 \pt^2}{\bar{s}} \right)\,,
\end{eqnarray}
where
\[ t = \frac{4 \pt^2}{\bar{s} (1 - y^2)},\quad \xi = - \frac{t}{2} +
   \sqrt{\frac{t^2}{4} + t}\,, \]
\[ \  \]\
and $t_{\max}$ is defined as the $t$ value corresponding to
$\xi_{\max}$, namely
\[ \xi_{\max} = 1 - \max \left\{ \frac{2 (1 + y) \bar{x}_1^2}{\sqrt{(1 +
   \bar{x}_1^2)^2 (1 - y)^2 + 16 y \bar{x}_1^2} + (1 - y) (1 - \bar{x}_1^2)},
   (y \rightarrow - y, 1 \rightarrow 2) \right\} . \]
The variables $\bar{x}_{1,2}$ are the
momentum fractions of the two initial-state
partons in the $\PhiB$ phase space, and are
defined as~\cite{Frixione:2007vw}
\begin{equation}
\bar{x}_1 \equiv x_1\,\sqrt{1-\xi}\sqrt{\frac{2-\xi (1+y)}{2-\xi
    (1-y)}}\,,\qquad \bar{x}_2 \equiv x_2\,\sqrt{1-\xi}\sqrt{\frac{2-\xi (1-y)}{2-\xi
    (1+y)}}\,,
\end{equation}
with $x_{1,2}$ being the momentum fractions of the two initial-state
partons in the $\PhiBJ$ phase space.
For most choices of the function $J_\ell (\PhiBJ)$ discussed in
section~\ref{sec:implementation}, the above integral must be evaluated
numerically via usual Monte Carlo techniques. However, for the choice
$J_\ell (\PhiBJ)=1$, we can perform the $y$ integration
analytically. We need the following elementary integral
\begin{eqnarray}
  I (y) & = & \int \mathd y \frac{1}{(1 - y^2) \left( 2 + \frac{K}{2 (1 -
  y^2)} - \sqrt{\frac{K}{1 - y^2} + \frac{K^2}{4 (1 - y^2)^2}} \right)} \nonumber \\
  & = & \frac{1}{2} \ln \left(  \frac{1 + y}{1 - y} \right) + \frac{1}{4}
  \ln \frac{\sqrt{K^{- 1} (1 - y^2) + \frac{1}{4}} + \frac{1}{2} + 2 K^{- 1}
  (1 - y)}{\sqrt{K^{- 1} (1 - y^2) + \frac{1}{4}} + \frac{1}{2} + 2 K^{- 1} (1
  + y)}\,.
\end{eqnarray}
The limits of integration have to be computed numerically, by finding
the minimum and maximum value such that the $\Theta$ function in
eq.~\eqref{eq:reweight} is 1. Denoting by $y_{1}$ and $y_2$ the lower
and upper integration boundaries, respectively, we simply find that
\begin{equation}
    (\Fcorr (\PhiBJ))^{-1}  = \frac{\pt}{4
    \pi^2}  (I (y_2) - I (y_1)) \,.
\end{equation}

Finally, we conclude this appendix by reporting the collinear
approximation for $J_\ell (\PhiBJ)$ discussed in
section~\ref{sec:implementation}. Its expressions are taken from
ref.~\cite{Nason:2006hfa} and adapted to the notation of this
appendix. For gluon-initiated processes, the different flavour configurations read
\begin{align}
\label{eq:APsplitting-gluon} 
 J_{qg} (\PhiBJ) =&\, C_F \frac{\as}{2\pi} \frac{1+\xi^2}{(1-\xi)(1-y)\xi} f_q^{[a]} f_g^{[b]}\,,\notag\\ 
 J_{gq} (\PhiBJ) =&\, C_F \frac{\as}{2\pi} \frac{1+\xi^2}{(1-\xi)(1+y)\xi} f_g^{[a]} f_q^{[b]}\,,\notag\\ 
 J_{gg} (\PhiBJ) =&\, 2 C_A \frac{\as}{2\pi} \left[\frac{1-\xi}{\xi}+\frac{\xi}{1-\xi}+\xi (1-\xi)\right]\frac{2}{(1-y^2)\xi} f_g^{[a]} f_g^{[b]}\,.
\end{align}
For quark-initiated reactions we have
\begin{align}
\label{eq:APsplitting-quark} 
J_{q\bar q} (\PhiBJ) =&\, C_F
                         \frac{\as}{2\pi}\left[\frac{1+(1-\xi)^2}{\xi}\right]\frac{2}{(1-y^2)\xi}f_q^{[a]} f_{\bar{q}}^{[b]}\,,
                         \notag\\ 
 J_{qg} (\PhiBJ) =&\, T_F \frac{\as}{2\pi}\frac{\xi^2+(1-\xi)^2}{(1+y)\xi}f_q^{[a]} f_g^{[b]}\,,\notag\\ 
 J_{gq} (\PhiBJ) =&\, T_F \frac{\as}{2\pi}\frac{\xi^2+(1-\xi)^2}{(1-y)\xi}f_g^{[a]} f_q^{[b]}\,.
\end{align}

\section{Resummation formulae}
\label{app:formulae}

In this section we report the expressions of the quantities appearing in the calculation
of the analytic transverse-momentum spectrum that we have used
throughout this article.

First of all we report our convention for the renormalisation-group equation of the
strong coupling:
\begin{equation}\label{eq:beta}
  \frac{\rd\as(\mu)}{\rd\ln \mu^2}
  =
  \beta(\as) \equiv
  -\as\left( \beta_0 \as +\beta_1\as^2 +\beta_2
  \as^3 + \dots\right),
\end{equation}
where the coefficients of the $\beta$-function are
\begin{eqnarray}
  \beta_0 &=& \frac{11 C_A - 2 n_f}{12\pi}\,,\qquad 
  \beta_1 = \frac{17 C_A^2 - 5 C_A n_f - 3 C_F n_f}{24\pi^2}\,,\\
  \beta_2 &=& \frac{2857 C_A^3+ (54 C_F^2 -615C_F C_A -1415 C_A^2)n_f
       +(66 C_F +79 C_A) n_f^2}{3456\pi^3}\,,
\end{eqnarray}
with
$C_A = N_c$, $C_F = \frac{N_c^2-1}{2N_c}$, $N_c = 3$, and the number of light flavours $n_f=5$.

The Sudakov radiator $\tilde{S}(\pt)$ in~\eqn{eq:Rdef}, with the accuracy
considered in this article, can be expressed as
\begin{equation}\label{eq:SudRadLLexp}
\tilde{S}(\pt) = - L g_1(\lambda) - g_2(\lambda) - \frac{\as(Q)}{\pi} g_3(\lambda)\,
\end{equation}
with $\lambda = \as(Q) \beta_0 \ln (Q/\pt)$, and
\begin{align}
  g_{1}(\lambda) =& \frac{A^{(1)}}{\pi\beta_{0}}\frac{2 \lambda +\ln (1-2 \lambda )}{2  \lambda }, \\
  g_{2}(\lambda) =& \frac{1}{2\pi \beta_{0}}\ln (1-2 \lambda )
  B^{(1)}
  -\frac{A^{(2)}}{4 \pi ^2 \beta_{0}^2}\frac{2 \lambda +(1-2
    \lambda ) \ln (1-2 \lambda )}{1-2
    \lambda} \notag\\
  &-A^{(1)} \frac{\beta_{1}}{4 \pi \beta_{0}^3}\frac{\ln
    (1-2 \lambda ) ((2 \lambda -1) \ln (1-2 \lambda )-2)-4
    \lambda}{1-2 \lambda}\,,
\end{align}

\begin{align}
  g_{3}(\lambda) =
  & B^{(1)}\frac{\beta_{1}}{2 \beta_{0}^2}\frac{2 \lambda
    +\ln (1-2 \lambda )}{1-2 \lambda}
    -\frac{1}{2 \pi\beta_{0}}\frac{\lambda}{1-2\lambda}\tilde{B}^{(2)}
  -\frac{A^{(3)}}{4 \pi ^2 \beta_{0}^2}\frac{\lambda ^2}{(1-2\lambda )^2} 
  \notag\\
  &   +A^{(2)} \frac{\beta_{1}}{4 \pi  \beta_{0}^3 }\frac{2 \lambda  (3
    \lambda -1)+(4 \lambda -1) \ln (1-2 \lambda )}{(1-2 \lambda
    )^2} \notag\\
  & +A^{(1)} \bigg(\frac{\lambda  \left(\beta_{0} \beta_{2} (1-3 \lambda
    )+\beta_{1}^2 \lambda \right)}{\beta_{0}^4 (1-2 \lambda)^2}
  +\frac{(1-2 \lambda) \ln (1-2 \lambda ) \left(\beta_{0} \beta_{2} 
    (1-2 \lambda )+2 \beta_{1}^2 \lambda \right)}{2\beta_{0}^4 (1-2 \lambda)^2} 
  \notag\\
  &\hspace{10mm}+\frac{\beta_{1}^2}{4 \beta_{0}^4}
  \frac{(1-4 \lambda ) \ln ^2(1-2 \lambda )}{(1-2 \lambda)^2}\bigg)\,.
\label{eq:gfunctions}
\end{align}
The resummation coefficient $\tilde{B}^{(2)}$ is defined as
according to eqs.~\eqref{eq:new_coeffs},~\eqref{eq:new_coeffs_2},
namely
\begin{align}
\tilde{B}^{(2)} =& B^{(2)} + 2\zeta_3 (A^{(1)})^2 + 2 \pi \beta_0
                   H^{(1)}\,.
\end{align}
For Higgs-boson production in gluon fusion, the coefficients $A^{(i)}$
and $B^{(i)}$ which enter the formulae above
are
\begin{align}
  A_{\rm ggH}^{(1)} =&\; 2 C_A,
  \notag\\
  \vspace{1.5mm}
  A_{\rm ggH}^{(2)} =&
  \left( \frac{67}{9}-\frac{\pi ^2}{3} \right) C_A^2
  -\frac{10}{9} C_A n_f,
  \notag\\
  \vspace{1.5mm}
  A_{\rm ggH}^{(3)} =&
   \left( -22 \zeta_3 - \frac{67 \pi^2}{27}+\frac{11 \pi^4}{90}+\frac{15503}{324} \right) C_A^3
  + \left( \frac{10 \pi^2}{27}-\frac{2051}{162} \right) C_A^2 n_f\notag\\
  &+ \left( 4 \zeta_3-\frac{55}{12} \right) C_A C_F n_f
  + \frac{50}{81} C_A n_f^2,
  \notag\\
  \vspace{1.5mm}
  B_{\rm ggH}^{(1)} =&
  -\frac{11}{3} C_A + \frac{2}{3}n_f,
  \notag\\
  \vspace{1.5mm}
  B_{\rm ggH}^{(2)} =&
  \left( \frac{11 \zeta _2}{6}-6 \zeta _3-\frac{16}{3} \right) C_A^2 
  + \left( \frac{4}{3}-\frac{\zeta _2}{3} \right) C_A n_f 
  + n_f C_F.
\end{align}
Similarly, for Drell-Yan production they read
\begin{align}
  A_{\rm DY}^{(1)} = &\; 2C_F,
  \notag\\
  \vspace{1.5mm}
  A_{\rm DY}^{(2)} =& \left( \frac{67}{9} - \frac{\pi^2}{3} \right)C_A C_F - \frac{10}{9}C_F n_f,
  \notag\\
  \vspace{1.5mm}
  A_{\rm DY}^{(3)} =&
  \left( \frac{15503}{324} - \frac{67\pi^2}{27} + \frac{11\pi^4}{90} - 22\zeta_3 \right)C_A^2 C_F
  + \left( -\frac{2051}{162} + \frac{10\pi^2}{27} \right)C_A C_F n_f
  \notag\\&
  + \left( -\frac{55}{12} + 4\zeta_3 \right) C_F^2 n_f
  + \frac{50}{81} C_F n_f^2,
  \notag\\
  \vspace{1.5mm}
  B_{\rm DY}^{(1)} =&
  -3 C_F,
  \notag\\
  \vspace{1.5mm}
  B_{\rm DY}^{(2)} =&
  \left(-\frac{17}{12} - \frac{11 \pi^2}{12} +  6 \zeta_3\right) C_A C_F+
  \left( -\frac{3}{4} + \pi^2 - 12\zeta_3 \right) C_F^2+
  \left( \frac{1}{6}+\frac{\pi^2}{6} \right) C_F n_f.
\end{align}
The expressions for the coefficients $A^{(i)}$ and $B^{(i)}$ are
extracted from \citere{deFlorian:2001zd,Becher:2010tm} for Higgs-boson
production and \citere{Davies:1984hs} for DY production.
The hard-virtual coefficient functions $H$ and $\tilde{H}$ up to two
loops are given by
\begin{align}
\label{eq:Hdef}
  H(Q) =&  \,1  +  \left( \frac{\as(Q)}{2\pi} \right)\, H^{(1)} + \left( \frac{\as(Q)}{2\pi} \right)^2\, H^{(2)},\notag\\
  \tilde{H}(Q) =&  \,1  +  \left( \frac{\as(Q)}{2\pi} \right)\, H^{(1)} + \left( \frac{\as(Q)}{2\pi} \right)^2\, \tilde{H}^{(2)},
\end{align}
with
\begin{align}
\label{eq:H-fun-Q}
  H_{{\rm ggH}}^{(1)} =&  C_A\left(5+\frac{7}{6}\pi^2\right)-3 C_F,\notag\\
  H_{{\rm ggH}}^{(2)} =&   \frac{5359}{54} + \frac{137}{6}\ln\frac{\mh{}^2}{m_T^2} 
+ \frac{1679}{24}\pi^2 + \frac{37}{8}\pi^4- \frac{499}{6}\zeta_3
              + C_A \Delta {\rm H}^{(2)} \,,\qquad {\rm for} \> n_f=5,
\end{align}
for Higgs-boson prodcution, and 
\begin{align}
\label{eq:DY-fun-Q}
  H_{{\rm DY}}^{(1)} =& C_F \left( -8 + \frac{7}{6}\pi^2 \right), \notag\\
  H_{{\rm DY}}^{(2)} =&  -\frac{57433}{972}+\frac{281}{162}\pi^2
               +\frac{22}{27}\pi^4 +\frac{1178}{27}\zeta_3+ C_F \Delta {\rm H}^{(2)} \,,\qquad {\rm for} \> n_f=5.
\end{align}
for the DY process. The extra term
\begin{equation}
  \Delta {\rm H}^{(2)} =\frac{16}{3} \pi
  \beta_0 \zeta_3,
\label{eq:delta_H2}
\end{equation}
is a feature of performing the resummation in momentum space, and does
not appear in the impact-parameter ($b$) space formulation of
transverse-momentum resummation (see ref.~\cite{Bizon:2017rah} for
details). The coefficient $\tilde{H}^{(2)}$, that appears in
eqs.~\eqref{eq:redefinition_final} and~\eqref{eq:luminosity} reads
\begin{equation}
\tilde{H}^{(2)} = H^{(2)} - 2\zeta_3 A^{(1)} B^{(1)}\,.
\end{equation}
Finally, we report the expansion of the collinear coefficient
functions $C_{ab}$, $\tilde{C}_{ab}$, $G_{ab}$
\begin{align}
\label{eq:coeff-fun}
  C_{ab}(z) =& \,\delta(1-z)\delta_{ab} + \left( \frac{\as(\mu)}{2\pi}
               \right) \,C_{ab}^{(1)}(z)+\left( \frac{\as(\mu)}{2\pi}
               \right)^2 \,C_{ab}^{(2)}(z),\notag\\
  \tilde{C}_{ab}(z) =& \,\delta(1-z)\delta_{ab} + \left(
                       \frac{\as(\mu)}{2\pi} \right)
                       \,C_{ab}^{(1)}(z)+\left( \frac{\as(\mu)}{2\pi}
                       \right)^2 \,\tilde{C}_{ab}^{(2)}(z),\notag\\
  G_{ab}(z) =& \,\left( \frac{\as(\mu)}{2\pi} \right) \,G_{ab}^{(1)}(z),
\end{align}
where $\mu$ is the same scale that enters parton densities. The
first-order expansion has been known for a long time and reads 
\begin{equation}
C_{ab}^{(1)}(z)= - \hat P_{ab}^{(0),\epsilon}(z) - \delta_{ab}\delta(1-z)\frac{\pi^2}{12},
\end{equation}
where $\hat P_{ab}^{(0),\epsilon}(z)$ is the $\mathcal{O}(\epsilon)$
part of the leading-order regularised splitting functions $\hat P_{ab}^{(0)}(z)$
\begin{align}
  &\hat P^{(0)}_{qq}(z)=C_F\left[\frac{1+z^2}{(1-z)_+}+\frac32\delta(1-z)\right],  &\hat P^{(0),\epsilon}_{qq}(z) = -C_F (1-z), \nonumber\\
  &\hat P^{(0)}_{qg}(z)=\frac12\left[z^2+(1-z)^2\right],   &\hat P^{(0),\epsilon}_{qg}(z) = -z(1-z),\ \ \ \nonumber\\
  &\hat P^{(0)}_{gq}(z)=C_F\frac{1+(1-z)^2}{z},   &\hat P^{(0),\epsilon}_{gq}(z) = -C_F z,\ \ \ \ \ \ \ \ \nonumber\\
  &\hat P^{(0)}_{gg}(z)=2C_A\left[\frac z{(1-z)_+}+\frac{1-z}z+z(1-z)\right]+2\pi\beta_0\delta(1-z),  & \hat P^{(0),\epsilon}_{gg}(z) = 0.\ \ \ \ \ \ \ \ \ \qquad
\end{align}
The second-order collinear coefficient functions $C_{ab}^{(2)}(z)$, as
well as the $G$ coefficients for gluon-fusion processes are obtained
in refs.~\cite{Catani:2011kr,Gehrmann:2014yya,Echevarria:2016scs},
while for quark-induced processes they are derived in
ref.~\cite{Catani:2012qa}. In the present work we extract their
expressions using the results of
refs.~\cite{Catani:2011kr,Catani:2012qa}. For gluon-fusion processes,
the $C^{(2)}_{gq}$ and $C^{(2)}_{gg}$ coefficients normalised as in
eq.~\eqref{eq:coeff-fun} are extracted from eqs.~(30) and~(32) of
ref.~\cite{Catani:2011kr}, respectively, where we use the hard
coefficients of eqs.~\eqref{eq:H-fun-Q} {\it without} the
momentum-space term $\Delta {\rm H}^{(2)}$ in the expression for the
$H^{(2)}(Q)$ coefficient.\footnote{Additionally, we have to do the
  replacement $H^{(1)}\to H^{(1)}/2$ and $H^{(2)}\to H^{(2)}/4$ to
  match the convention of refs.~\cite{Catani:2011kr,Catani:2012qa}.}
The coefficient $G^{(1)}$ is taken from eq.~(13) of
ref.~\cite{Catani:2011kr}. Similarly, for quark-initiated processes,
we extract $C^{(2)}_{qg}$ and $C^{(2)}_{qq}$ from eqs.~(32) and~(34)
of ref.~\cite{Catani:2012qa}, respectively, where we use the hard
coefficients from eqs.~\eqref{eq:DY-fun-Q} {\it without} the
momentum-space term $\Delta {\rm H}^{(2)}$ in the expression for the
$H^{(2)}(Q)$ coefficient. The remaining quark coefficient functions
$C^{(2)}_{q\bar{q}}$, $C^{(2)}_{q\bar{q}'}$ and $C^{(2)}_{qq'}$ are
extracted from eq.~(35) of the same article.
The coefficient $\tilde{C}^{(2)}(z)$, that appears in
eqs.~\eqref{eq:redefinition_final} and~\eqref{eq:luminosity} finally
reads
\begin{equation}
\tilde{C}^{(2)}(z) = C^{(2)}(z) - 2 \zeta_3 A^{(1)} \hat{P}^{(0)}(z)\,.
\end{equation}

\section{Explicit expression for the $[D(\pt)]^{(3)}$ term}
\label{app:D3formulae}
The $[D(\pt)]^{(3)}$ term necessary to achieve NNLO accuracy is
defined in eq.~\eqref{eq:D3}, its expression reads
\begin{align}
  [D(\pt)]^{(3)} &= 
  -\left[\frac{\mathd \tilde{S}(\pt)}{\mathd \pt}\right]^{(1)}[{\cal L}(\pt)]^{(2)}
  -\left[\frac{\mathd \tilde{S}(\pt)}{\mathd \pt}\right]^{(2)}[{\cal
    L}(\pt)]^{(1)}\notag\\
& -\left[\frac{\mathd \tilde{S}(\pt)}{\mathd \pt}\right]^{(3)}[{\cal L}(\pt)]^{(0)}
                   + \left[\frac{\mathd {\cal L}(\pt)}{\mathd \pt}\right]^{(3)} \\
      & = 
   \frac{2}{\pt} \left( A^{(1)} \ln
  \frac{Q^2}{\pt^2} + B^{(1)} \right) [{\cal L}(\pt)]^{(2)}     + \frac{2}{\pt} \left( A^{(2)} \ln \frac{Q^2}{\pt^2} + \tilde{B}^{(2)} \right)
  [{\cal L}(\pt)]^{(1)} \notag\\
& + \frac{2}{\pt}  A^{(3)} \ln \frac{Q^2}{\pt^2} \,[{\cal L}(\pt)]^{(0)}+ \left[  \frac{\mathd {\cal L}(\pt)}{\mathd \pt}
  \right]^{(3)},\nonumber
\end{align}
where the resummation coefficients are reported in appendix~\ref{app:formulae},
while the expansion of the luminosity factors reads
\begin{align}
[{\cal L}(\pt)]^{(0)} & = \sum_{c,
c'}\frac{\mathd|M^{\scriptscriptstyle\rm F}|_{cc'}^2}{\mathd\Phi_{\rm
                        B}}\,f_c^{[a]}f_{c'}^{[b]}\,,
\end{align}
\begin{align}
[{\cal L}(\pt)]^{(1)} & = \sum_{c,
c'}\frac{\mathd|M^{\scriptscriptstyle\rm F}|_{cc'}^2}{\mathd\Phi_{\rm
                        B}}\bigg\{H^{(1)}f_c^{[a]}f_{c'}^{[b]} +
                        (C^{(1)}\otimes f)_c^{[a]}f_{c'}^{[b]} + 
                        f_{c}^{[a]} (C^{(1)}\otimes f)_{c'}^{[b]}\bigg\}\,,
\end{align}
\begin{align}
[{\cal L}(\pt)]^{(2)}  = \sum_{c,
c'}\frac{\mathd|M^{\scriptscriptstyle\rm F}|_{cc'}^2}{\mathd\Phi_{\rm
                        B}}&\bigg\{\tilde{H}^{(2)}f_c^{[a]}f_{c'}^{[b]} +
                        (\tilde{C}^{(2)}\otimes f)_c^{[a]}f_{c'}^{[b]} + 
                        f_{c}^{[a]} (\tilde{C}^{(2)}\otimes
                        f)_{c'}^{[b]}\notag\\
& + H^{(1)} (C^{(1)}\otimes f)_c^{[a]}f_{c'}^{[b]} +
  H^{(1)}f_{c}^{[a]} (C^{(1)}\otimes f)_{c'}^{[b]}\notag\\
& +  (C^{(1)}\otimes f)_c^{[a]}(C^{(1)}\otimes f)_{c'}^{[b]}  
+  (G^{(1)}\otimes f)_c^{[a]}(G^{(1)}\otimes f)_{c'}^{[b]} 
\bigg\}\,,
\end{align}
\begin{align}
\left[  \frac{\mathd {\cal L}(\pt)}{\mathd \pt}
  \right]^{(3)} = &\sum_{c,
c'}\frac{\mathd|M^{\scriptscriptstyle\rm F}|_{cc'}^2}{\mathd\Phi_{\rm
                        B}}\frac{2}{\pt}\bigg\{\tilde{H}^{(2)}\left[(\hat{P}^{(0)}\otimes
                             f)_c^{[a]}f_{c'}^{[b]} +  f_{c}^{[a]}
                             (\hat{P}^{(0)}\otimes f)_{c'}^{[b]}
                             \right]\notag\\
& + H^{(1)}\Big[(\hat{P}^{(1)}\otimes
                             f)_c^{[a]}f_{c'}^{[b]} +  f_{c}^{[a]}
                             (\hat{P}^{(1)}\otimes f)_{c'}^{[b]}
                             \notag\\
& +  (C^{(1)}\otimes f)_c^{[a]}(\hat{P}^{(0)}\otimes f)_{c'}^{[b]} +
  (\hat{P}^{(0)}\otimes f)_c^{[a]}(C^{(1)}\otimes
  f)_{c'}^{[b]}\notag\\
& +  f_c^{[a]}(\hat{P}^{(0)}\otimes C^{(1)}\otimes f)_{c'}^{[b]} +
  (\hat{P}^{(0)}\otimes C^{(1)}\otimes f)_c^{[a]}f_{c'}^{[b]}\Big]\notag\\
& + (\hat{P}^{(2)}\otimes
                             f)_c^{[a]}f_{c'}^{[b]} +  f_{c}^{[a]}
                             (\hat{P}^{(2)}\otimes
  f)_{c'}^{[b]}\notag\\
& + (\tilde{C}^{(2)}\otimes f)_c^{[a]}(\hat{P}^{(0)}\otimes f)_{c'}^{[b]} +
  (\hat{P}^{(0)}\otimes f)_c^{[a]}(\tilde{C}^{(2)}\otimes f)_{c'}^{[b]}
  \notag\\
& + f_c^{[a]}(\hat{P}^{(0)}\otimes \tilde{C}^{(2)}\otimes f)_{c'}^{[b]} +
  (\hat{P}^{(0)}\otimes \tilde{C}^{(2)}\otimes f)_c^{[a]}f_{c'}^{[b]}
  \notag\\
& + (C^{(1)}\otimes f)_c^{[a]}(\hat{P}^{(1)}\otimes f)_{c'}^{[b]} +
  (\hat{P}^{(1)}\otimes f)_c^{[a]}(C^{(1)}\otimes f)_{c'}^{[b]}
  \notag\\
& + f_c^{[a]}(\hat{P}^{(1)}\otimes C^{(1)}\otimes f)_{c'}^{[b]} +
  (\hat{P}^{(1)}\otimes C^{(1)}\otimes f)_c^{[a]}f_{c'}^{[b]}
  \notag\\
& + (C^{(1)}\otimes f)_c^{[a]}(\hat{P}^{(0)}\otimes C^{(1)}\otimes f)_{c'}^{[b]} +
  (\hat{P}^{(0)}\otimes C^{(1)}\otimes f)_c^{[a]}(C^{(1)}\otimes
  f)_{c'}^{[b]}\notag\\
& + (G^{(1)}\otimes f)_c^{[a]}(\hat{P}^{(0)}\otimes G^{(1)}\otimes f)_{c'}^{[b]} +
  (\hat{P}^{(0)}\otimes G^{(1)}\otimes f)_c^{[a]}(G^{(1)}\otimes
  f)_{c'}^{[b]}\notag\\
& - 4\beta_0 \pi \Big[\tilde{H}^{(2)}f_c^{[a]}f_{c'}^{[b]} +
                        (\tilde{C}^{(2)}\otimes f)_c^{[a]}f_{c'}^{[b]} + 
                        f_{c}^{[a]} (\tilde{C}^{(2)}\otimes
  f)_{c'}^{[b]}\notag\\
& + H^{(1)}  (C^{(1)}\otimes f)_c^{[a]}f_{c'}^{[b]} + H^{(1)}
  f_{c}^{[a]} (C^{(1)}\otimes f)_{c'}^{[b]}\notag\\
& +  (C^{(1)}\otimes f)_c^{[a]} (C^{(1)}\otimes f)_{c'}^{[b]} +  (G^{(1)}\otimes f)_c^{[a]} (G^{(1)}\otimes f)_{c'}^{[b]}\Big]\notag\\
& -4 \beta_1\pi^2 \left[ H^{(1)}f_c^{[a]}f_{c'}^{[b]} +
                        (C^{(1)}\otimes f)_c^{[a]}f_{c'}^{[b]} + 
                        f_{c}^{[a]} (C^{(1)}\otimes f)_{c'}^{[b]}\right]
\bigg\}\,.
\end{align}

\section{Scale dependence of the \minnlo{} formula}
\label{app:scaledep}
In this appendix we discuss the renormalisation and
factorisation scale dependence of the \minnlo{}
formula~\eqref{eq:master}.
Our starting formula is
\begin{align}
\label{eq:mainsigma}
  \frac{\mathd\sigma}{\mathd\PhiB\mathd \pt} &= \frac{\mathd}{\mathd \pt}
     \bigg\{ \exp[-\tilde{S}(\pt)] {\cal L}(\PhiB, \pt)\Bigg\} + R_f(\PhiB, \pt)\,,
\end{align}
where all ingredients are introduced in \sct{sec:procedure}. 
The scales appearing
in the strong coupling constant and in the parton densities, $\muR$ and $\muF$, are
set to $\pt$.
After integration over the transverse momentum we get
\begin{equation} \frac{\mathd \sigma}{\mathd \PhiB} = \mathcal{L} (\PhiB, Q) + \int \mathd
  \pt R_f (\PhiB, \pt)\,,
\end{equation}
that corresponds to the inclusive NNLO cross section at fixed kinematics
of the colour singlet system \F{}. 
The scale dependence is introduced by evaluating the PDFs and $\as$ 
at $\muR=\KR\pt$ and $\muF=\KF\pt$, and by adding appropriate scale-compensating 
$\KF$ and $\KR$ dependent terms in $\mathcal{L}$ and $R_f$, thus redefining
\begin{eqnarray}
  \mathcal{L} (\PhiB, \pt) \rightarrow \mathcal{L} (\PhiB, \pt, \KR, \KF) +
  \mathcal{O} (\as^3)\,,
  \\
  R_f (\PhiB, \pt) \rightarrow R_f (\PhiB, \pt, \KR, \KF) +
  \mathcal{O} (\as^3)\,.
\end{eqnarray}
Accordingly, \eqn{eq:mainsigma} becomes
\begin{equation}
\label{eq:mainsigmawithscales}
  \frac{\mathd\sigma}{\mathd\PhiB\mathd \pt} = \frac{\mathd}{\mathd \pt}
     \bigg\{ \exp[-\tilde{S}(\pt)] {\cal L}(\PhiB, \pt, \KR, \KF)\Bigg\} + R_f(\PhiB, \pt, \KR, \KF)\,,
\end{equation}
and includes all relevant scale-dependent terms at NNLO.

Formula~(\ref{eq:mainsigmawithscales}) retains its NNLO accuracy
whether or not we include also scale-dependent 
terms in the Sudakov form factor  $\tS$. However, in order to make contact
with the \POWHEG{} formula, the scale dependence in $\tS$ must be included.
In fact, if we take the derivative in \eqn{eq:mainsigmawithscales} we obtain
\begin{eqnarray*}
  \frac{\mathd \sigma}{\mathd \PhiB \mathd \pt} & = & \exp [- \tS (\pt)] \left\{
  - \frac{\mathd \tS (\pt)}{\mathd \pt}  \mathcal{L} (\PhiB, \pt, \KR, \KF) +
  \frac{\mathd \mathcal{L} (\PhiB, \pt, \KR, \KF)}{\mathd \pt}  \right.\\
                                                    & + & \left. \exp [\tS (\pt)] R_f (\PhiB, \pt, \KR, \KF) \right\}
 \\
  & = & \exp [- \tS (\pt)] \left\{ - \frac{\mathd \tS (\pt)}{\mathd \pt} 
  \mathcal{L} (\PhiB, \pt, \KR, \KF) + \frac{\mathd \mathcal{L} (\PhiB, \pt,
  \KR, \KF)}{\mathd \pt}  \right.\\
  & + &   \left(1+\abarmu{\KR\pt} [\tilde{S}(\pt)]^{(1)}\right) R_f^{(1)} (\PhiB, \pt, \KR, \KF) \\
  & + & \left. \left(\abarmu{\KR\pt}\right)^2 R_f^{(2)} (\PhiB, \pt, \KR, \KF) \right\}+\order{\as^3}\,.
\end{eqnarray*}
Since in \POWHEG{} we do not have access separately to the terms arising from the derivative of $\tS$,
and all terms in the curly bracket are evaluated with the same scale choice, we must make sure that
also the derivative of $\tS$ is given in terms of $\as (\KR\pt)$. This is achieved by
writing $\tS$ as
\begin{equation}
\tilde{S}(\pt) = 2\int_{\pt}^{Q}\frac{\mathd q}{q}
                    \left(A(\as(\KR q),\KR)\ln\frac{Q^2}{q^2} +
                    \tilde{B}(\as(\KR q),\KR)\right),
\end{equation}
where the $A$ and $B$ coefficients include scale-compensating terms in such a way 
that they are
formally independent upon $\KR$ when summed up to all orders in perturbation
theory. It is easy to see that, with this replacement, the form of $\tS$ given in
\eqn{eq:SudRadLLexp} remains the same provided that the $A^{(i)}$ and  $B^{(i)}$ coefficients
are replaced by the $\KR$ dependent ones, and that $\as(Q)$ is replaced by $\as(\KR Q)$.

We now present in detail the formulae needed to implement the scale variation.
We start by discussing the $\mathcal{L}$ factor, defined in eq.~\eqref{eq:luminosity}.
The coefficients $H^{(1)}$ and $\tilde{H}^{(2)}$ become
\begin{align}
H^{(1)}(\KR) =& H^{(1)} + (2\pi\beta_0) n_B \ln \KR^2\,, \notag\\
\tilde{H}^{(2)}(\KR) =& \tilde{H}^{(2)} + 4 \, n_B \left( \frac{1+n_{B} }{2}  \pi^2\beta_0^2 \ln^2 \KR^2+ \pi^2 \beta_1
  \ln \KR^2\right)\notag\\
& + 2 \,H^{(1)} \left(1+n_B\right) \pi\beta_0  \ln \KR^2 \,,
\end{align}
with $n_B$ being the $\as$ power of the Born cross section for
the production of the colour singlet \F{}. The coefficient functions
$C$ receive the following scale dependence:
\begin{align}
C^{(1)}(z,& \KF) = C^{(1)}(z) -
                                    \hat{P}^{(0)}(z)\ln \KF^{2},\notag\\
\tilde{C}^{(2)}(z,& \KF, \KR) = \tilde{C}^{(2)}(z) +
                                    \pi\beta_0 \hat{P}^{(0)}(z)\left(
                                    \ln^2\KF^{2} -
                                   2 \ln \KF^{2}
                                    \ln \KR^{2}\right) -
                                    \hat{P}^{(1)}(z)\ln \KF^{2}\notag\\
& + \frac{1}{2}(\hat{P}^{(0)}\otimes \hat{P}^{(0)})(z) \ln^2\KF^{2} -
  (\hat{P}^{(0)} \otimes C^{(1)})(z) \ln \KF^{2} + 2\pi\beta_0
  C^{(1)}(z) \ln \KR^{2}\,,
\end{align}
while $G$ (which is present only in the case of gluon-induced reactions) remains unchanged.

We then consider the Sudakov radiator $\tilde{S}$, defined in
eq.~\eqref{eq:Rdef}. We change the scale of the strong coupling in its
integrand~\eqref{eq:Rdef} from $\pt$ to $\KR\pt$, and modify the
$A$ and $B$ coefficients as follows\footnote{We stress that, formally,
  the perturbative coefficient $A^{(3)}$ gives a subleading
  contribution to the NNLO cross section, and it is included in
  $[D(\pt)]^{(3)}$ to ensure consistency with the Sudakov radiator
  $\tilde{S}$. Since its scale dependence would add information beyond
  the desired perturbative order, we explicitly decide to omit it in
  our implementation.}
\begin{align}
A^{(2)}(\KR) =& A^{(2)} + (2\pi \beta_0) A^{(1)} \ln \KR^2,\notag\\ 
\tilde{B}^{(2)}(\KR) =& \tilde{B}^{(2)}  + (2\pi \beta_0) B^{(1)} \ln \KR^2 +
                (2\pi\beta_0)^2\,n_B \ln \KR^2\,.
\end{align}
The term proportional to $n_B$ (the power of $\as$ at LO),
is induced by the presence of $H^{(1)}$ in the $\tilde{B}^{(2)}$
coefficient, that in turn originates from evaluating the hard virtual
corrections at $\pt$ in the factor ${\cal L}$, see
eq.~\eqref{eq:new_coeffs_2}.

The scale dependence also propagates into the constituents of the
$[D(\pt)]^{(3)}$ term, whose $\as^3$ prefactor in
eq.~\eqref{eq:master} is evaluated at $\KR\pt$. Besides the
dependence in the coefficients reported above (which is understood in
the equation that follows), $[D(\pt)]^{(3)}$ acquires additional
explicit scale-dependent terms:
\begin{align}
[D(\pt)]^{(3)}(\KF,\KR) &= [D(\pt)]^{(3)}\notag\\
& - \sum_{c,
c'}\frac{\mathd|M^{\scriptscriptstyle\rm F}|_{cc'}^2}{\mathd\Phi_{\rm
                        B}}\frac{4\pi}{\pt}\bigg\{2\pi\beta_1 \left(f_c^{[a]}(\hat{P}^{(0)}\otimes f)_{c'}^{[b]} +
  (\hat{P}^{(0)}\otimes f)_c^{[a]}f_{c'}^{[b]}\right) \ln\frac{\KF^2}{\KR^2}\notag\\
& + \beta_0\bigg( H^{(1)}(\KR)\left(f_c^{[a]}(\hat{P}^{(0)}\otimes f)_{c'}^{[b]} +
  (\hat{P}^{(0)}\otimes f)_c^{[a]}f_{c'}^{[b]}\right)\notag\\
&+ 2 \left(f_c^{[a]}(\hat{P}^{(1)}\otimes f)_{c'}^{[b]} +
  (\hat{P}^{(1)}\otimes f)_c^{[a]}f_{c'}^{[b]}\right)\notag\\
&+ (C^{(1)}(\KF)\otimes f)_c^{[a]}(\hat{P}^{(0)}\otimes f)_{c'}^{[b]} +
  (\hat{P}^{(0)}\otimes f)_c^{[a]}(C^{(1)}(\KF)\otimes f)_{c'}^{[b]}
  \notag\\
& + f_c^{[a]}(\hat{P}^{(0)}\otimes C^{(1)}(\KF)\otimes f)_{c'}^{[b]} +
  (\hat{P}^{(0)}\otimes C^{(1)}(\KF)\otimes f)_c^{[a]}f_{c'}^{[b]}\bigg)
  \ln\frac{\KF^2}{\KR^2}\notag\\
& - 2\pi\beta_0^2 \left(f_c^{[a]}(\hat{P}^{(0)}\otimes f)_{c'}^{[b]} +
  (\hat{P}^{(0)}\otimes f)_c^{[a]}f_{c'}^{[b]}\right) \ln^2\frac{\KF^2}{\KR^2}\bigg\}\,.
\end{align}

\section{Considerations from impact-parameter space formulation}
\label{app:bspace}
In this section, we derive the form of the starting
equation~\eqref{eq:start} using the impact-parameter space formulation of 
transverse-momentum resummation. We start
from the formula
\begin{align}
  \frac{\mathd\sigma(\pt)}{\mathd\PhiB}  =
 \pt \int_0^{\infty} db J_1(b\,\pt) \, e^{-S(b_0/b)}{\cal L}_b(b_0/b)\,,
\end{align}
where
\begin{equation}
  S(b_0/b) =  - \ln(Qb/b_0) g_1(\lambda_b) - g_2(\lambda_b) -
  \frac{\as}{\pi} \bar{g}_3(\lambda_b)\,,
\end{equation}
and $\lambda_b = \as(Q)\beta_0 \ln(Qb/b_0)$,
$b_0 = 2 e^{-\gamma_E}$. The $g_i$ functions are analogous to those
used in momentum space~\eqref{eq:gfunctions}, and~\cite{Bozzi:2005wk}
\begin{equation}
\bar{g}_3 \equiv g_3 + \frac{2\zeta_3 (A^{(1)})^2}{2 \pi\beta_{0}}\frac{\lambda_b}{1-2\lambda_b}\,.
\end{equation}
The factor ${\cal L}_b$ is defined as
\begin{align} 
{\cal L}_b(b_0/b)&=\sum_{c,
c'}\frac{\mathd|M^{\scriptscriptstyle\rm F}|_{cc'}^2}{\mathd\PhiB} \sum_{i, j}
\bigg\{\left(C^{[a]}_{c i}\otimes f_i^{[a]}\right) \bar{H}(b_0/b)
\left(C^{[b]}_{c' j}\otimes f_j^{[b]}\right) \notag\\ & +
\left(G^{[a]}_{c i}\otimes f_i^{[a]}\right) \bar{H}(b_0/b) \left(G^{[b]}_{c'
j}\otimes f_j^{[b]}\right)\bigg\}\,,
\label{eq:luminosity_bspace}
\end{align}
where $\bar{H}$ is identical to $H$ of eq.~\eqref{eq:Hdef}, with the
only difference being that the $\bar{H}^{(2)}$ coefficient does not
contain the term $\Delta {\rm H}^{(2)}$~\eqref{eq:delta_H2}.

We evaluate the $b$ integral by expanding $b_0/b$ about $\pt$ in the
integrand. While this procedure is known to generate a geometric
singularity in the $\pt$ space resummation, in this article we are
only interested in retaining ${\cal O}(\as^2)$ accuracy and
therefore this is not an issue for the present discussion. We follow
the appendix of ref.~\cite{Banfi:2012jm}, and by neglecting terms that
contribute beyond ${\cal O}( \as^2)$, we obtain
\begin{align}
  \frac{\mathd\sigma(\pt)}{\mathd\PhiB}
&  =e^{-S(\pt)}\left\{{\cal L}_b(\pt) \left(1-\frac{1}{2}
  S''(\pt) \partial^2_{S'} + \frac{1}{6} S'''(\pt) \partial^3_{S'}\right) + \frac{1}{2}
  S''(\pt) \frac{\mathd {\cal
  L}_b(\pt)}{\mathd \ln(Q/\pt)} \partial^3_{S'} \right\}\notag\\
& \times e^{-\gamma_E
  S'}\frac{\Gamma(1-\frac{S'}{2})}{\Gamma(1+\frac{S'}{2})} + {\cal O}(\as^3(Q))
\,,
\end{align}
where $S'''(\pt) = \mathd S''(\pt)/\mathd \ln(Qb/b_0)$.  After
performing the derivatives, we observe that, retaining
${\cal O}(\as^2)$ accuracy, we can approximate the above equation as
follows
\begin{align}
  \label{eq:ptBsigma}
  \frac{\mathd\sigma(\pt)}{\mathd\PhiB}
 & =e^{-S(\pt)}\bigg\{{\cal L}_b(\pt) \left(1-\frac{\zeta_3}{4}
  S''(\pt) S'(\pt) + \frac{\zeta_3}{12} S'''(\pt)\right) \notag\\
& - \frac{\zeta_3}{4}\frac{\as(\pt)}{\pi} S''(\pt) \hat{P}
  \otimes {\cal L}_b(\pt)  \bigg\} + {\cal O}(\as^3(Q))
\,.
\end{align}
We directly observe that the two terms proportional to $S''$ are
analogous to those produced in the last line of
eq.~\eqref{eq:starting-2}. These two terms can be incorporated in the
master formula via the replacements~\eqref{eq:redefinition_final}. On
the other hand, the term proportional to $S'''$ is a new feature of
the $b$-space formulation, and it is not present in the momentum space
formulation. Using the expression
\begin{equation}
S'''(\pt) = 32 A^{(1)} \pi \beta_0 \frac{\as^2}{(2\pi)^2} + {\cal O}(\as^3)\,
\end{equation}
we observe that the new ${\cal O}(\as^2)$ constant term $8/3\,
\zeta_3 A^{(1)} \pi \beta_0 $ can be absorbed into the coefficient
$\bar{H}^{(2)}$ as
\begin{equation}
\bar{H}^{(2)} \rightarrow H^{(2)} = \bar{H}^{(2)} + \frac{8}{3}\,\zeta_3 A^{(1)} \pi
\beta_0\,.
\label{eq:new_coeffs_3}
\end{equation}
This is precisely the difference between the $\bar{H}^{(2)}$
coefficient (defined in $b$ space) and the $H^{(2)}$ coefficient
present in the momentum-space formulation.
Therefore, the ${\cal O}(\as^2)$ expansion of eq.~\eqref{eq:ptBsigma}
coincides with that of eq.~\eqref{eq:start}.

\newpage
\addcontentsline{toc}{section}{References}
\bibliography{MiNNLO}

\providecommand{\href}[2]{#2}\begingroup\raggedright\begin{thebibliography}{10}

\bibitem{Aad:2012tfa}
{\scshape ATLAS} collaboration, G.~Aad et~al., \emph{{Observation of a new
  particle in the search for the Standard Model Higgs boson with the ATLAS
  detector at the LHC}},
  \href{http://dx.doi.org/10.1016/j.physletb.2012.08.020}{\emph{Phys. Lett.}
  {\bfseries B716} (2012) 1--29},
  [\href{https://arxiv.org/abs/1207.7214}{{\ttfamily 1207.7214}}].

\bibitem{Chatrchyan:2012xdj}
{\scshape CMS} collaboration, S.~Chatrchyan et~al., \emph{{Observation of a new
  boson at a mass of 125 GeV with the CMS experiment at the LHC}},
  \href{http://dx.doi.org/10.1016/j.physletb.2012.08.021}{\emph{Phys. Lett.}
  {\bfseries B716} (2012) 30--61},
  [\href{https://arxiv.org/abs/1207.7235}{{\ttfamily 1207.7235}}].

\bibitem{Aaboud:2017ffb}
{\scshape ATLAS} collaboration, M.~Aaboud et~al., \emph{{Measurement of the
  Drell-Yan triple-differential cross section in $pp$ collisions at $\sqrt{s} =
  8$ TeV}}, \href{http://dx.doi.org/10.1007/JHEP12(2017)059}{\emph{JHEP}
  {\bfseries 12} (2017) 059},
  [\href{https://arxiv.org/abs/1710.05167}{{\ttfamily 1710.05167}}].

\bibitem{Sirunyan:2017igm}
{\scshape CMS} collaboration, A.~M. Sirunyan et~al., \emph{{Measurement of
  differential cross sections in the kinematic angular variable $\phi^*$ for
  inclusive Z boson production in pp collisions at $\sqrt{s}=$ 8 TeV}},
  \href{http://dx.doi.org/10.1007/JHEP03(2018)172}{\emph{JHEP} {\bfseries 03}
  (2018) 172}, [\href{https://arxiv.org/abs/1710.07955}{{\ttfamily
  1710.07955}}].

\bibitem{Aaboud:2017svj}
{\scshape ATLAS} collaboration, M.~Aaboud et~al., \emph{{Measurement of the
  $W$-boson mass in pp collisions at $\sqrt{s}=7$ TeV with the ATLAS
  detector}},
  \href{http://dx.doi.org/10.1140/epjc/s10052-017-5475-4}{\emph{Eur. Phys. J.}
  {\bfseries C78} (2018) 110},
  [\href{https://arxiv.org/abs/1701.07240}{{\ttfamily 1701.07240}}].

\bibitem{Hamilton:2012np}
K.~Hamilton, P.~Nason and G.~Zanderighi, \emph{{MINLO: Multi-Scale Improved
  NLO}}, \href{http://dx.doi.org/10.1007/JHEP10(2012)155}{\emph{JHEP}
  {\bfseries 10} (2012) 155},
  [\href{https://arxiv.org/abs/1206.3572}{{\ttfamily 1206.3572}}].

\bibitem{Hamilton:2012rf}
K.~Hamilton, P.~Nason, C.~Oleari and G.~Zanderighi, \emph{{Merging H/W/Z + 0
  and 1 jet at NLO with no merging scale: a path to parton shower + NNLO
  matching}}, \href{http://dx.doi.org/10.1007/JHEP05(2013)082}{\emph{JHEP}
  {\bfseries 05} (2013) 082},
  [\href{https://arxiv.org/abs/1212.4504}{{\ttfamily 1212.4504}}].

\bibitem{Frederix:2015fyz}
R.~Frederix and K.~Hamilton, \emph{{Extending the MINLO method}},
  \href{http://dx.doi.org/10.1007/JHEP05(2016)042}{\emph{JHEP} {\bfseries 05}
  (2016) 042}, [\href{https://arxiv.org/abs/1512.02663}{{\ttfamily
  1512.02663}}].

\bibitem{Nason:2004rx}
P.~Nason, \emph{{A New method for combining NLO QCD with shower Monte Carlo
  algorithms}},
  \href{http://dx.doi.org/10.1088/1126-6708/2004/11/040}{\emph{JHEP} {\bfseries
  11} (2004) 040}, [\href{https://arxiv.org/abs/hep-ph/0409146}{{\ttfamily
  hep-ph/0409146}}].

\bibitem{Hamilton:2008pd}
K.~Hamilton, P.~Richardson and J.~Tully, \emph{{A Positive-Weight
  Next-to-Leading Order Monte Carlo Simulation of Drell-Yan Vector Boson
  Production}},
  \href{http://dx.doi.org/10.1088/1126-6708/2008/10/015}{\emph{JHEP} {\bfseries
  10} (2008) 015}, [\href{https://arxiv.org/abs/0806.0290}{{\ttfamily
  0806.0290}}].

\bibitem{Bahr:2008pv}
M.~Bahr et~al., \emph{{Herwig++ Physics and Manual}},
  \href{http://dx.doi.org/10.1140/epjc/s10052-008-0798-9}{\emph{Eur. Phys. J.}
  {\bfseries C58} (2008) 639--707},
  [\href{https://arxiv.org/abs/0803.0883}{{\ttfamily 0803.0883}}].

\bibitem{Hoeche:2009rj}
S.~Hoeche, F.~Krauss, S.~Schumann and F.~Siegert, \emph{{QCD matrix elements
  and truncated showers}},
  \href{http://dx.doi.org/10.1088/1126-6708/2009/05/053}{\emph{JHEP} {\bfseries
  05} (2009) 053}, [\href{https://arxiv.org/abs/0903.1219}{{\ttfamily
  0903.1219}}].

\bibitem{Hoche:2010kg}
S.~Hoche, F.~Krauss, M.~Schonherr and F.~Siegert, \emph{{NLO matrix elements
  and truncated showers}},
  \href{http://dx.doi.org/10.1007/JHEP08(2011)123}{\emph{JHEP} {\bfseries 08}
  (2011) 123}, [\href{https://arxiv.org/abs/1009.1127}{{\ttfamily 1009.1127}}].

\bibitem{Alioli:2013hqa}
S.~Alioli, C.~W. Bauer, C.~Berggren, F.~J. Tackmann, J.~R. Walsh and S.~Zuberi,
  \emph{{Matching Fully Differential NNLO Calculations and Parton Showers}},
  \href{http://dx.doi.org/10.1007/JHEP06(2014)089}{\emph{JHEP} {\bfseries 06}
  (2014) 089}, [\href{https://arxiv.org/abs/1311.0286}{{\ttfamily 1311.0286}}].

\bibitem{Hoeche:2014aia}
S.~Höche, Y.~Li and S.~Prestel, \emph{{Drell-Yan lepton pair production at
  NNLO QCD with parton showers}},
  \href{http://dx.doi.org/10.1103/PhysRevD.91.074015}{\emph{Phys. Rev.}
  {\bfseries D91} (2015) 074015},
  [\href{https://arxiv.org/abs/1405.3607}{{\ttfamily 1405.3607}}].

\bibitem{Hamilton:2013fea}
K.~Hamilton, P.~Nason, E.~Re and G.~Zanderighi, \emph{{NNLOPS simulation of
  Higgs boson production}},
  \href{http://dx.doi.org/10.1007/JHEP10(2013)222}{\emph{JHEP} {\bfseries 10}
  (2013) 222}, [\href{https://arxiv.org/abs/1309.0017}{{\ttfamily 1309.0017}}].

\bibitem{Hoche:2014dla}
S.~Höche, Y.~Li and S.~Prestel, \emph{{Higgs-boson production through gluon
  fusion at NNLO QCD with parton showers}},
  \href{http://dx.doi.org/10.1103/PhysRevD.90.054011}{\emph{Phys. Rev.}
  {\bfseries D90} (2014) 054011},
  [\href{https://arxiv.org/abs/1407.3773}{{\ttfamily 1407.3773}}].

\bibitem{Karlberg:2014qua}
A.~Karlberg, E.~Re and G.~Zanderighi, \emph{{NNLOPS accurate Drell-Yan
  production}}, \href{http://dx.doi.org/10.1007/JHEP09(2014)134}{\emph{JHEP}
  {\bfseries 09} (2014) 134},
  [\href{https://arxiv.org/abs/1407.2940}{{\ttfamily 1407.2940}}].

\bibitem{Alioli:2015toa}
S.~Alioli, C.~W. Bauer, C.~Berggren, F.~J. Tackmann and J.~R. Walsh,
  \emph{{Drell-Yan production at NNLL'+NNLO matched to parton showers}},
  \href{http://dx.doi.org/10.1103/PhysRevD.92.094020}{\emph{Phys. Rev.}
  {\bfseries D92} (2015) 094020},
  [\href{https://arxiv.org/abs/1508.01475}{{\ttfamily 1508.01475}}].

\bibitem{Astill:2016hpa}
W.~Astill, W.~Bizon, E.~Re and G.~Zanderighi, \emph{{NNLOPS accurate associated
  HW production}}, \href{http://dx.doi.org/10.1007/JHEP06(2016)154}{\emph{JHEP}
  {\bfseries 06} (2016) 154},
  [\href{https://arxiv.org/abs/1603.01620}{{\ttfamily 1603.01620}}].

\bibitem{Astill:2018ivh}
W.~Astill, W.~Bizoń, E.~Re and G.~Zanderighi, \emph{{NNLOPS accurate
  associated HZ production with $ H\to b\overline{b} $ decay at NLO}},
  \href{http://dx.doi.org/10.1007/JHEP11(2018)157}{\emph{JHEP} {\bfseries 11}
  (2018) 157}, [\href{https://arxiv.org/abs/1804.08141}{{\ttfamily
  1804.08141}}].

\bibitem{Re:2018vac}
E.~Re, M.~Wiesemann and G.~Zanderighi, \emph{{NNLOPS accurate predictions for
  $W^+W^-$ production}},
  \href{http://dx.doi.org/10.1007/JHEP12(2018)121}{\emph{JHEP} {\bfseries 12}
  (2018) 121}, [\href{https://arxiv.org/abs/1805.09857}{{\ttfamily
  1805.09857}}].

\bibitem{Hamilton:2016bfu}
K.~Hamilton, T.~Melia, P.~F. Monni, E.~Re and G.~Zanderighi, \emph{{Merging WW
  and WW+jet with MINLO}},
  \href{http://dx.doi.org/10.1007/JHEP09(2016)057}{\emph{JHEP} {\bfseries 09}
  (2016) 057}, [\href{https://arxiv.org/abs/1606.07062}{{\ttfamily
  1606.07062}}].

\bibitem{Grazzini:2016ctr}
M.~Grazzini, S.~Kallweit, S.~Pozzorini, D.~Rathlev and M.~Wiesemann,
  \emph{{$W^+W^-$ production at the LHC: fiducial cross sections and
  distributions in NNLO QCD}},
  \href{http://dx.doi.org/10.1007/JHEP08(2016)140}{\emph{JHEP} {\bfseries 08}
  (2016) 140}, [\href{https://arxiv.org/abs/1605.02716}{{\ttfamily
  1605.02716}}].

\bibitem{Grazzini:2017mhc}
M.~Grazzini, S.~Kallweit and M.~Wiesemann, \emph{{Fully differential NNLO
  computations with MATRIX}},
  \href{http://dx.doi.org/10.1140/epjc/s10052-018-5771-7}{\emph{Eur. Phys. J.}
  {\bfseries C78} (2018) 537},
  [\href{https://arxiv.org/abs/1711.06631}{{\ttfamily 1711.06631}}].

\bibitem{Monni:2016ktx}
P.~F. Monni, E.~Re and P.~Torrielli, \emph{{Higgs Transverse-Momentum
  Resummation in Direct Space}},
  \href{http://dx.doi.org/10.1103/PhysRevLett.116.242001}{\emph{Phys. Rev.
  Lett.} {\bfseries 116} (2016) 242001},
  [\href{https://arxiv.org/abs/1604.02191}{{\ttfamily 1604.02191}}].

\bibitem{Bizon:2017rah}
W.~Bizon, P.~F. Monni, E.~Re, L.~Rottoli and P.~Torrielli,
  \emph{{Momentum-space resummation for transverse observables and the Higgs
  $p_\perp$ at N$^3$LL+NNLO}},
  \href{https://arxiv.org/abs/1705.09127}{{\ttfamily 1705.09127}}.

\bibitem{Frixione:2007vw}
S.~Frixione, P.~Nason and C.~Oleari, \emph{{Matching NLO QCD computations with
  Parton Shower simulations: the POWHEG method}},
  \href{http://dx.doi.org/10.1088/1126-6708/2007/11/070}{\emph{JHEP} {\bfseries
  11} (2007) 070}, [\href{https://arxiv.org/abs/0709.2092}{{\ttfamily
  0709.2092}}].

\bibitem{Alioli:2010xd}
S.~Alioli, P.~Nason, C.~Oleari and E.~Re, \emph{{A general framework for
  implementing NLO calculations in shower Monte Carlo programs: the POWHEG
  BOX}}, \href{http://dx.doi.org/10.1007/JHEP06(2010)043}{\emph{JHEP}
  {\bfseries 06} (2010) 043},
  [\href{https://arxiv.org/abs/1002.2581}{{\ttfamily 1002.2581}}].

\bibitem{Catani:2010pd}
S.~Catani and M.~Grazzini, \emph{{QCD transverse-momentum resummation in gluon
  fusion processes}},
  \href{http://dx.doi.org/10.1016/j.nuclphysb.2010.12.007}{\emph{Nucl. Phys.}
  {\bfseries B845} (2011) 297--323},
  [\href{https://arxiv.org/abs/1011.3918}{{\ttfamily 1011.3918}}].

\bibitem{Gatheral:1983cz}
J.~G.~M. Gatheral, \emph{{Exponentiation of Eikonal Cross-sections in
  Nonabelian Gauge Theories}},
  \href{http://dx.doi.org/10.1016/0370-2693(83)90112-0}{\emph{Phys. Lett.}
  {\bfseries 133B} (1983) 90--94}.

\bibitem{Frenkel:1984pz}
J.~Frenkel and J.~C. Taylor, \emph{{Nonabelian Eikonal Exponentiation}},
  \href{http://dx.doi.org/10.1016/0550-3213(84)90294-3}{\emph{Nucl. Phys.}
  {\bfseries B246} (1984) 231--245}.

\bibitem{Frixione:1995ms}
S.~Frixione, Z.~Kunszt and A.~Signer, \emph{{Three jet cross-sections to
  next-to-leading order}},
  \href{http://dx.doi.org/10.1016/0550-3213(96)00110-1}{\emph{Nucl. Phys.}
  {\bfseries B467} (1996) 399--442},
  [\href{https://arxiv.org/abs/hep-ph/9512328}{{\ttfamily hep-ph/9512328}}].

\bibitem{Parisi:1979se}
G.~Parisi and R.~Petronzio, \emph{{Small Transverse Momentum Distributions in
  Hard Processes}},
  \href{http://dx.doi.org/10.1016/0550-3213(79)90040-3}{\emph{Nucl. Phys.}
  {\bfseries B154} (1979) 427--440}.

\bibitem{Dasgupta:2018nvj}
M.~Dasgupta, F.~A. Dreyer, K.~Hamilton, P.~F. Monni and G.~P. Salam,
  \emph{{Logarithmic accuracy of parton showers: a fixed-order study}},
  \href{http://dx.doi.org/10.1007/JHEP09(2018)033}{\emph{JHEP} {\bfseries 09}
  (2018) 033}, [\href{https://arxiv.org/abs/1805.09327}{{\ttfamily
  1805.09327}}].

\bibitem{Butterworth:2015oua}
J.~Butterworth et~al., \emph{{PDF4LHC recommendations for LHC Run II}},
  \href{http://dx.doi.org/10.1088/0954-3899/43/2/023001}{\emph{J. Phys.}
  {\bfseries G43} (2016) 023001},
  [\href{https://arxiv.org/abs/1510.03865}{{\ttfamily 1510.03865}}].

\bibitem{Ball:2014uwa}
{\scshape NNPDF} collaboration, R.~D. Ball et~al., \emph{{Parton distributions
  for the LHC Run II}},
  \href{http://dx.doi.org/10.1007/JHEP04(2015)040}{\emph{JHEP} {\bfseries 04}
  (2015) 040}, [\href{https://arxiv.org/abs/1410.8849}{{\ttfamily 1410.8849}}].

\bibitem{Alioli:2010qp}
S.~Alioli, P.~Nason, C.~Oleari and E.~Re, \emph{{Vector boson plus one jet
  production in POWHEG}},
  \href{http://dx.doi.org/10.1007/JHEP01(2011)095}{\emph{JHEP} {\bfseries 01}
  (2011) 095}, [\href{https://arxiv.org/abs/1009.5594}{{\ttfamily 1009.5594}}].

\bibitem{Campbell:2012am}
J.~M. Campbell, R.~K. Ellis, R.~Frederix, P.~Nason, C.~Oleari and C.~Williams,
  \emph{{NLO Higgs Boson Production Plus One and Two Jets Using the POWHEG BOX,
  MadGraph4 and MCFM}},
  \href{http://dx.doi.org/10.1007/JHEP07(2012)092}{\emph{JHEP} {\bfseries 07}
  (2012) 092}, [\href{https://arxiv.org/abs/1202.5475}{{\ttfamily 1202.5475}}].

\bibitem{Buckley:2014ana}
A.~Buckley, J.~Ferrando, S.~Lloyd, K.~Nordström, B.~Page, M.~Rüfenacht,
  M.~Schönherr and G.~Watt, \emph{{LHAPDF6: parton density access in the LHC
  precision era}},
  \href{http://dx.doi.org/10.1140/epjc/s10052-015-3318-8}{\emph{Eur. Phys. J.}
  {\bfseries C75} (2015) 132},
  [\href{https://arxiv.org/abs/1412.7420}{{\ttfamily 1412.7420}}].

\bibitem{Salam:2008qg}
G.~P. Salam and J.~Rojo, \emph{{A Higher Order Perturbative Parton Evolution
  Toolkit (HOPPET)}},
  \href{http://dx.doi.org/10.1016/j.cpc.2008.08.010}{\emph{Comput. Phys.
  Commun.} {\bfseries 180} (2009) 120--156},
  [\href{https://arxiv.org/abs/0804.3755}{{\ttfamily 0804.3755}}].

\bibitem{Gehrmann:2001pz}
T.~Gehrmann and E.~Remiddi, \emph{{Numerical evaluation of harmonic
  polylogarithms}},
  \href{http://dx.doi.org/10.1016/S0010-4655(01)00411-8}{\emph{Comput. Phys.
  Commun.} {\bfseries 141} (2001) 296--312},
  [\href{https://arxiv.org/abs/hep-ph/0107173}{{\ttfamily hep-ph/0107173}}].

\bibitem{Harlander:2002wh}
R.~V. Harlander and W.~B. Kilgore, \emph{{Next-to-next-to-leading order Higgs
  production at hadron colliders}},
  \href{http://dx.doi.org/10.1103/PhysRevLett.88.201801}{\emph{Phys.Rev.Lett.}
  {\bfseries 88} (2002) 201801},
  [\href{https://arxiv.org/abs/hep-ph/0201206}{{\ttfamily hep-ph/0201206}}].

\bibitem{Anastasiou:2002yz}
C.~Anastasiou and K.~Melnikov, \emph{{Higgs boson production at hadron
  colliders in NNLO QCD}},
  \href{http://dx.doi.org/10.1016/S0550-3213(02)00837-4}{\emph{Nucl.Phys.}
  {\bfseries B646} (2002) 220--256},
  [\href{https://arxiv.org/abs/hep-ph/0207004}{{\ttfamily hep-ph/0207004}}].

\bibitem{Ravindran:2003um}
V.~Ravindran, J.~Smith and W.~L. van Neerven, \emph{{NNLO corrections to the
  total cross-section for Higgs boson production in hadron hadron collisions}},
  \href{http://dx.doi.org/10.1016/S0550-3213(03)00457-7}{\emph{Nucl.Phys.}
  {\bfseries B665} (2003) 325--366},
  [\href{https://arxiv.org/abs/hep-ph/0302135}{{\ttfamily hep-ph/0302135}}].

\bibitem{Ravindran:2002dc}
V.~Ravindran, J.~Smith and W.~Van~Neerven, \emph{{Next-to-leading order QCD
  corrections to differential distributions of Higgs boson production in
  hadron-hadron collisions}},
  \href{http://dx.doi.org/10.1016/S0550-3213(02)00333-4}{\emph{Nucl. Phys.}
  {\bfseries B634} (2002) 247--290},
  [\href{https://arxiv.org/abs/hep-ph/0201114}{{\ttfamily hep-ph/0201114}}].

\bibitem{Hamberg:1990np}
R.~Hamberg, W.~L. van Neerven and T.~Matsuura, \emph{{A complete calculation of
  the order $\alpha_{s}^{2}$ correction to the Drell-Yan $K$ factor}},
  \href{http://dx.doi.org/10.1016/S0550-3213(02)00814-3,
  10.1016/0550-3213(91)90064-5}{\emph{Nucl. Phys.} {\bfseries B359} (1991)
  343--405}.

\bibitem{vanNeerven:1991gh}
W.~L. van Neerven and E.~B. Zijlstra, \emph{{The $O(\alpha_s^2)$ corrected
  Drell-Yan $K$ factor in the DIS and MS scheme}},
  \href{http://dx.doi.org/10.1016/j.nuclphysb.2003.12.019,
  10.1016/0550-3213(92)90078-P}{\emph{Nucl. Phys.} {\bfseries B382} (1992)
  11--62}.

\bibitem{Anastasiou:2003yy}
C.~Anastasiou, L.~J. Dixon, K.~Melnikov and F.~Petriello, \emph{{Dilepton
  rapidity distribution in the Drell-Yan process at NNLO in QCD}},
  \href{http://dx.doi.org/10.1103/PhysRevLett.91.182002}{\emph{Phys. Rev.
  Lett.} {\bfseries 91} (2003) 182002},
  [\href{https://arxiv.org/abs/hep-ph/0306192}{{\ttfamily hep-ph/0306192}}].

\bibitem{Melnikov:2006kv}
K.~Melnikov and F.~Petriello, \emph{{Electroweak gauge boson production at
  hadron colliders through O($\alpha_{s}^{2}$)}},
  \href{http://dx.doi.org/10.1103/PhysRevD.74.114017}{\emph{Phys. Rev.}
  {\bfseries D74} (2006) 114017},
  [\href{https://arxiv.org/abs/hep-ph/0609070}{{\ttfamily hep-ph/0609070}}].

\bibitem{Anastasiou:2003ds}
C.~Anastasiou, L.~J. Dixon, K.~Melnikov and F.~Petriello, \emph{{High precision
  QCD at hadron colliders: Electroweak gauge boson rapidity distributions at
  NNLO}}, \href{http://dx.doi.org/10.1103/PhysRevD.69.094008}{\emph{Phys. Rev.}
  {\bfseries D69} (2004) 094008},
  [\href{https://arxiv.org/abs/hep-ph/0312266}{{\ttfamily hep-ph/0312266}}].

\bibitem{Grazzini:2008tf}
M.~Grazzini, \emph{{NNLO predictions for the Higgs boson signal in the H --->
  WW ---> lnu lnu and H ---> ZZ ---> 4l decay channels}},
  \href{http://dx.doi.org/10.1088/1126-6708/2008/02/043}{\emph{JHEP} {\bfseries
  02} (2008) 043}, [\href{https://arxiv.org/abs/0801.3232}{{\ttfamily
  0801.3232}}].

\bibitem{Catani:2009sm}
S.~Catani, L.~Cieri, G.~Ferrera, D.~de~Florian and M.~Grazzini, \emph{{Vector
  boson production at hadron colliders: a fully exclusive QCD calculation at
  NNLO}}, \href{http://dx.doi.org/10.1103/PhysRevLett.103.082001}{\emph{Phys.
  Rev. Lett.} {\bfseries 103} (2009) 082001},
  [\href{https://arxiv.org/abs/0903.2120}{{\ttfamily 0903.2120}}].

\bibitem{Sjostrand:2014zea}
T.~Sjöstrand, S.~Ask, J.~R. Christiansen, R.~Corke, N.~Desai, P.~Ilten,
  S.~Mrenna, S.~Prestel, C.~O. Rasmussen and P.~Z. Skands, \emph{{An
  Introduction to PYTHIA 8.2}},
  \href{http://dx.doi.org/10.1016/j.cpc.2015.01.024}{\emph{Comput. Phys.
  Commun.} {\bfseries 191} (2015) 159--177},
  [\href{https://arxiv.org/abs/1410.3012}{{\ttfamily 1410.3012}}].

\bibitem{Banfi:2012jm}
A.~Banfi, P.~F. Monni, G.~P. Salam and G.~Zanderighi, \emph{{Higgs and Z-boson
  production with a jet veto}},
  \href{http://dx.doi.org/10.1103/PhysRevLett.109.202001}{\emph{Phys. Rev.
  Lett.} {\bfseries 109} (2012) 202001},
  [\href{https://arxiv.org/abs/1206.4998}{{\ttfamily 1206.4998}}].

\bibitem{Cacciari:2008gp}
M.~Cacciari, G.~P. Salam and G.~Soyez, \emph{{The anti-$k_t$ jet clustering
  algorithm}},
  \href{http://dx.doi.org/10.1088/1126-6708/2008/04/063}{\emph{JHEP} {\bfseries
  04} (2008) 063}, [\href{https://arxiv.org/abs/0802.1189}{{\ttfamily
  0802.1189}}].

\bibitem{Nason:2006hfa}
P.~Nason and G.~Ridolfi, \emph{{A Positive-weight next-to-leading-order Monte
  Carlo for Z pair hadroproduction}},
  \href{http://dx.doi.org/10.1088/1126-6708/2006/08/077}{\emph{JHEP} {\bfseries
  08} (2006) 077}, [\href{https://arxiv.org/abs/hep-ph/0606275}{{\ttfamily
  hep-ph/0606275}}].

\bibitem{deFlorian:2001zd}
D.~de~Florian and M.~Grazzini, \emph{{The Structure of large logarithmic
  corrections at small transverse momentum in hadronic collisions}},
  \href{http://dx.doi.org/10.1016/S0550-3213(01)00460-6}{\emph{Nucl. Phys.}
  {\bfseries B616} (2001) 247--285},
  [\href{https://arxiv.org/abs/hep-ph/0108273}{{\ttfamily hep-ph/0108273}}].

\bibitem{Becher:2010tm}
T.~Becher and M.~Neubert, \emph{{{Drell-Yan} Production at Small $q_T$,
  Transverse Parton Distributions and the Collinear Anomaly}},
  \href{http://dx.doi.org/10.1140/epjc/s10052-011-1665-7}{\emph{Eur. Phys. J.}
  {\bfseries C71} (2011) 1665},
  [\href{https://arxiv.org/abs/1007.4005}{{\ttfamily 1007.4005}}].

\bibitem{Davies:1984hs}
C.~T.~H. Davies and W.~J. Stirling, \emph{{Nonleading Corrections to the
  Drell-Yan Cross-Section at Small Transverse Momentum}},
  \href{http://dx.doi.org/10.1016/0550-3213(84)90316-X}{\emph{Nucl. Phys.}
  {\bfseries B244} (1984) 337--348}.

\bibitem{Catani:2011kr}
S.~Catani and M.~Grazzini, \emph{{Higgs Boson Production at Hadron Colliders:
  Hard-Collinear Coefficients at the NNLO}},
  \href{http://dx.doi.org/10.1140/epjc/s10052-012-2013-2,
  10.1140/epjc/s10052-012-2132-9}{\emph{Eur. Phys. J.} {\bfseries C72} (2012)
  2013}, [\href{https://arxiv.org/abs/1106.4652}{{\ttfamily 1106.4652}}].

\bibitem{Gehrmann:2014yya}
T.~Gehrmann, T.~Luebbert and L.~L. Yang, \emph{{Calculation of the transverse
  parton distribution functions at next-to-next-to-leading order}},
  \href{http://dx.doi.org/10.1007/JHEP06(2014)155}{\emph{JHEP} {\bfseries 06}
  (2014) 155}, [\href{https://arxiv.org/abs/1403.6451}{{\ttfamily 1403.6451}}].

\bibitem{Echevarria:2016scs}
M.~G. Echevarria, I.~Scimemi and A.~Vladimirov, \emph{{Unpolarized Transverse
  Momentum Dependent Parton Distribution and Fragmentation Functions at
  next-to-next-to-leading order}},
  \href{http://dx.doi.org/10.1007/JHEP09(2016)004}{\emph{JHEP} {\bfseries 09}
  (2016) 004}, [\href{https://arxiv.org/abs/1604.07869}{{\ttfamily
  1604.07869}}].

\bibitem{Catani:2012qa}
S.~Catani, L.~Cieri, D.~de~Florian, G.~Ferrera and M.~Grazzini, \emph{{Vector
  boson production at hadron colliders: hard-collinear coefficients at the
  NNLO}}, \href{http://dx.doi.org/10.1140/epjc/s10052-012-2195-7}{\emph{Eur.
  Phys. J.} {\bfseries C72} (2012) 2195},
  [\href{https://arxiv.org/abs/1209.0158}{{\ttfamily 1209.0158}}].

\bibitem{Bozzi:2005wk}
G.~Bozzi, S.~Catani, D.~de~Florian and M.~Grazzini, \emph{{Transverse-momentum
  resummation and the spectrum of the Higgs boson at the LHC}},
  \href{http://dx.doi.org/10.1016/j.nuclphysb.2005.12.022}{\emph{Nucl. Phys.}
  {\bfseries B737} (2006) 73--120},
  [\href{https://arxiv.org/abs/hep-ph/0508068}{{\ttfamily hep-ph/0508068}}].

\end{thebibliography}\endgroup
\bibliographystyle{JHEP}

\end{document}